\documentclass[a4paper,fleqn,usenatbib]{mnras_arXiv}

\usepackage{mathptmx}
\usepackage[T1]{fontenc}
\usepackage{ae,aecompl}

\usepackage{graphicx}	% Including figure files
\usepackage{amsmath}	% Advanced maths commands
\usepackage{amssymb}	% Extra maths symbols
\usepackage{mathtools}
\usepackage{tikz}
\newcommand{\ii}{\mathrm{i}}
\DeclareMathOperator{\e}{e}

\DeclareMathOperator{\st}{ \quad s. t. \quad}

\title[RV Data Analysis with Compressed Sensing]{Radial Velocity Data Analysis with Compressed Sensing Techniques}

\author[N. C. Hara et al.]{
Nathan C. Hara,$^{1}$\thanks{E-mail:nathan.hara@obspm.fr}
G. Bou\'e,$^{1}$
J. Laskar$^{1}$ and 
A. C. M Correia$^{2,1}$
\\
$^{1}$ ASD/IMCCE, CNRS-UMR8028, Observatoire de Paris,  PSL, UPMC, 77 Avenue Denfert-Rochereau, 75014 Paris, France\\
$^{2}$ CIDMA, Departamento de F\'isica, Universidade de Aveiro, Campus de Santiago, 3810-193 Aveiro, Portugal\\
}

\pubyear{2016}

\begin{document}
\label{firstpage}
\pagerange{\pageref{firstpage}--\pageref{lastpage}}
\maketitle

\begin{abstract}
We present a novel approach for analysing radial velocity data that combines two features: all the planets are searched at once and the algorithm is fast.
This is achieved by utilizing compressed sensing techniques, which are modified to be compatible with the Gaussian processes framework. The resulting tool can be used like a Lomb-Scargle periodogram and has the same aspect but with much fewer peaks due to aliasing.
The method is applied to five systems with published radial velocity data sets: HD 69830, HD 10180, 55 Cnc, GJ 876 and a simulated very active star. The results are fully compatible with previous analysis, though obtained more straightforwardly. We further show that 55 Cnc e and f could have been respectively detected and suspected in early measurements from the Lick observatory and Hobby-Eberly Telescope available in 2004, and that frequencies due to dynamical interactions in GJ 876 can be seen.
\end{abstract}


%%%%%%%%%%%%%%%%%%%%%%%%%%%%%%%%%%%%%%%%%%%%%%%%%%

%%%%%%%%%%%%%%%%% BODY OF PAPER %%%%%%%%%%%%%%%%%%

\section{Introduction}
\subsection{Overview}

Determining the content of radial velocity data is a challenging task. There might be several companions to the star, unpredictable instrumental effects as well as astrophysical jitter. Fitting separately the different features of the model might distort the residual and prevent from finding small planets, as pointed out for instance by~\cite{angladaescude2010,tuomi2012}. There might even be cases where, due to aliasing and noise, the tallest peak of the periodogram is a spurious one while being statistically significant. To overcome those issues, recent approaches privilege the fitting of the whole model at once. 
In those cases, the usual framework is the maximization of an \textit{a posteriori} probability distribution. In order to avoid being trapped in a suboptimal solution, random searches such as Monte Carlo Markov Chain (MCMC) methods or genetic algorithm are used~\citep[e.g.][]{gregory2011,segransan2011}. The goal of this paper is to suggest an alternative method using convex optimization, therefore offering a unique minimum and faster algorithms.

To do so, we will not try to find directly the orbital parameters of the planets but to unveil the true spectrum of the underlying continuous signal, which is equivalent. The power spectrum is often estimated with a Lomb-Scargle periodogram~\citep{ lomb1976,scargle} or generalizations~\citep{ferrazmello1981,cumming1999,zechmeister}. However, as said above the estimation of the power spectrum with one frequency at a time has severe drawbacks. To improve the estimate, we introduce an \textit{a priori} information: the representation of exoplanetary signal in the Fourier domain is sparse. In other words, the number of sine functions needed to represent the signal is small compared to the number of observations.  The Keplerian models are not the only ones to verify this assumptions, stable planetary systems are quasi-periodic as well~\citep[e.g.][]{laskar1993}. By doing so, the periodogram can be efficiently cleaned (see figures~\ref{hd69830},\ref{hd10180},\ref{rvsurvey55cnc},\ref{rvsurvey_gj876},\ref{rvsurvey_challenge}).
 
 The field of signal processing devoted to the study of sparse signals is often referred to as  ``Compressed Sensing'' or ``Compressive Sampling''~\citep{donoho2006_2, candes2006} -- though it is sometimes restricted to sampling strategies based on sparsity of the signal. The related methods show very good performances and are backed up by solid theoretical results. For instance, Compressed Sensing techniques allow to recover exactly a spectrum while sampling it at a much lower rate than the Nyquist frequency~\citep{mishali2008, tropp2009}. Its use was advocated to improve the scientific data transmission in space-based astronomy~\citep{bobin2008}. Sparse recovery techniques are also used in image processing~\cite[e.g.][]{starck2005}.

It seems relevant to add to that list a few techniques developed by astronomers to retrieve harmonics in a signal. In the next section, we show that even though the term ``sparsity" is not explicitly used~\citep[except in][]{bourguignon2007}, some of the existing techniques have an equivalent in the Compressed Sensing literature. After those remarks on our framework, the paper is organized as follows: in section~\ref{methods}, the theoretical background and the associated algorithms are presented. Section~\ref{implementation} presents in detail the procedure we developed for analysing radial velocity data. This one is applied section~\ref{results} to simulated observations and four real radial velocity data sets: HD 69830, HD 10180, 55 Cnc and GJ 876 and to a simulated very active star. The performance of the method is discussed section~\ref{discussion} and conclusions are drawn section~\ref{conclusion}.

\subsection{Previous work}
\label{previouswork}

The goal of this paper is to devise a method to efficiently analyse radial velocity data. As it builds upon the retrieval of harmonics, the discussion will focus on spectral synthesis of unevenly sampled data~\citep[see][for surveys]{kay1981,schwarzenberg1998,babustoica2010}.

First let us consider the methods that are efficient to spot one harmonic at a time. The first statistical analysis is given by~\cite{schuster1898}. However, the statistical properties of Schuster's periodogram only hold when the measurements are equispaced in time.
When this is not the case, one can use Lomb-Scargle periodogram~\citep{lomb1976,scargle} or its generalisation consisting in adding a constant to the model~\citep{ ferrazmello1981,cumming1999,reegen2007, zechmeister}. 
More recently,~\cite{mortier2015} derived a Bayesian periodogram associated to the maximum of an \textit{a posteriori} distribution. 
Also, \cite{cumming2004} and~\cite{otoole2009} define the Keplerian periodogram, which measures the $\chi^2$ of residuals after the fit of a Keplerian curve.  
One can remark that ``Keplerian'' vectors defined by $P,e,\omega$ and $M_0$ 
form a family of vectors in which the sparsity of exoplanetary signals is enhanced.

These methods can be applied iteratively to retrieve several harmonics. In the context of radial velocity data processing, one searches for the peak of maximum power, then the corresponding signal is subtracted and the search is performed again. This procedure is  very close to CLEAN~\citep{roberts1987}, which relies on the same principle of maximum correlation and subtraction. One of the first general algorithm exploiting sparsity of a signal in a given set of vectors~\citep[Matching Pursuit,][]{mallatzhang1993} relies on the same iterative process. This method was formerly known as Forward Stepwise Regression~\citep[e.g.][]{bellman1975}. To limit the effects of error propagation in the residuals, one can use the Orthogonal Matching Pursuit algorithm~\citep{pati1993,troppgilbert2007}. In that case, when an harmonic is found to have maximum correlation with the residuals, it is not directly subtracted. The next residual is computed as the original signal minus the fit of all the frequencies found so far. The CLEANest algorithm~\citep{foster1995}, and  Frequency Map Analysis~\citep{laskar1988,laskar1992,laskar1993,laskar2003}, though developped earlier, are particular cases of this algorithm. To analyse radial velocity data, \cite{baluev2009} and \cite{angladaescude2012} introduce what they call respectively the ``residual periodogram'' and the ``recursive periodogram'', which can be seen as pushing that logic one step further. The principle is to re-fit at each trial frequency the previous Keplerian signals plus a sine at the considered frequency.

Besides the matching pursuit procedures, there are two other popular algorithms in the Compressed Sensing literature: convex relaxations~\citep[e.g.][]{tibshirani1994,chen1998, starck2005} and iteratively re-weighted least squares (IRWLS)~\citep[e.g.][]{focuss,donoho2006_2,candesromberg2006,daubechies2010}. In the context of astronomy, \cite{bourguignon2007} implements a convex relaxation method using $\ell_1$ norm weighting (see equation~\eqref{lpnorm}) to find periodicity in unevenly sampled signals and \cite{babu2010} presents an IRWLS algorithm named IAA to analyse radial velocity.
 
The methods presented above are apparently very different, yet they can be viewed as a way to bypass the brute force minimization of
\begin{equation}
\arg \min \limits_{K,\omega,\phi} \quad \sum\limits_{i=1}^{m} \left(y(t_i)-\sum\limits_{j=1}^k K_j\cos (\omega_j t_i+\phi_j) \right)^2
\label{p1}
\end{equation} 
 where $y(t)$ is a vector made of $m$ measurements, and $x^\star = \arg \min f(x)$ denotes the element such that $f(x^\star)=\min f(x)$ for a function $f$. This problem is very similar to ``best $k$-term approximation'', and its link to compressed sensing has been studied in~\cite{cohen2009} in the noise-free case. Solving that problem is suggested by~\cite{baluev2013} under the name of ``multi-frequency periodograms''. However, finding that minimum by discretizing the values of $(K_j,\omega_j,\phi_j)_{j=1..k}$ depends exponentially on the number of parameters, and the multi-frequency periodograms could hardly handle more than three or four sines with conventional methods. However, with parallel progamming on GPUs one can handle up to $\approx$25 frequencies depending on the number of measurements~\citep{baluev2013_freqdecomposer}. \cite{jenkins2014} explicitly mentions the above problem and suggests a tree-like algorithm to explore the frequency space. They analyse GJ 876 with their procedure and find six significant harmonics, which we confirm section~\ref{gj876_sixsines}.  

Let us mention that searching for a few sources of periodicity in a signal is not always done with the Fourier space. When the shape of the repeating signal or the noise structure are not well known, other tests might be more robust. A large part of those methods consists in computing the autocorrelation function or folding the data at a certain period and look for correlation. See \cite{engelbrecht2013} for a survey or~\cite{zucker2015,zucker2016} in the context of radial velocity measurements. 
Finally, we point out that the use of the sparsity of the signal is not specific to Compressed Sensing. The number of planets in a model is often selected via likelihood ratio tests. A model with an additional planet must yield a significant improvement of the evidence. In general the model with $k+1$ planets $\mathcal{M}_{k+1}$ is selected over a model with $k$ planet if $\rm{Pr}\{y(t) | \mathcal{M}_{k+1}\}/\rm{Pr}\{y(t)|\mathcal{M}_{k}\}$ is greater than 150~\citep[see][]{tuomi2014}, $y(t)$ being the observations. Indeed, adding more parameters to the model automatically decreases the $\chi^2$ of the residuals. Putting a minimum improvement of the $\chi^2$ acts against overly complicated models.

The discussion above points that searching planets one after another is already in the compressed sensing paradigm: this iterative procedure is close to the orthogonal matching pursuit algorithm. 
\cite{donoho2006} shows that for a wide range of signals, this algorithm is outperformed by $\ell_1$ relaxation methods. 
Does this claim still applies to radial velocity signals ?  In this paper, this question is not treated in full generality, but we show the interest of $\ell_1$ relaxation on several examples. To address that question more directly, it is shown appendix~\ref{appendix_wrongpeak} that in some cases, the tallest peak of the periodogram is spurious but $\ell_1$ minimization prevents from being mislead.

%The progress made in astronomy and compressed sensing show a parallel trend. We wish to take it a step further and tailor the tools of compressed sensing for analysing radial velocity measurements.

\section{Methods}
\label{methods}
\subsection{Minimization problem}
\label{generalprinciples}

Techniques based on sparsity are thought to enforce the "Occam's razor" principle: the simplest explanation is the best. To apply that principle we must have an idea of ``how'' the signal is simple. In the compressed sensing framework (or compressive sampling), this is done by selecting a set of vectors $\mathcal{A} =(a_j(t))_{j \in I}$ such that the signal to be analysed $y(t)$ is represented by a linear combination of a few elements of $\mathcal{A}$. Such a set is often called the ''dictionary'' and can be finite or not (the set of indices $I$ can be finite or infinite). It is here made of vectors $a_{-\omega}(t) = \e^{-\ii\omega t }$ and $a_{\omega}(t) =\e^{\ii \omega t }$ where $t$ is the array of measurement times. 
%Ideally one would like to fit any finite set of vectors of the dictionary to the data, and select the best compromise between the fit residuals and the sparsity of the solution. This c

Before going into the details, let us define some quantities.
\begin{itemize}
\item $y(t)$ denotes the vector of observations at times $t=t_1...t_m$, $ y(t) \in \mathbb{R}^m$ for radial velocity data sets.
\item The $\ell_p$ norm of a complex or real vector $x$ with $n$ components is defined as 
\begin{align}
\label{lpnorm}
\|x\|_{\ell_p} := \left( \sum\limits_{k=1}^n |x_k|^p \right)^{1/p}
\end{align} 
for $p>0$. In particular  $\|x \|_{\ell_1}$ is the sum of absolute values of the vector components and $\|x \|_{\ell_2} = \sqrt{\sum\limits_{k=1}^n {|x_k|}^2}$ is the usual Euclidian norm. When $p=0$, $\|x \|_{\ell_0}$ is the number of non-zero components of $x$.
\item For a function $f$ defined on a set $E$, $\arg \min\limits_{x\in E} f(x)$ is the element for which the minimum is attained, that is $x^\star$ of $E$ such that $f(x^{\star}) =\min\limits_{ x \in E} f(x)$. We denote by the superscript $\star$ the solution of the minimization problem under consideration. In all the cases considered here except~\eqref{p1} and~\eqref{l0}, the minimum is attained as we consider convex functions on convex sets.
\end{itemize} 
 
Let us consider combinations of $S$ elements of the dictionary $(a_j(t))_{j=1..S}$ and their corresponding amplitudes $x_j$. To enhance the sparsity of the representation, one can think of solving 
\begin{align}
\label{l0}
\arg \min\limits_{\mathclap{\scriptsize \begin{array}{c}
a_j(t) \in \mathcal{A} \\ S \in \mathbb{C} \end{array} \normalsize}} \; \; \; S \quad \mathrm{s.t.} \quad  \left\| \sum_{j=1}^S x_j a_j(t) -y(t) \right\|_{\ell_2} \leqslant \epsilon
\end{align}
that is finding the smallest number of elements of $\mathcal{A}$ required to approximate $y(t)$ with a certain tolerance $\epsilon$. This one is \textit{a priori} a combinatorial problem which seems unsolvable if $\mathcal{A}$ is infinite or of an exponential complexity if the dictionary is finite. In the latter case $\mathcal{A}$ can be viewed as an $m \times n$ matrix $A$. In that case, on can re-write equation~\eqref{l0} like:
\begin{align}
x^\star =\arg \min_{ x \in \mathbb{C}^n \ } \|x\|_{\ell_0} \st  \left\| Ax-y(t)\right\|_{\ell_2} \leqslant \epsilon.
\end{align}
 This problem is in general combinatorial~\citep{ge2011}, therefore computationally intractable. Fortunately, when replacing the $\ell_0$ norm  by the $\ell_1$ norm,
\begin{align}
\label{l1}
x^\star =\arg \min_{ x \in \mathbb{C}^n \ } \|x\|_{\ell_1} \st  \left\| Ax-y(t)\right\|_{\ell_2} \leqslant \epsilon
\end{align}
the problem becomes convex and still enhances sparsity efficiently. In the signal processing litterature,  this problem is referred to as Basis Pursuit Denoising~\citep{chen1998}, and is sometimes denoted by $\mathrm{BP}_{\epsilon}$. At this point one can ask what is lost by considering~\eqref{l1} instead of~\eqref{l0}. Let us cite a few results -- among many: when $y(t)$ is noise free, \cite{donoho2006_2} shows that under certain hypotheses the solution to~\eqref{l1} is equal to the solution of~\eqref{l0}; more generally, denoting by $y_t=Ax_t$ the true signal, such that $y=y_t + e$, $e$ being the error, there is a theoretical bound on $\|Ax^\star-y_t \|_{\ell_2} $~\citep{candes2006}. One can also obtain constraints on $\|x^\star- x_t\|_{\ell_2}$ or conditions to have $\mathrm{supp}(x^\star) \subset \mathrm{supp}(x_t)_{\ell_2}$ where $\mathrm{supp}(x)$ is the set of indices where $x$ is non-zero~\citep[e.g.][]{donoho2006}. In summary, there are results guaranteeing the performance for de-noising, compression and also for inverse problems, the search for planets being a particular case of the latter.
%This latter type of results, named ``support recovery'' is particularly of interest in our case, as it may guarantee that the signals

These results apply to a finite dictionary $\mathcal{A}$, but the periods of the planets could be anywhere:  $\mathcal{A}$ is infinite for our purposes.
We will eventually go back to solving a modified version of the discrete problem~\eqref{l1} and smooth its solution with a moving average. Beforehand, we will present next section what seems to be the most relevant theoretical background for our studies, ``atomic norm minimization'', in particular used in ``super-resolution theory''~\citep{candesfernandez2012}. This one will give guidelines to improve our procedure.

\subsection{Atomic norms minimization}

If $\mathcal{A}$ is infinite, the $\ell_1$ norm cannot be used straightforwardly. \cite{chandrasekharan2010} suggests to use an ``atomic norm'' that extends~\eqref{l1} to infinite dictionaries.
Practical methods to solve the new minimization problem are designed in \cite{candesfernandez2012a} and~\cite{tangbhaskar2013}.
The atomic norm $\| y \|_{\mathcal{A}}$, of $y \in \mathbb{R}^m$ or $\mathbb{C}^m$  defined for a dictionary $\mathcal{A}$ is the smallest $\ell_1$ norm of a combination of vectors of the dictionary reproducing $y$: 
\begin{align}
\|y\|_{\mathcal{A}} &= \inf \left\{ \sum\limits_j |x_j|, y = \sum\limits_j x_j  a_j(t) \right\}
\end{align}
If the observations were not noisy, computing the atomic norm of $y$ would be sufficient. As this is obviously not the case, the following problem is considered. 
\begin{equation}
 u^\star= \underset{u \in \mathbb{C}^m}{ \arg \min}\quad \|u-y(t)\|_{\ell_2}^2 + \lambda \|u\|_{\mathcal{A}}
  \label{ANDNlambda}
\end{equation}
where $\lambda$ is a positive real number fixed according to the noise. This problem is often referred to as Atomic Norm De-Noising. The coefficient $\lambda$ can be interpreted as a Lagrange multiplier, and this problem  can be seen as maximizing a posterior likelihood with a prior on $u$. The quantities we are interested in are the dictionary elements $a_j^{\star}$ and the coefficients $x^\star$ selected by the minimization, where $u^\star = \sum_{j=1}^{S^\star} x_j^{\star} a_j^\star(t)$.

 \subsection{More complex noise models}
 \label{complexnoisemodels}
If exoplanetary signals are arguably a sum of sines plus noise, the noise variance is not constant. Even more, the noise might not be independent nor Gaussian. Recent papers as~\cite{tuomi2013} or~\cite{rajpaul2015} stress that the detection efficiency and robustness improves as the noise model becomes more realistic. \cite{aigrain2011} suggests to consider the RV time series as Gaussian processes: the noise $n(t)$ is then characterized by its covariance matrix $V$ which is such that $V_{kl} = \mathbb{E}\{n(t_k)n(t_l) \}$, $\mathbb{E}$ being the mathematical expectancy. When the noise is stationary, by definition there exists a covariance function $R$ such that $V_{kl} = R(|t_l-t_k|)$, therefore choosing $V$ is equivalent to choosing $R$. This approach is similar to~\cite{sulis2016}, which normalizes the periodogram by the power spectrum of the stationary part of the stellar noise. The similarity comes from the fact that the power spectrum  of the noise is  $P(\omega)= |\mathcal{F}(R)|^2$ where $\mathcal{F}$ denotes the Fourier transform.

Here, the noise is assumed to be Gaussian of covariance matrix $V$. In that case, the logarithm of the likelihood is~(e.g. \cite{baluev2011} equation 21, \cite{pelat})
\small
\begin{align}
\ln(L) = 
-\frac{m}{2}\ln(2\pi)-\frac{1}{2} \rm{det}(V)-\frac{1}{2} (y-Ax)^TV^{-1}(y-Ax)
\end{align}
\normalsize
where the subscript $T$ denotes the matrix transpose.
Assuming the matrix $V$ is fixed, we wish to minimize $(y-Ax)^TV^{-1}(y-Ax)$. If $V^{-1}$ admits a square root, then $W$ is chosen such that $W^2 = V^{-1}$. This is the case when $V$ is symmetric positive definite, which is the case for covariance matrices of stationary processes. Consequently, $\|W(Ax-y) \|_{\ell_2}^2 = (y-Ax)^TV^{-1}(y-Ax)$ is always ensured for Gaussian noises. We then obtain the minimization:
\begin{equation}
  \underset{u \in \mathbb{C}^m}{ \arg \min}\quad \|W(u-y(t))\|_{\ell_2}^2 + \lambda \|u\|_{\mathcal{A}}.
  \label{ANDNwlambda}
\end{equation}
Handling problem~\eqref{l1} with correlated measurements and noise has been investigated by~\cite{arildsen2014}. However to the best of our knowledge the formulation above is not mentioned in the literature, thus we will briefly discuss its features. 

 The ability of problem~\eqref{l1} to unveil the true non zero coefficients of $x$ improves as the so-called mutual coherence of matrix $A$ diminishes~\citep{donoho2006}. This one is defined as the maximum correlation between two column-vectors of $A$. We here consider a weight matrix, but we can go back to the previous problem by noting that $W(Ax-y)$ can be re-written $A'x-y'$ where $A' = WA$ and $y'=Wy$. If we now consider two column vectors of $A'$, $a'_1 = Wa_1$ and $a'_2=Wa_2$, their correlation is $a_1'^T a_2' = a_1 W^T W a_2 = a_1 V^{-1} a_2$. In other words introducing a matrix $W$ only comes down to changing the scalar product. This should not be surprising. The matched filter technique~\citep{kay1993} proposes to detect a model $x$ in a signal $s=x+n$ where $n$ is a noise of covariance matrix $V$ if $xV^{-1}s\leqslant \gamma$ where $\gamma$ is a threshold. This means if the correlation is sufficient for a non trivial scalar product. 

  In the case of an independent Gaussian noise, its covariance matrix $V$ is diagonal and its elements are $\sigma_k^2$, where $\sigma_k$  is the measurement error at time $t_k$.  $W$ is defined as $V^{-1/2}$ so is a diagonal matrix of elements $w_{kk} = 1/\sigma_k$. Therefore,  $a_1'^T a'_2 = a_1 W^T W a_2 = \sum_{k=1}^n \frac{a_1(t_k) a_2(t_k)}{\sigma_k^2}$. This is compatible with the behaviour we intuitively expect: the less precise is the measurement, the lesser the correlation between the signals matter through the weighting by $\sigma_k$. 

 Unfortunately, having a non identically independent distributed (i.i.d) Gaussian noise model biases the estimates of the true signals as it acts as a frequency filter. Whether this bias prevents from having the benefits of a correct noise model is discussed in appendix~\ref{appendix_noise}. We show that choosing an appropriate weight matrix $W$ indeed allows to see signals that would be buried in the red noise otherwise. 
 
 \section{Implementation}
 \label{implementation}
\subsection{Overview}

As said above, stable planetary systems are quasi-periodic. This means in particular that radial velocity measurements are well approximated by a linear combination of a few vectors $\e^{-\ii \omega t}$ and $\e^{ \ii\omega t}$.
The minimization problem~\eqref{ANDNlambda} seems therefore well suited for searching for exoplanets. This section is concerned with the numerical resolution, and the numerous issues it raises: the numerical scheme to be used, the choice of the algorithm parameters and the evaluation of the confidence in a detection.

Solving~\eqref{ANDNlambda} is done either by reformulating it as a quadratic program~\citep{candesfernandez2012a,tangbhaskar2013,chen2013} or by discretizing the dictionary~\citep{tangbhaskar2013bis}. The first one necessitates to see the sampling  as a regularly spaced one with missing samples. As the measurement times are far from being equispaced in the considered applications, the required time discretization results in large matrices. Therefore, the second approach is used. 

Let us pick a set of frequencies equispaced with interval $\Delta \omega$, $\Omega = \{ \omega_k= k \Delta \omega, k=0..n \}$ and a $m\times 2n$ matrix $A$ whose columns are $\e^{-\ii\omega_k t}$ and $\e^{\ii \omega_k t}$. In that case~\eqref{ANDNwlambda} reduces to:
\begin{align}
\label{lassolambdaw}
  \underset{x \in \mathbb{C}^{2n}}{ \arg \min}\quad \|W(Ax-y)\|_{\ell_2}^2 + \lambda \|x\|_{\ell_1}
\end{align}
Which is often referred to as the LASSO problem when $W$ is the identity matrix.
As the parameter $\lambda$ is not so easy to tune, an equivalent formulation of discretized~\eqref{ANDNwlambda} is chosen, 
\begin{equation}
\label{BPDNepsilonw}
\tag{$11,\mathrm{BP}_{\epsilon,W}$}
\addtocounter{equation}{1}
  x^\star =\underset{x \in \mathbb{C}^{2n}}{ \arg \min} \quad \|x\|_{\ell_1} \quad \text{s. t.} \quad \|W(Ax-y)\|_{\ell_2}\leqslant \epsilon 
\end{equation}
where $\epsilon$ is a positive number. By ``equivalent'', we mean there exists a $\lambda_{\epsilon}$ such that the solution of~\eqref{lassolambdaw} is equal to the solution of \eqref{BPDNepsilonw}~\citep{rockafellar1970}. As this problem will often be referred to, we add to the equation number $\mathrm{BP}_{\epsilon,W}$ in the rest of the text, BP standing for Basis Pursuit.
There are several codes written to solve~\eqref{l1}. The existing codes we have tested for analysing radial velocity data sets are: $\ell_1$-magic~\citep{candesromberg2006}, SparseLab~\citep{donoho2006_2}, NESTA~\citep{becker2011}, CVX~\citep{grantboyd2008}, Spectral Compressive Sampling~\citep{duarte2013} and SPGL1~\citep{spgl1}. The latter gave the best results in general for exoplanetary data and consequently is the one we selected (the code can be downloaded from this~\href{https://www.math.ucdavis.edu/~mpf/spgl1/supplement.html}{link} \footnote{https://www.math.ucdavis.edu/$\sim$mpf/spgl1/supplement.html}).
%encadré SPGL1 ?
%The algorithm requires some tuning, such as the choice of the weighting matrix $W$, of the tolerance $\epsilon$, and the frequency grid $\Omega$. These choices are in fact interconnected and detailed section~\ref{tuning}. 

The solution of~\eqref{BPDNepsilonw} offers an estimate for the periods, but the efficiency of the method can be improved by using a moving average on $x^\star$, to approximate better~\eqref{ANDNwlambda}.
Indeed if a sine of frequency $\omega_0$ and amplitude $K$ is in the signal, corollary 1~\citep{tangbhaskar2013bis}  shows that the solution of~\eqref{l1} $x^\star$ verifies 
\begin{align}
K \quad \approx \quad \sum\limits_{\mathclap{\scriptsize \begin{array}{c}
|\omega_k| \in [\omega_0-\eta,\omega_0+\eta] \end{array}}} x^\star(\omega_k)
\end{align}
 rather than $|x(\omega_0)| \approx K $. The coefficients $x^\star(\omega_k)$ are added up for $\omega_k$ lying in a certain interval of length $2\eta$ (see section~\ref{postprocessing}).

Finally, the confidence in the detection must be estimated. Problem~\eqref{BPDNepsilonw} selects significant frequencies in the data, but the estimates of their amplitude is biased due to the $\ell_1$ norm minimization. To obtain unbiased amplitudes, we first check that the peaks are not aliases of each others. Then the most significant peaks are fitted until non significant residuals are obtained (see section~\ref{significance}).

In summary, the method follows a seven step process:
\begin{enumerate}
\item Pre-process the data: remove the mean in radial velocity data or an estimate of the stellar noise.
%These ones can be harmonic ($\cos{\omega t},\sin \omega t$) or Keplerian ($\frac{r}{a}\e^{i\nu( t)}$ where $\nu$ is the true anomaly).
\item Choose the discrete grid $\Omega$, tolerance $\epsilon$, weighting matrix $W$ and the width $\eta$ of the interval over which the result of~\eqref{BPDNepsilonw} is averaged.
\item Define the dictionary $\mathcal{A}$ and normalize the columns of $WA$. 
\item Run the program solving the convex optimization~\eqref{BPDNepsilonw} to obtain $x^\star$.
\item Denoting $\Omega = [\omega_{\mathrm{min}},\omega_{\mathrm{max}}]$ for each frequency  $\omega \in \{\omega_{\mathrm{min}}+\eta,...,\omega_{\mathrm{max}}-\eta \}$, sum up the amplitudes of $x^\star(\omega')$ from $\omega' \in [-\omega-\eta,-\omega+\eta] \cup [\omega-\eta,\omega+\eta] $ to obtain a smoothed figure $x^\sharp$.
\item Plot $x^\sharp$ as a function of the frequencies or the periods.
\item Evaluate the significance of the main peaks (figure~\ref{rvsurveyfap}).
\end{enumerate}
Each of these steps are detailed in the following sections.

\subsection{Optimization routine}
\label{optimizationroutine}
Many solvers can handle~\eqref{l1}, however, their precision and speed vary. Among the solvers tested, SPGL1~\citep{spgl1} gives the best results in general. 
 This one has several user-defined parameters such as a stopping criterion that must be tuned. For a given tolerance, this one is  $ \frac{|\|Ax - y\|_{\ell_2}- \epsilon|}{\max \left(1,\|Ax - y\|_{\ell_2}\right)}  < $ tol. The default parameters seem acceptable, in particular tol=$10^{-4}$. 
 
 \subsection{Dictionary $A$}
\label{dictionary}
To estimate the spectrum, a natural choice for the columns of matrix $A$ is  $(\e^{- \ii \omega t} , \e^{\ii \omega t}) $. However, the data might not contain only planetary signals. In the case of a binary star, a linear  trend $t$ and a quadratic term $t^2$ are added. If the star is active the ancillary measurements are also added. 
%As the signal is real-valued, $(\cos { \omega t}, \sin {\omega t})_{\omega \in \Omega}$ could have been a possible choice. However,

The method described in section~\ref{implementation} is applicable to a wider range of dictionary.  As the timespan of the observations is in general a few years, the signal might be more sparsely represented either by Poisson terms ($(a_0+a_1t+a_2t^2+...)\cos(\omega t + \phi)$) or Keplerian motions. In the latter case, column vectors would be of the form $\frac{r}{a}\e^{\ii \nu(t)}$ where $\nu(t)$ is a vector of true anomalies depending on the period $P$, eccentricity $e$ and initial mean anomaly $M_0$ (or any combination of three variables that cover all possible orbits). Unfortunately, the size of $A$ increases exponentially with the number of parameters describing the dictionary elements (here $P, e, M_0$).

\subsection{Pre-processing}
\label{preprocessing}
Theoretical results in~\cite{tangbhaskar2013bis} guarantee that the solution to~\eqref{l1} will be close to~\eqref{ANDNlambda} as the discretization gets finer, provided the dictionary is continuous. As linear trends or stellar activity related signals are not sine, removing these from the data before solving~\eqref{BPDNepsilonw} is crucial. The mean, a linear trend and estimates of the stellar noise can be fitted and removed. We reckon this is contrary to the philosophy of fitting the whole model at once. However, the vectors fitted are included again in the dictionary which allows to mitigate the distortions induced by their removal. 

Secondly, to make the precision of the SPGL1 solver independent from the value of $Wy$, the weighted observations $Wy$ are normed by $\|W y\|_{\ell_2}$, the columns of the matrix $WA$ are also normed. Denoting by $y' = \frac{1}{\epsilon} Wy/\|W y\|_{\ell_2}$ and $A'= \frac{1}{\epsilon}( WA_k / \|W A_k\|_{\ell_2})_{k=1..n}$, we set in input of the solver:
\begin{align}
\arg \min\limits_{x \in \mathbb{C}^n} \|x\|_{\ell_1} \st \left\| A'x-y' \right\|_{\ell_2} \leqslant 1,
\end{align}
to always be in the same kind of use of the solver and ensure the accuracy of the result does not depends on its units. Going back to the correct units in the post-processing step is described section~\ref{postprocessing}.

\subsection{Tuning}
\label{tuning}
\smallskip
\noindent
\textbf{Choice of $W$}: We have seen section~\ref{complexnoisemodels} the weight matrix $W$ is characterized by the covariance function $R$ via $W_{kl} = R(|t_k-t_l|)$.
Several forms for the covariance functions were suggested~\citep[e.g.][]{rajpaul2015}. Here we only consider exponential covariances, that are 
\begin{align}
\begin{split}
R(\Delta t) &= \sigma_R^2 \e^{-\frac{|\Delta t|}{\tau}}, \quad \Delta t \neq 0 \\
R(0) &= \sigma_W^2 +\sigma_R^2 
\end{split}
\label{corrnoise}
\end{align}
%Where $\sigma_R$ is the additional variance due to the red noise, $\tau$ a characteristic time and $\sigma_W$ an addit
where the subscripts $W$ and $R$ stand respectively for white and red.
As the red and white noise are here supposed independent, the covariance  function of their sum is the sum of their covariance functions.
Therefore, the matrix $W$ is such that its diagonal terms are $V_{kk}= \sigma_k^2 + \sigma_W^2 + \sigma_R^2$ and $V_{kl} = \sigma_R^2 \e^{-\frac{|t_k-t_l|}{\tau}}$ for $k \neq l$. 

\smallskip
\noindent
\textbf{Choice of $\Omega$}: We have two parameters  to choose: the grid span and the grid spacing. For the first one we take 1.5 cycles/day as a default value but it is also advisable to re-do the analysis for 0.95 cycles/day, as discussed in the examples sections~\ref{methods}.
 We ensure that if the signal is made of sinusoids (a.k.a. it is quasi-periodic), there exists at least one vector $x$ verifying $\|W(Ax-y)\|_{\ell_2}<\epsilon$ that has the correct $\ell_0$ norm. Let us consider a signal made of $p$ pure sinusoids sampled at times $t = (t_k)_{k=1..m}$, $y(t) = \sum\limits_{j=1}^p c_j \e^{\ii \omega_j t} $. Assuming the frequencies on the grid are regularly spaced with step $\Delta \omega$, this leads to the condition (see~\ref{mindeltaomega} for calculation details):
\begin{equation}
\label{omegabound}
\Delta \omega \leqslant \frac{4}{T} \arcsin \frac{\epsilon}{ 2\sqrt{\sum\limits_{j=1}^p |c_j|^2} \sqrt{\sum\limits_{k=1}^m   \frac{1}{\sigma_k^2}}  }.
\end{equation}
Let us note that the values of $c_j$ are \textit{a priori} unknown, so the term $\sqrt{\sum_{j=1}^p |c_j|^2}$ has to be approximated. Supposing the signal is made of sinusoids plus small noise, $\sqrt{\sum_{j=1}^p |c_j|^2} \approx \|y \|_{\ell_2}/\sqrt{m}$.
Furthermore, it must be ensured that all possible significant frequencies are in the signal. 

\smallskip
\noindent
The choice of the grid spacing can be based on other criteria: \cite{stoicababu2012} suggests to choose a spacing such that the ``practical rank" of matrix $M_{kl}=\e^{\ii \Delta \omega (t_k-t_l)}$ is equal to one. This term designates the number of singular values above a certain  threshold. Here the condition states that only one singular value is non negligible.
Let us also mention that one can perform the reconstruction with different grids and average out the results. However, this approach does not practically generate better results than using a finer grid.

\smallskip

\noindent
\textbf{Choice of $\epsilon$}: The error is due to two sources: grid discretization which gives an error $\epsilon_{\mathrm{grid}}$ and noise, which yields $\epsilon_{\mathrm{noise}}$. Supposing the noise is Gaussian, denoting by $y_t$ the underlying non noisy observations, $\|W(y_t-y)\|^2_{\ell_2}$ as a function of random variable $y=y_t+n$ follows a $\chi^2$ distribution with $m$ degrees of freedom. Denoting its cumulative distribution function (CDF) by $F_{\chi^2_m}$, the probability $1-\alpha$ that the true signal $y_t$ is in the set $\{ y', \|W(y'-y)\|^2_{\ell_2} \leq \epsilon_{\mathrm{noise}} \}$ is:

\begin{equation}
\label{eqchi2}
F_{\chi^2_m} (\epsilon_{\mathrm{noise}}^2) = 1-\alpha
\end{equation} 
The bound $\epsilon_{\mathrm{noise}}$ is determined according the equation above for a small $\alpha$.
Once $\epsilon_{\mathrm{noise}}$ is chosen, rearranging equation~\eqref{omegabound} gives a minimal value of $\epsilon_{\mathrm{grid}}$ that ensures a signal with a correct $\ell_0$ norm exists,

 \begin{equation}
\label{epsilonbound}
\epsilon_{\mathrm{grid}} = 2 \sqrt{\sum\limits_{j=1}^p |c_j|^2} \sqrt{\sum\limits_{k=1}^m   \frac{1}{\sigma_k^2}}  \sin \frac{\Delta \omega T_{\mathrm{obs}}}{4}.
\end{equation}
%\epsilon^c_{grid} =  2\sqrt{m}\| \hat{y_c} \|_{\ell_1} \sin \frac{\Delta \omega T_{\mathrm{obs}}}{4}

An alternative is to set $\epsilon$ to zero and let the algorithm find a representation for the noise, which will not be sparse. In that case one must obviously not perform the re-normalization of the columns of $WA$ by $\epsilon$ of section~\ref{preprocessing}. Below a certain amplitude, a ``forest'' of peaks would be seen on the $\ell_1$-periodogram. This has the advantage to give an estimation of the noise structure. However, this method is more sensitive to the solver inner uncertainties and requires more time, it was not retained for this work.

\smallskip

\noindent
\textbf{Choice of $\eta$}: See next section.

\subsection{Post-processing}
\label{postprocessing}
Once the solution to~\eqref{BPDNepsilonw} is computed, the spectrum $x^\star$ is filtered with a moving average.
We expect from discretization~\eqref{ANDNwlambda} that the frequencies might leak to close frequencies. Indeed, the amplitude of the solution to~\eqref{BPDNepsilonw} might be untrustworthy. When the signal is made of several frequencies, the solution might over-estimate the one with the greatest amplitude, and under-estimate the others; this problem arises especially when less than a hundred observations are available. To mitigate this effect, one can sum up the contribution of subsequent frequencies and estimate the amplitude of the resulting signal. If $x^\star$ is the solution to~\eqref{BPDNepsilonw}, denoting by $x^\star(\omega)$ the coefficient corresponding to frequency $\omega$, we compute
\begin{align} 
\label{haty}
\hat{y}_\omega(t)  = \|Wy \|_{\ell_2} \quad \sum\limits_{\mathclap{\footnotesize \begin{array}{c}\omega' \in \Omega \\
 \omega-\eta\leqslant |\omega'| \leqslant \omega+\eta\end{array}}} \quad \frac{x^\star(\omega') a_{\omega'}(t)}{\|Wa_{\omega'}(t)\|_{\ell_2}}
\end{align}
Where $a_{\omega'}(t)$ is the column of $A$ corresponding to frequency $\omega'$. The terms $\|Wy\|_{\ell_2}$ and $1/\|Wa_{\omega'}(t)\|_{\ell_2}$ appear because the columns of $WA$ and the weighted observations $Wy$ were normalized in step~\ref{preprocessing}.
The vector $\hat{y}_\omega(t)$, $t=t_1..t_m$  is approximately a sine function, the new estimation of the signal power is:
\begin{align}
\label{xsharp}
x^\sharp(\omega) = \max\limits_{t_1..t_m} |\hat{y}_\omega(t_k) |.
\end{align}
 Other estimates are possible, such as the power of a sine at frequency $\omega$ fitted on $\hat{y}(\omega)$. 
  Though the choice of $\eta$ is heuristic, corollary 1 of~\cite{tangbhaskar2013bis} is used as a guideline. It indeed states that the summed amplitudes of coefficients of $x^\star$ within a certain distance $\eta_0$ from the actual peak in the signal tend to the appropriate value as the discretization step tends to zero. In the proof, they choose $\epsilon$ such that the balls of width $\eta_0$ centred around the  true peaks have a null intersection. Thus, it seems reasonable to select $\eta$ as the largest interval within which the probability to distinguish frequencies is low. Values such as $\approx 0.5\pi/T_{\mathrm{obs}}$ to $ \pi/T_{\mathrm{obs}}$ are robust in practice.

\subsection{Significance and uncertainties}
\subsubsection{Detection threshold}

It is simple to associate a ``global" false alarm probability (FAP) to the $\ell_1$-periodogram similar to the classical FAP of the Lomb-Scargle periodogram~\citep[][eq. 14]{scargle}.  Let us consider the probability that ``$x=0$ is not a solution knowing the signal is pure independent Gaussian noise". Denoting this probability $\tilde{\alpha}$, following notations of section~\ref{generalprinciples},  $\epsilon^2 = {F_{\chi^2}}^{-1}(1- \tilde{\alpha}) $. As $\epsilon \approx \epsilon_{\mathrm{noise}}$, the value of $\tilde{\alpha}$ is close to the user-defined parameter $\alpha$.  In the Lomb-Scargle case the FAP obeys: ``if the maximum of the periodogram is $z$ then the FAP is $\beta(z)$'', where $\beta$ is an increasing function of $z$ (often taken as  $\beta(z)=1-(1-\e^{-z})^M$ where $M$ is a parameter fitted with numerical simulations~\cite{scargle,hornebaliunas1986,cumming2004}). Here the formulation is ``If the solution to~\eqref{BPDNepsilonw} is not zero then a signal has been detected with a FAP lower or equal to  $\alpha$".

\subsubsection{Statistical significance of a peak}

The discussion above points out similarities with the FAP defined for periodograms. This one and the global FAP share in particular that they only allow to reject the hypothesis that the signal is pure Gaussian noise of covariance matrix $W$. However, the problem is rather to determine if a given peak indicates a true underlying periodicity, and if this one is due to a planet.

In that scope, our goal is to test if the harmonics spotted by the $\ell_1$-periodogram are statistically significant. Ultimately, one can use statistical hypothesis testing, which can be time consuming. To quickly assess the significance of the peaks, two methods seem to be efficient:
\begin{itemize}
\item  Re-sampling: Taking off randomly 10-20\% of the data and re-computing the $\ell_1$-periodogram. The peaks that show great variability should not be trusted.
\item Using the formulae of the  ``residual/recursive periodograms''~\cite{cumming2004,baluev2008,baluev2009,baluev2015,angladaescude2012}.
\end{itemize}
The first case is easy to code and has the advantage to implement implicitly a time-frequency analysis. Indeed, we might expect from stellar variability some wavelet like contributions: a signal with a certain frequency arises and then vanishes. The timespan of observation might be short enough so that feature is mistaken for a truly sinusoidal component. By taking off some of the measurements we can see if the amplitude of a given frequency varies through time. However, this method requires to re-compute the $\ell_1$-periodogram several times and might not be suited for systems with numerous measurements.

\subsubsection{Model}
\label{modeldef}

As the re-sampling approach is straightforward to code, we will now focus on the recursive periodogram formulae. These ones should be useful for readers more interested in speed than comprehensiveness. In this section, the relevant signal models are defined. We consider that the signal is of the form
\begin{align}
f_{K}(\theta_{0},({\theta_{K}}_j)_{j=1..np}) & = \mathrm{non \; planetary}(\theta_{0}) + \sum\limits_{j=1}^{n_p}\mathrm{Keplerian}_j({\theta_{K}}_j)
\end{align}
or
\begin{align}
 f_{C}(\theta_{0},({\theta_{C}}_j)_{j=1..np})  & = \mathrm{non \; planetary}(\theta_{0}) + \sum\limits_{j=1}^{n_p}\mathrm{Circular}_j({\theta_{C}}_j)
\label{model2}
\end{align}
That is a sum of a model accounting for non planetary effects $\mathrm{non \; planetary}(\theta_{0})$, $\theta_{0}$ being a real vector with $n_\theta$ components, and a sum of Keplerian or circular curves depending on five resp. three parameters, ${\theta_{K}}_j = (k_j,h_j,P_j,A_j,B_j)$ and ${\theta_{C}}_j = (P_j,A_j,B_j)$
\begin{align}
\mathrm{Keplerian}(\theta_K)& =  A \dot{U}(k,h,P) + B \dot{V}(k,h,P) \label{model:1}\\
\mathrm{Circular}(\theta_C) &= A \cos(\frac{2\pi t}{P}) + B \sin(\frac{2\pi t}{P}) 
\label{model:2}
\end{align}
Where $k=e\cos \varpi$, $h = e\sin \varpi$, $\varpi=\omega+\Omega$ is the sum of the argument of periastron and right ascension at ascending node, $U,V$ are the position on the orbital plane rotated by angle $\varpi$. These variables are chosen to avoid poor determination of the eccentricity and time at periastron for low eccentricities.
 
 We compare subsequently the $\chi^2$ of residuals of a model with $n_p$ and $n_p+1$ planets. In practice, one selects the tallest peak of the $\ell_1$-periodogram, and uses this frequency to initialize a least-square fit of a circular or Keplerian orbit. Then the two tallest peaks are selected and so on. 

To clarify the meaning of the computed FAP, let us define the recursive periodogram, depending on a frequency $\omega$. We denote the $\chi^2$ of the residuals by:
\begin{align}
&\chi^{2}_{K,C}(\theta_{0}^{\mathrm{fit}},\theta_{n_p}^{\mathrm{fit}},\omega) = \notag \\
& \left[y-f_{K,C}\left(\theta_{0}^{\mathrm{fit}},\theta_{n_p}^{\mathrm{fit}},\omega^{\mathrm{fit}}\right)\right]^T V^{-1}\left[y-f_{K,C}\left(\theta_{0}^{\mathrm{fit}},\theta_{n_p}^{\mathrm{fit}},\omega^{\mathrm{fit}}\right)\right] \\
&\chi^{2}_{K,C}(\theta_{0}^{\mathrm{fit}},\theta_{n_p}^{\mathrm{fit}}) = \notag\ \\ 
& \left[y-f_{K,C}\left(\theta_{0}^{\mathrm{fit}},\theta_{n_p}^{\mathrm{fit}}\right)\right]^T V^{-1}\left[y-f_{K,C}\left(\theta_{0}^{\mathrm{fit}},\theta_{n_p}^{\mathrm{fit}}\right)\right]
\end{align}
$f_{K,C}\left(\theta_{0}^{\mathrm{fit}},\theta_{n_p}^{\mathrm{fit}},\omega^{\mathrm{fit}}\right)$ is the model fitted depending on the non planetary effects $\theta_{0}$, the (Keplerian or circular) $\theta_{n_p} = ({\theta_{K,C}}_j)_{j=1..n_p}$ parameters of $n_p$ planets plus a circular or Keplerian orbit initialized at frequency $\omega$.
$V$ designates the covariance matrix of the noise model ($V^{-1} = W^2$ with the notations above). This one is often assumed to be diagonal but this is not necessary as all the properties of those periodograms come from the fact that they are likelihood ratios.
 The model fit can be done linearly~\citep{baluev2008} or non-linearly~\citep{angladaescude2012}. By linear we mean that among the five or three parameters defined equations~\eqref{model:1},\eqref{model:2}, only $(A_j)_{j=1..n_p +1}$ and $(B_j)_{j=1..n_p +1}$ are fitted and the non planetary effects are modelled linearly: there exists a matrix $\phi$ such that $\mathrm{non \; planetary}(\theta_0) = \phi \theta_0$. In the second option, the orbital elements of previously selected planets, the non-planetary effects and the signal at the trial frequency are re-adjusted non linearly for each trial frequency.
 
 \subsubsection{FAP formulae for recursive periodograms}
\label{significance}
%We take expression ``$z_1$'' in equation 2  of~\cite{baluev2008} as a definition of the  Keplerian or circular recursive periodogram.
Recursive periodogram is a term that refers to a general concept for comparing the residuals of a model with or without a signal at a given frequency. Here we specialize the formulae we use. 
Denoting by $P_C(\omega)$ and $P_K(\omega)$ in the circular resp. Keplerian case.
\begin{align}
P_C(\omega) &= N \frac{\chi^2_{C}(n_p,\omega) - \chi^2_{C}(n_p)}{\chi^2_{C}(n_p)} \\
P_K(\omega) & = \frac{1}{2}\left(\chi^2_{K}(n_p) - \chi^2_{K}(n_p,\omega)\right))
\end{align}
Where $N = m - 2n_p - n_{\theta}$
The circular case is expression ``$z_1$'' in equation 2  of~\cite{baluev2008}, and the Keplerian one is expression ``$z$'' in equation 4  of~\cite{baluev2015}. In what follows only the circular case will be used.

The quantity we are interested in is the probability that a selected peak is not a planet. We here use the FAP as a proxy for that quantity:
\begin{align}
\mathrm{FAP}(Z) = \mathrm{Pr}\left\{ \max\limits_{\omega \in [0,\omega_{\mathrm{max}}]} P(\omega) > Z  \; \middle| \;\mathrm{non \; planetary \;effects},n_p \right\}
\end{align}
Where $\omega_{\mathrm{max}}$ is the maximum frequency of the periodogram that has been scanned. This FAP is the probability to obtain a peak at least as high as $Z$ while there is only non planetary effects and $n_p$ planets.
\cite{baluev2008} has computed tight bounds for that quantity in case of a circular model and a linear fit (corresponding to subscript $C$), which we reproduce here:
\begin{align}
\mathrm{FAP}(z,\omega_{\mathrm{max}}) \approx W \gamma \left( \frac{2z}{ N_{\mathcal{H}}} \right)^{\frac{1}{2}} \left( 1 - \frac{2z}{N_{\mathcal{H}}}\right)^{\frac{N_{\mathcal{H}}+1}{2}}
\end{align}
where $N_{\mathcal{H}}$ is the number of degrees of freedom of the model without the sine at frequency $\omega$, $\gamma = \Gamma(N_{\mathcal{H}}/2)/\Gamma((N_{\mathcal{H}}+3/)2) $, $\Gamma$ being the Euler $\Gamma$ function, and $W =  \omega_{\mathrm{max}} \sqrt{(\bar{t^2}-\bar{t}^2)/\pi}$, $t$ being the array of measurement times and $\bar{t}$ is the mean value of $t$.
We have also tried the exact expression of the so-called Davies bound provided by equations 8, B5 and B7 of~\cite{baluev2008}, but the results were very similar to the simpler formula.
In the case of Keplerian periodogram, we used equation 21 and 24 from~\cite{baluev2015}.

%As said section~\ref{previouswork}, recursive periodograms can be seen as a more precise but more expensive version of Frequency Analysis or CLEANest or Orthogonal Matching Pursuit. They are very powerful techniques that complement well the $\ell_1$-periodogram. 
Again, we emphasize that the interest of the present method is to select candidates for future observations or unveiling signals unseen on periodograms. The FAP formulae used here do not guarantee the planetary origin of a signal. For robust results statistical hypothesis testing~\citep[e.g.][]{diaz2016} can be used.

\section{Results}
\label{results}

\subsection{Algorithm tuning}

For all the systems analysed in the following sections, the figures called $\ell_1$-periodogram represent $x^\sharp(\omega)$ as defined in equation~(\eqref{xsharp}) plotted versus periods. The name $\ell_1$-periodogram was chosen to avoid the confusion with the generalized Lomb-Scargle periodogram defined by~\cite{zechmeister}. In each case, the algorithm is tuned in the following way:
\begin{itemize}
\item The problem~\eqref{BPDNepsilonw} is solved with SPGL1~\citep{spgl1}
\item The solution of SPGL1 is averaged on an interval $\eta=2\pi/(3T_{\mathrm{obs}})$ according to section~\ref{tuning}.
\item The grid spacing is chosen according to equation~\eqref{omegabound}.
\end{itemize} 
The importance of the grid span and the tolerance $\epsilon$ will be discussed in the examples. 

\begin{figure*}
\noindent
\centering
\begin{tikzpicture}
\path (-.3,0) node[above right]{\includegraphics[width=8.3cm]{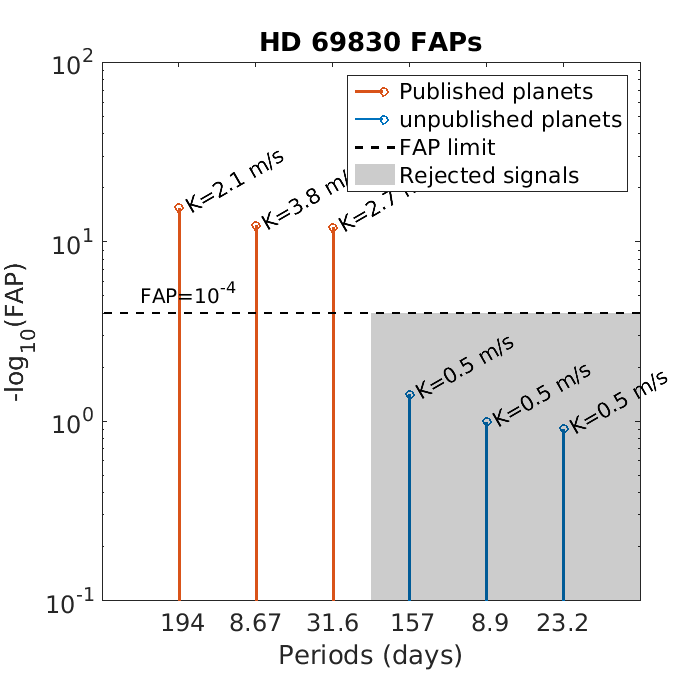}};%hd69830_fratio.png
\path (1.4,6.8) node[above right]{a)};
\path (9,.3) node[above right]{\includegraphics[width=8cm]{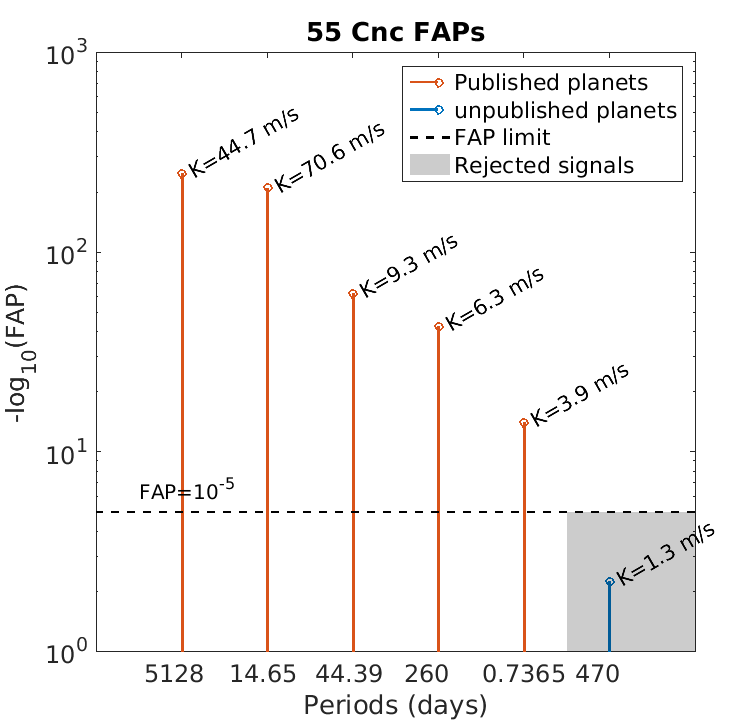}};%55cnc_fratio.png
\path (10.2,6.8) node[above right]{b)};
\begin{scope}[yshift=-8cm]
\path (0,0) node[above right]{\includegraphics[width=8cm]{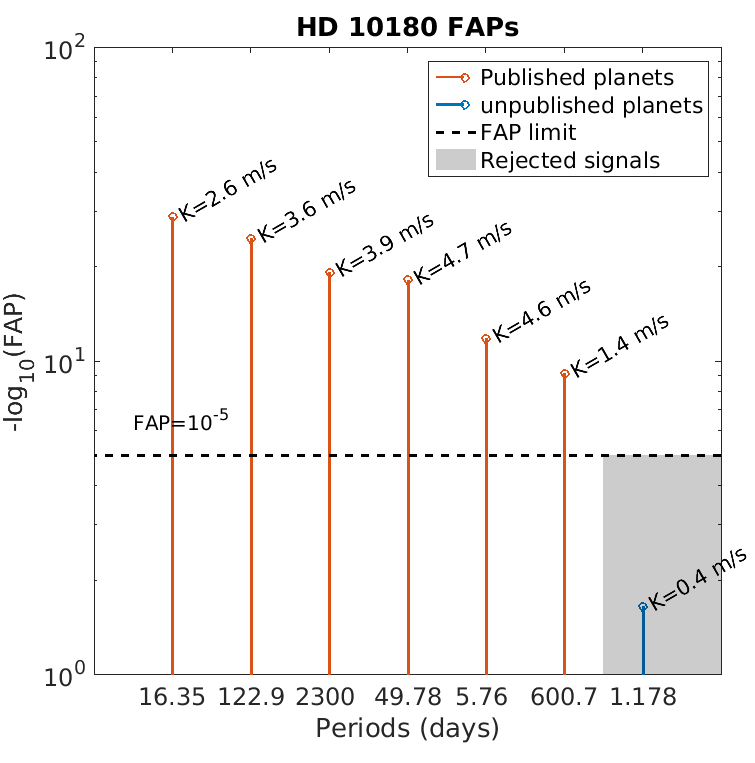}};%hd69830_fratio.png
\path (1.2,6.8) node[above right]{c)};
\path (9,0.15) node[above right]{\includegraphics[width=8cm]{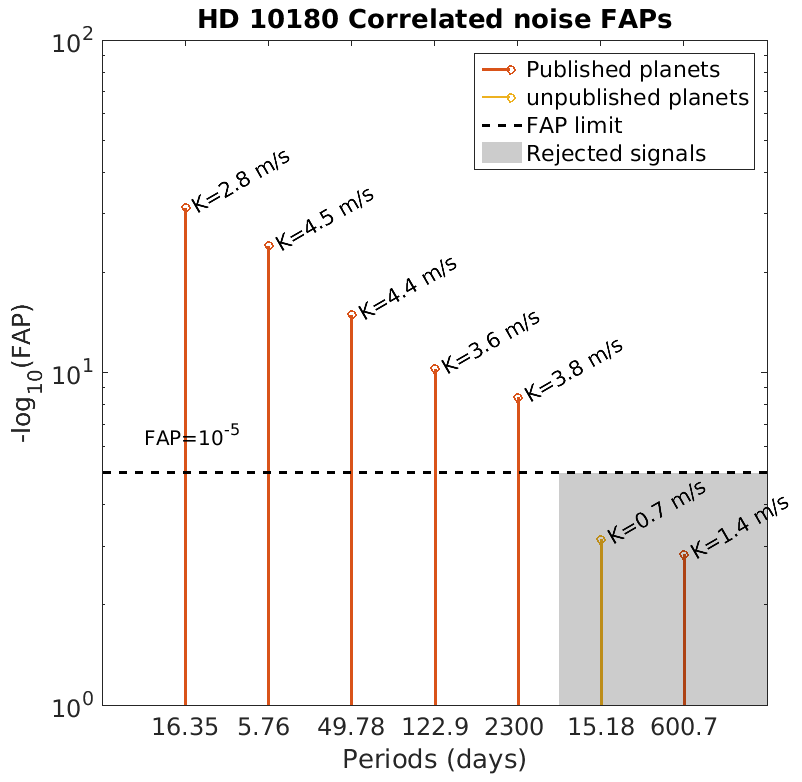}};%55cnc_fratio.png
\path (10.2,6.8) node[above right]{d)};
\end{scope}
\begin{scope}[yshift=-16cm]
\path (0,0.5) node[above right]{\includegraphics[width=8cm]{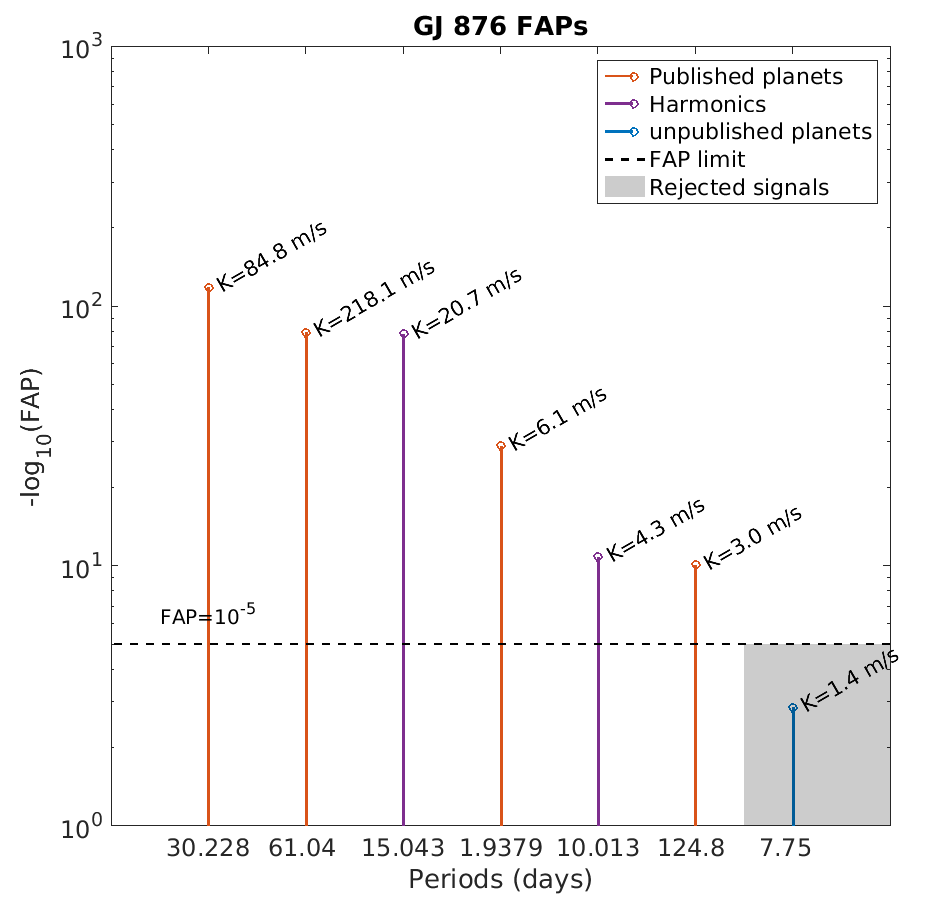}};%hd69830_fratio.png
\path (1.2,6.8) node[above right]{e)};
\path (9.15,0.3) node[above right]{\includegraphics[width=7.9cm]{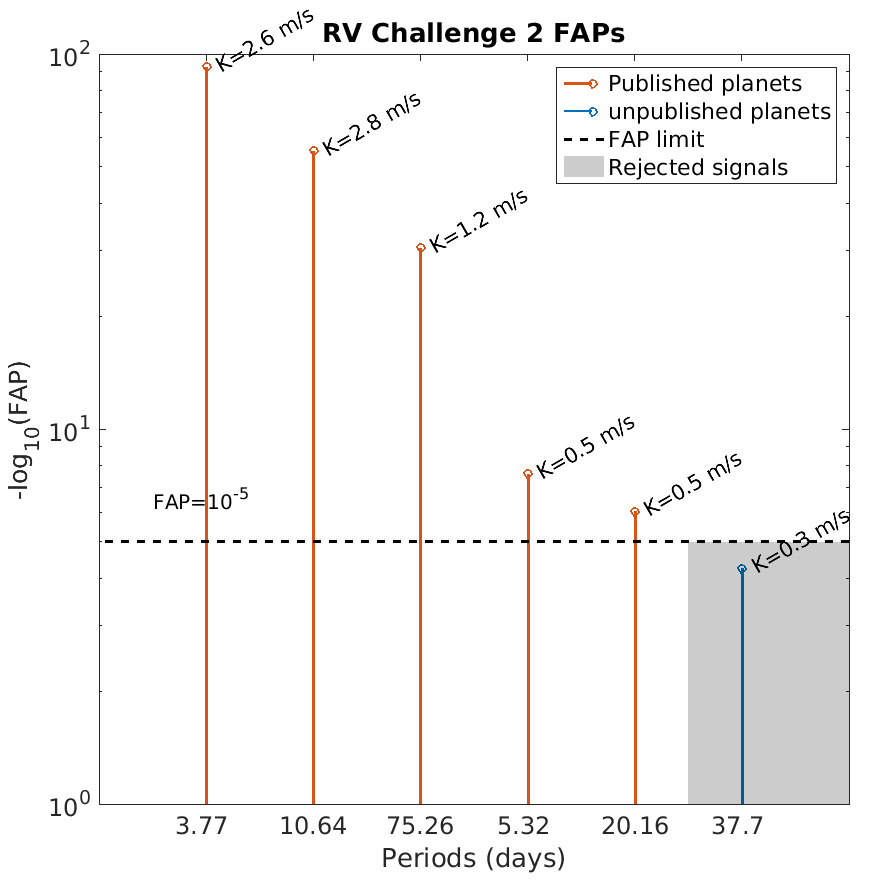}};%55cnc_fratio.png
\path (10.2,6.8) node[above right]{f)};
\end{scope}
\end{tikzpicture}
\caption{Peak amplitudes and associated FAPs for the four systems analysed}
\label{rvsurveyfap}
\end{figure*}

The FAPs are computed according to the procedure described section~\ref{significance} and are  represented figure~\ref{rvsurveyfap} with decreasing FAP. The ticks in abscissa correspond to the period of the signals and the flag to their semi-amplitude after a non-linear least square fit.

In the following, we will present our results for HD 69830, HD 10180, 55 Cnc, GJ 876, and a simulated very active star from the RV Challenge~\citep{dumusque2016}.
For each system, the Generalized Lomb-Scargle periodogram is plotted along with the $\ell_1$-periodogram.

 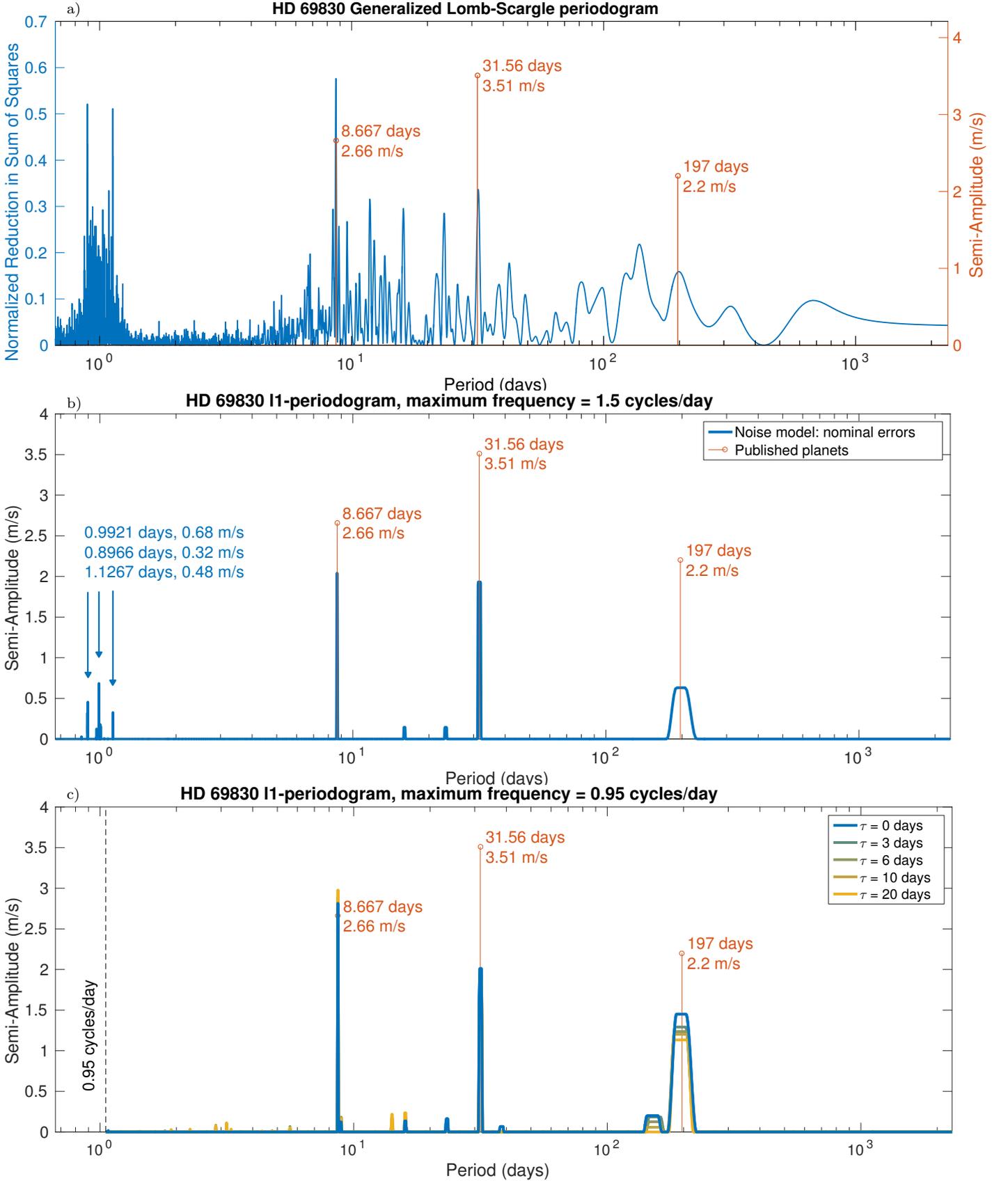
\begin{figure*}
\noindent
\centering
\hspace{-0.45cm}
\begin{tikzpicture}
\path (0,0) node[above right]{\includegraphics[scale=0.43]{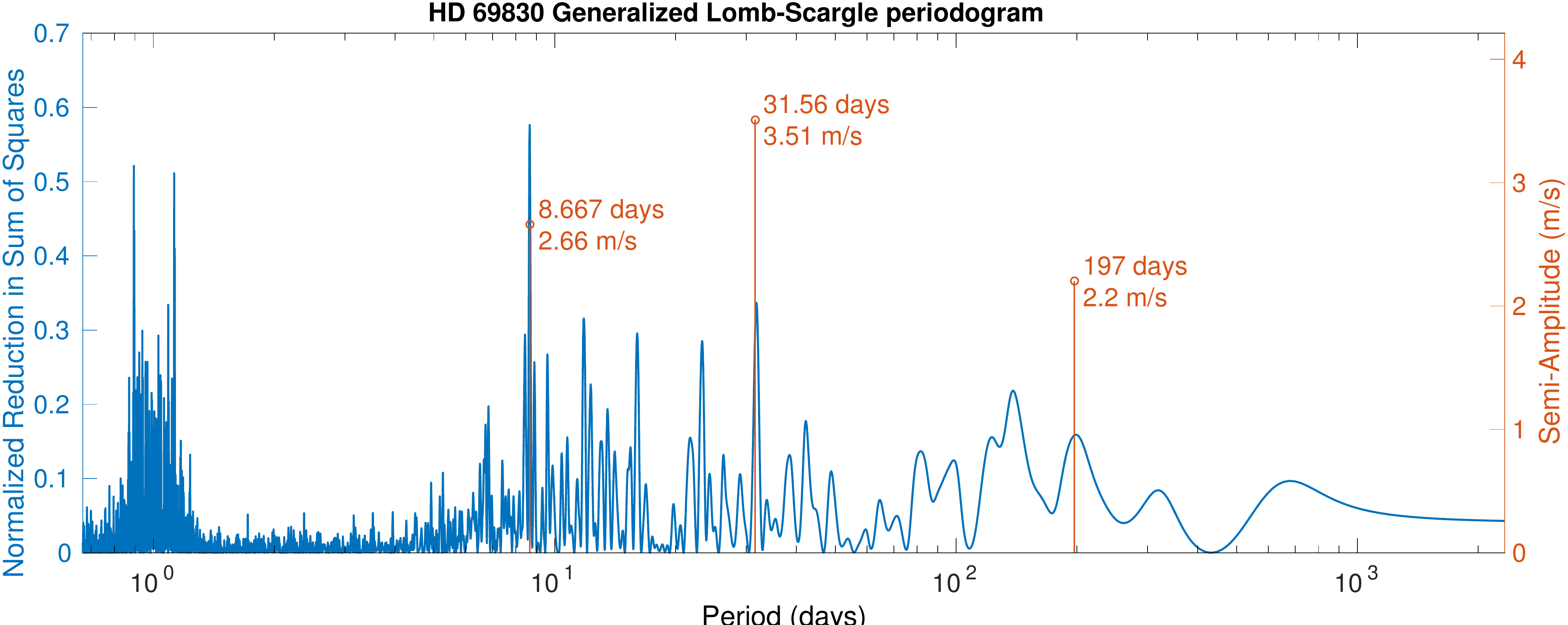}};
\path (1.2,7.1) node[above right]{a)};
\begin{scope}[yshift=-7.6cm]
\path (0,0) node[above right]{\includegraphics[scale=0.43]{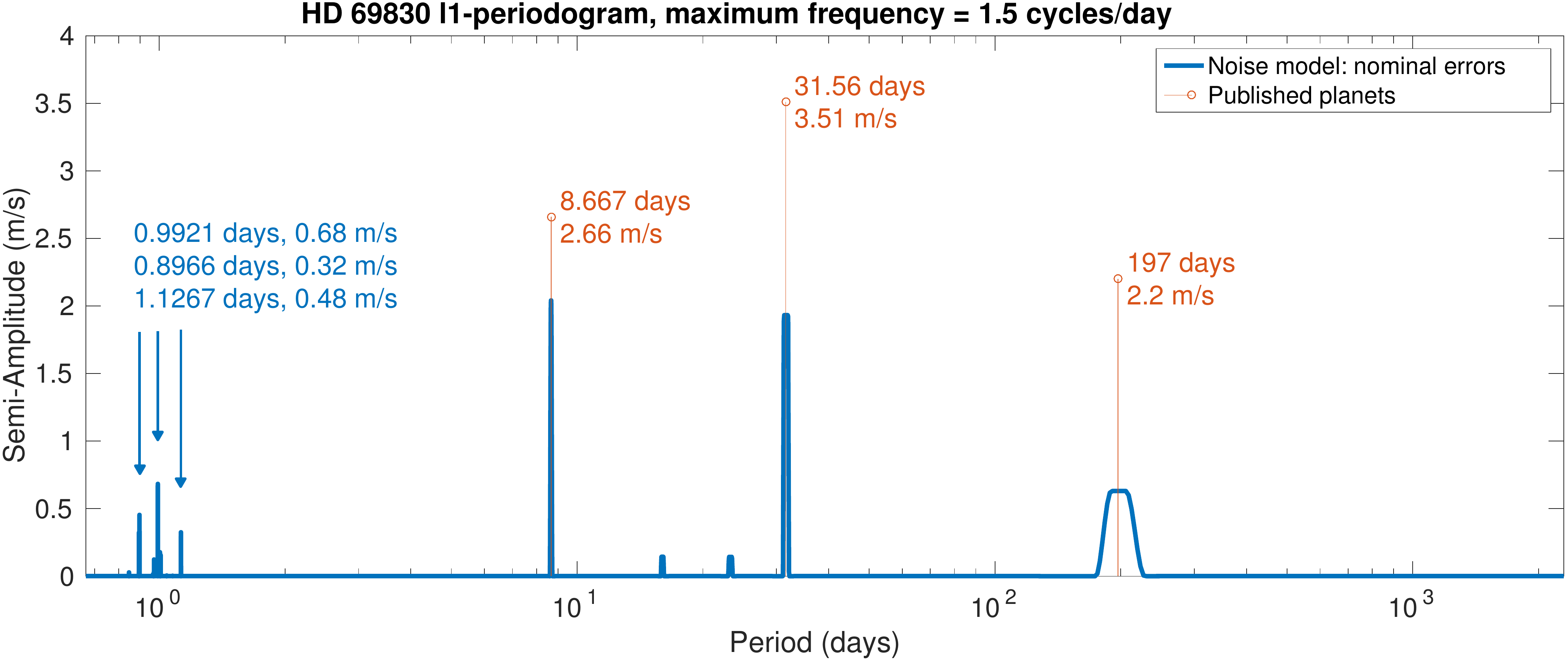}};
\path (1.2,7.2) node[above right]{b)};
\end{scope}
\begin{scope}[yshift=-15.05cm]
\path (0,0) node[above right]{\includegraphics[scale=0.43]{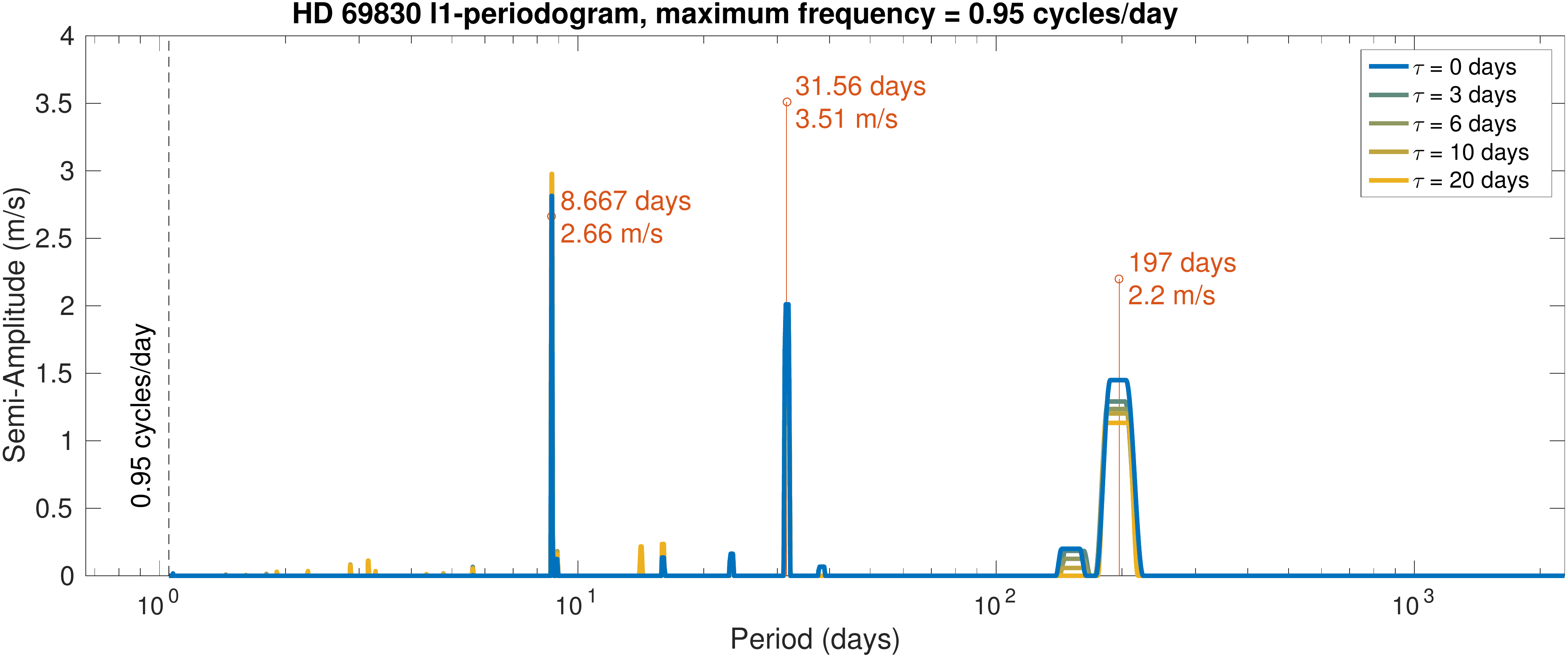}};
\path (1.2,7.2) node[above right]{c)};
\end{scope}
\end{tikzpicture}
\caption{Generalized Lomb-Scargle periodogram and $\ell_1$-periodogram of HD 69830 in blue, published planets are represented by the red stems. The frequency span used for figures b anc c are respectively 1.5 and 0.95 cycles/day.
The other signals mentioned section~\ref{hd69830} are spotted by the blue arrows. For all the noise model considered for matrix $W$, $\sigma_W =0$, $\sigma_R = 1$ m.s\textsuperscript{-1}.}
\label{hd69830}
\end{figure*}

\begin{figure*}
\noindent
\centering
\hspace{-0.65cm}
\begin{tikzpicture}
\path (0,0) node[above right]{\includegraphics[scale=0.43]{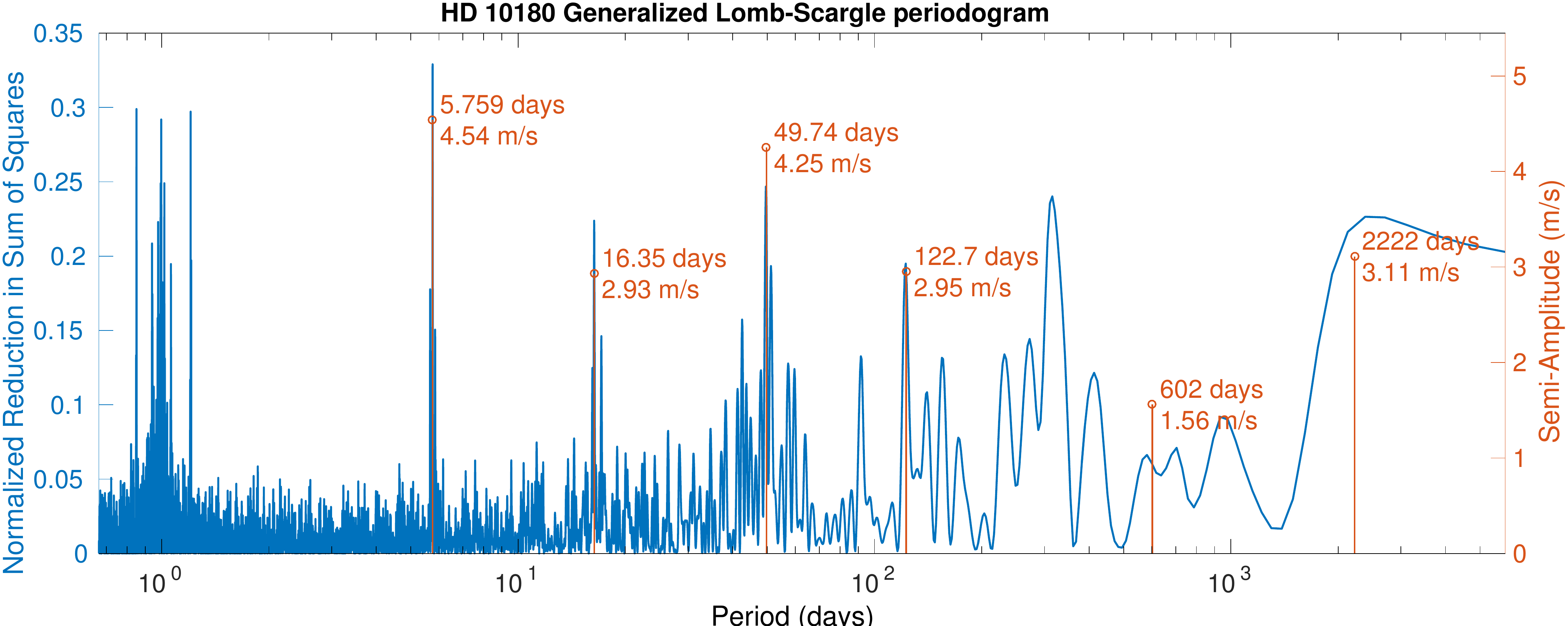}};
\path (1.2,7.2) node[above right]{a)};
\begin{scope}[yshift=-7.6cm]
\path (0.5,0) node[above right]{\includegraphics[scale=0.43]{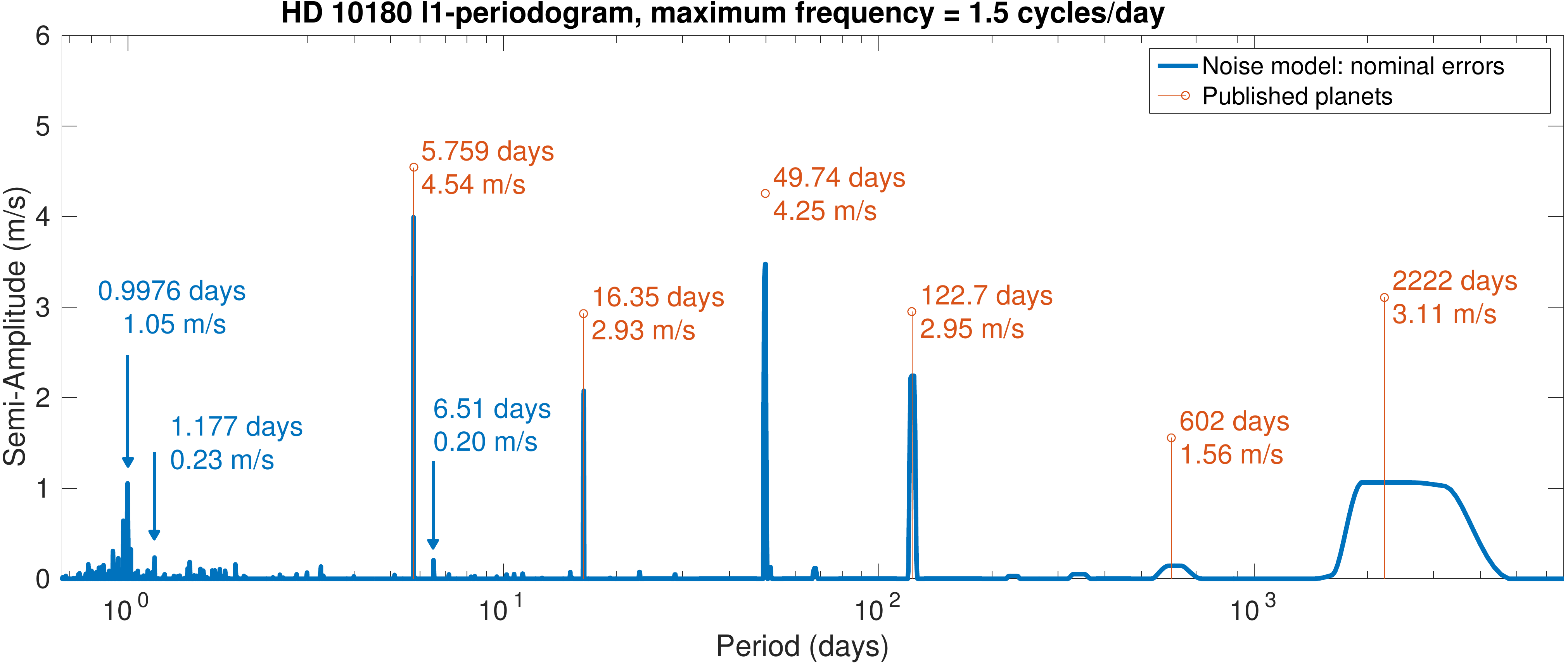}};
\path (1.2,7.2) node[above right]{b)};
\end{scope}
\begin{scope}[yshift=-15.05cm]
\path (0.5,0) node[above right]{\includegraphics[scale=0.43]{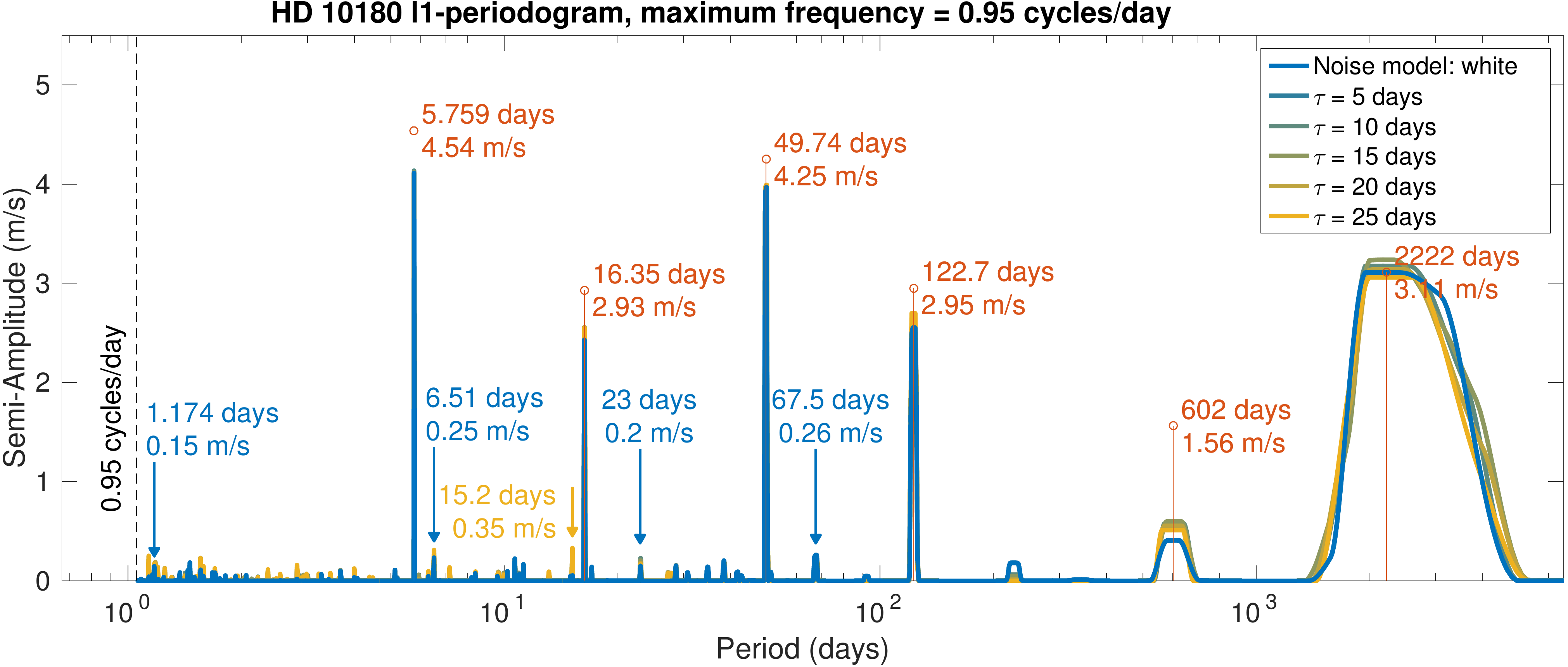}};
\path (1.2,7.2) node[above right]{c)};
\end{scope}
\end{tikzpicture}
\caption{GLS and $\ell_1$-periodograms of HD 10180 data set with mean subtracted. The red stems have the periods and amplitude of published planets. The other signals mentioned section~\ref{hd10180} are spotted by the blue arrows. For all the noise model considered for matrix $W$, $\sigma_W =0$, $\sigma_R = 1$ m.s\textsuperscript{-1}.}
\label{hd10180}
\end{figure*}
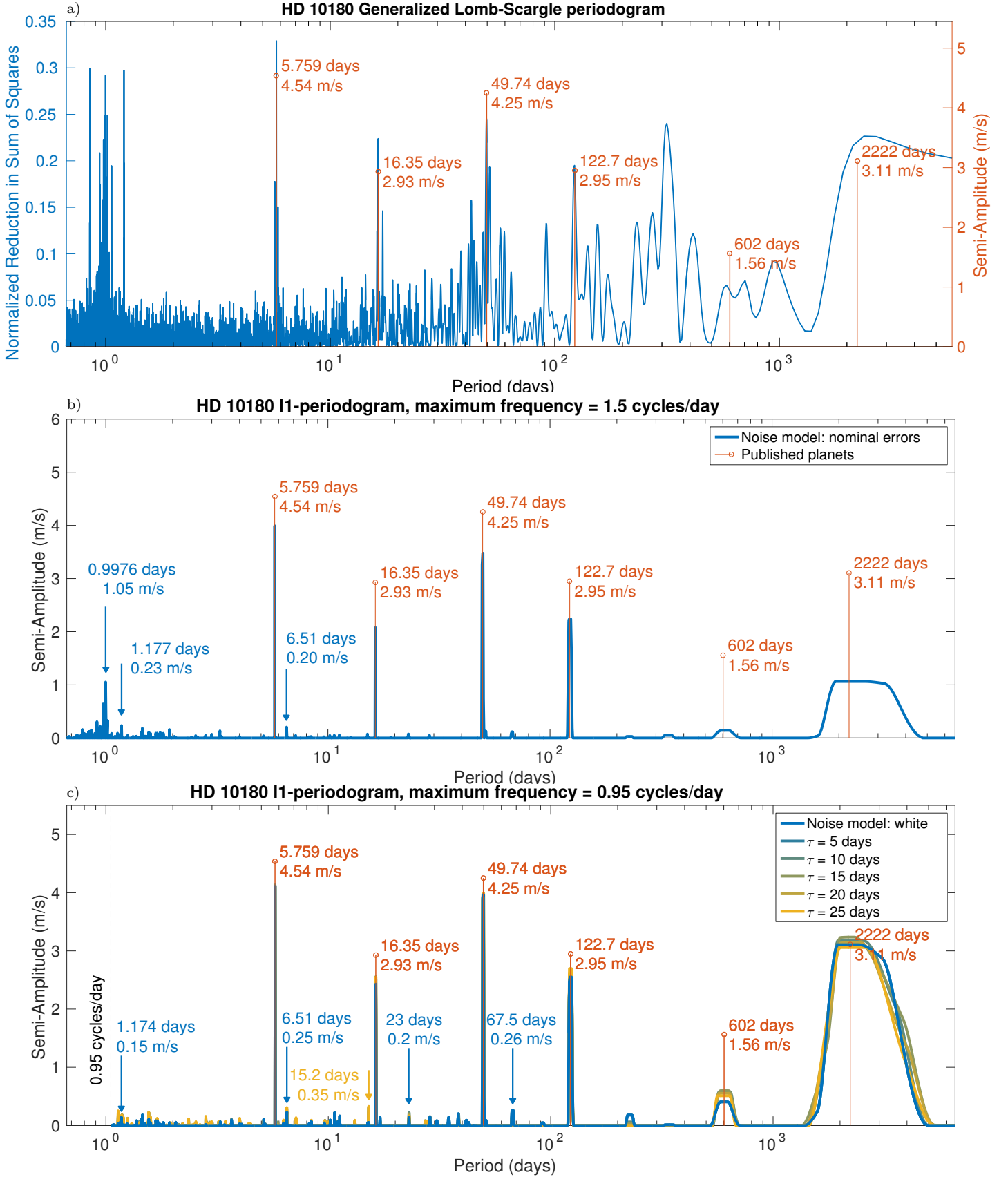

\begin{figure*}
\noindent
\centering
\hspace{-0.45cm}
\begin{tikzpicture}
\path (0,0) node[above right]{\includegraphics[scale=0.43]{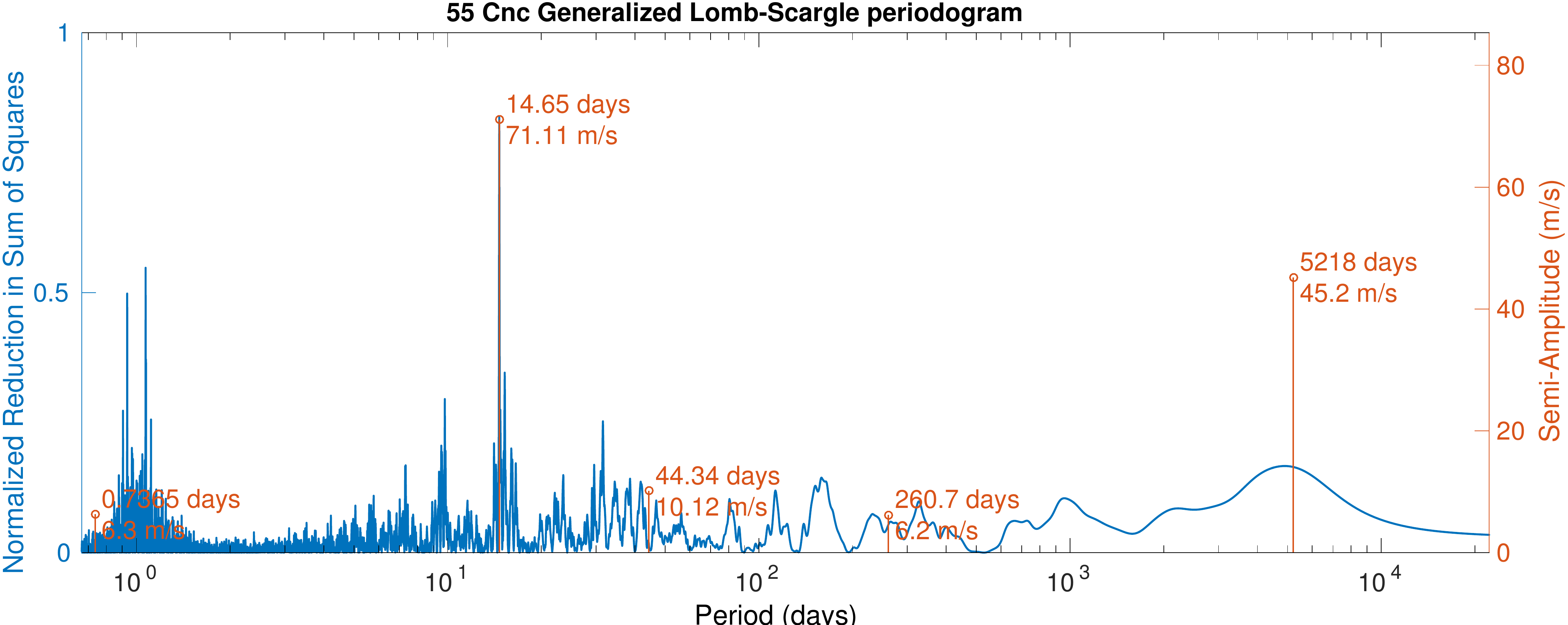}};
\path (1.2,7.2) node[above right]{a)};
\begin{scope}[yshift=-7.6cm]
\path (0.1,0) node[above right]{\includegraphics[scale=0.43]{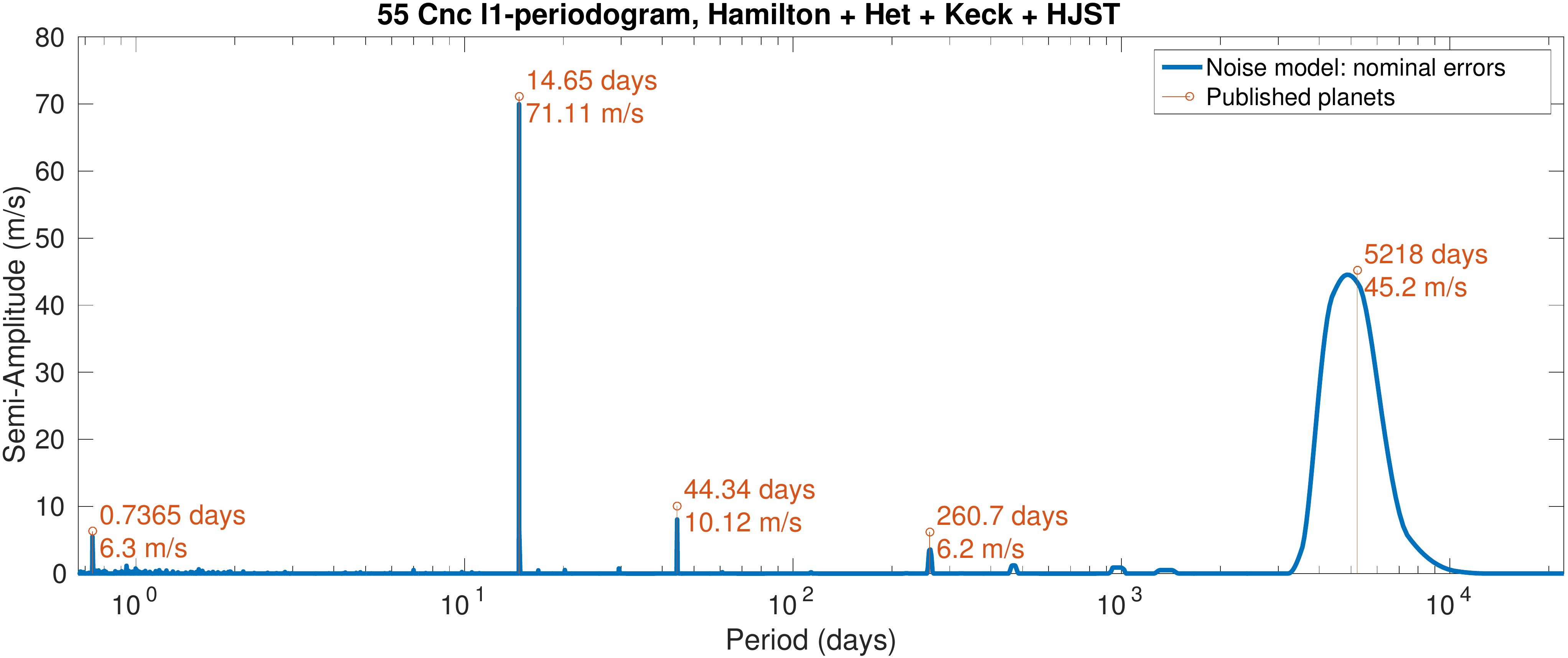}};
\path (1.2,7.2) node[above right]{b)};
\end{scope}
\begin{scope}[yshift=-15.05cm]
\path (0.3,0) node[above right]{\includegraphics[scale=0.43]{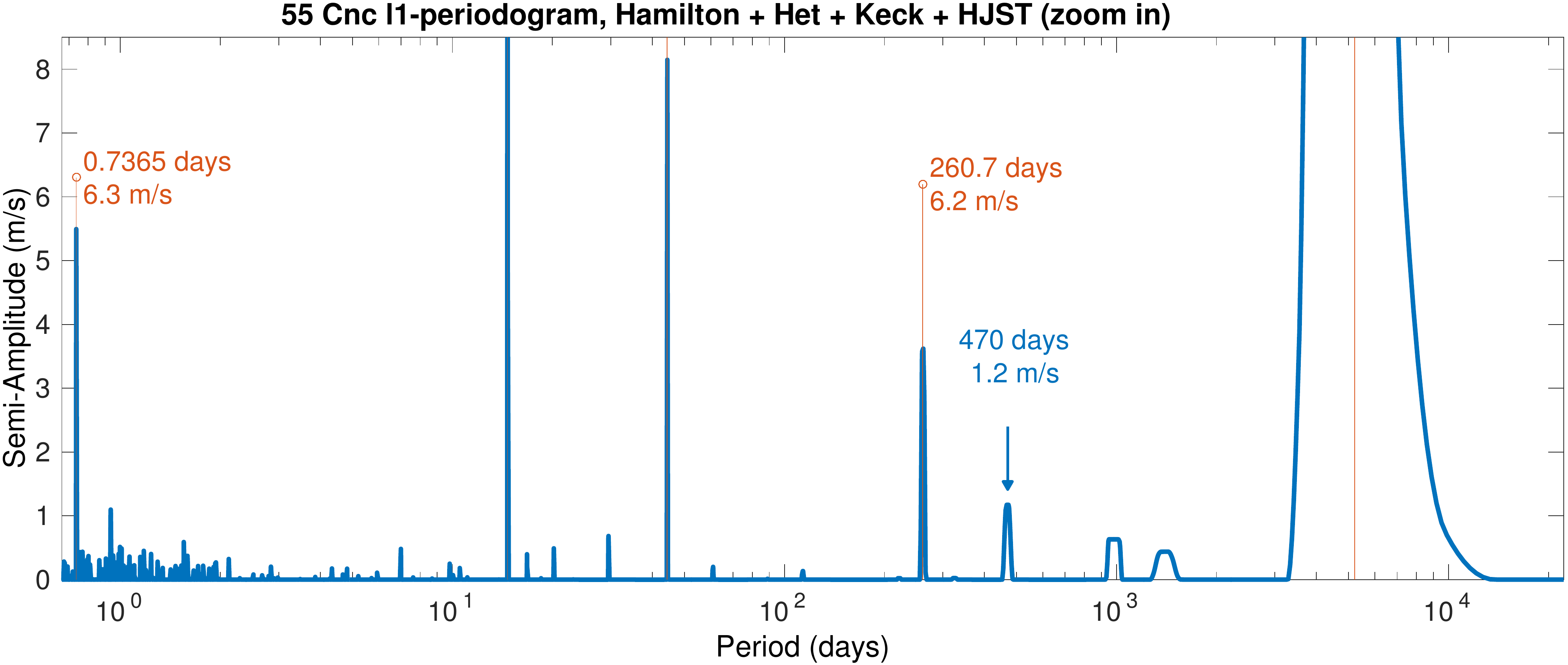}};
\path (1.2,7.2) node[above right]{c)};
\end{scope}
\end{tikzpicture}
\caption{GLS and $\ell_1$-periodograms of 55 Cnc data set with mean subtracted. The red stems have the periods and amplitude of published planets. The other signals mentioned section~\ref{55cnc} are indicated by the blue arrows.}
\label{rvsurvey55cnc}
\end{figure*}
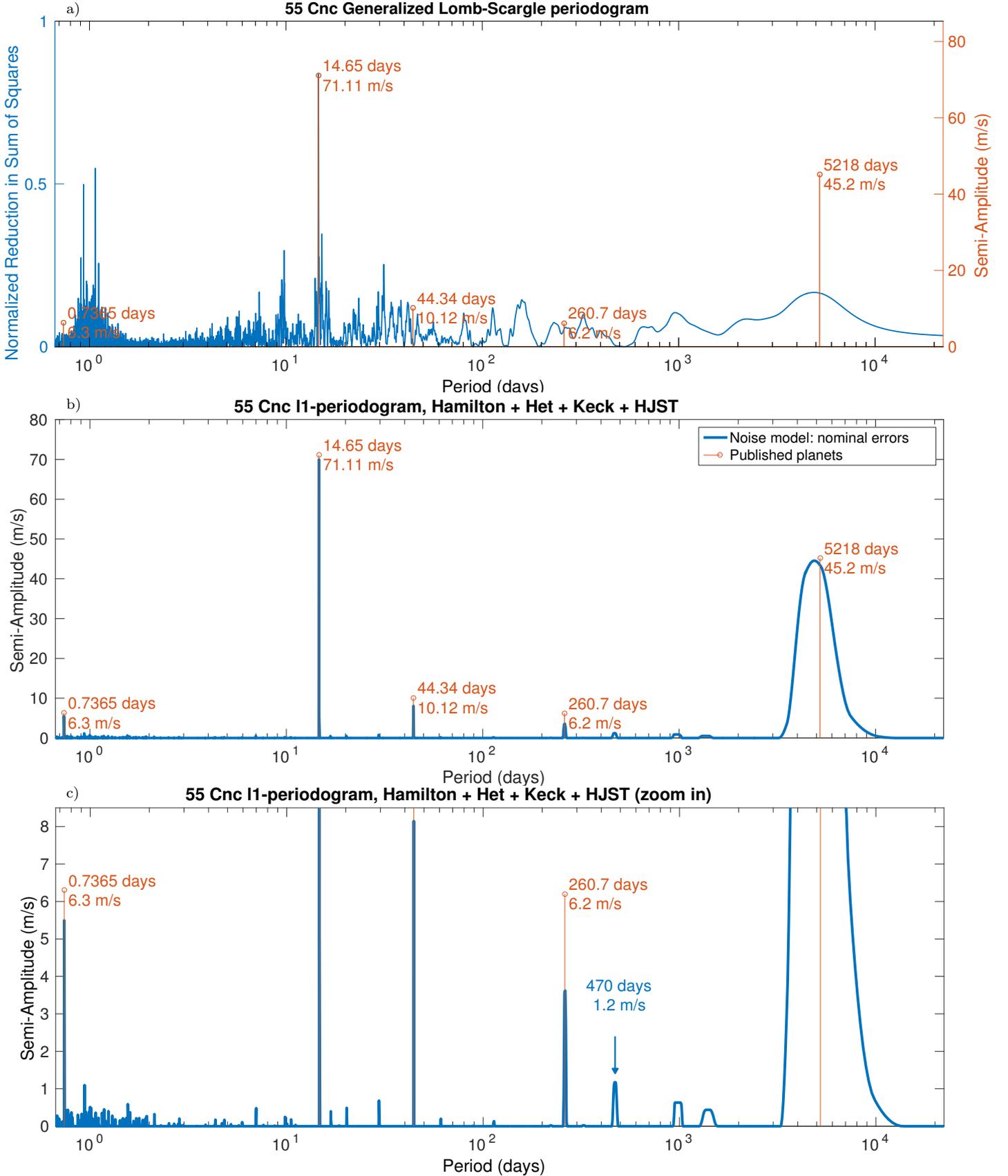

\begin{figure*}
\noindent
\centering
\hspace{-0.52cm}
\begin{tikzpicture}
\path (0.1,0) node[above right]{\includegraphics[scale=0.43]{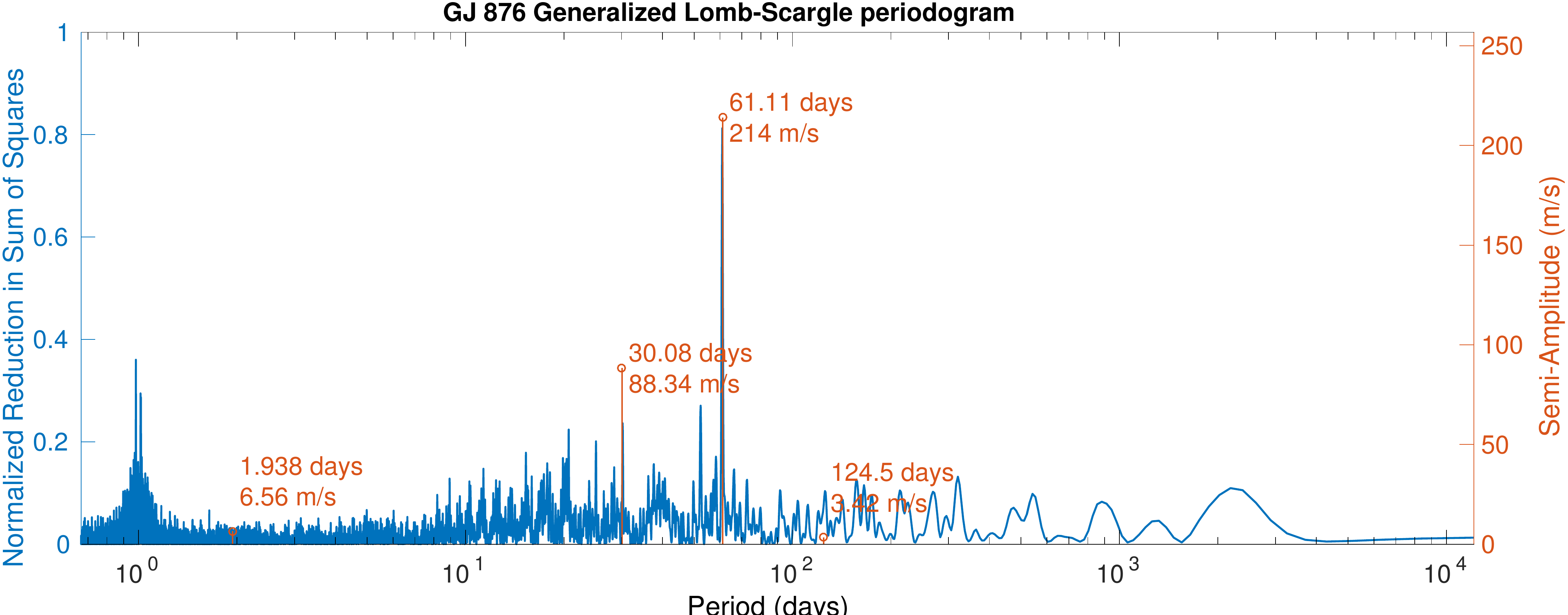}};
\path (1.2,7.2) node[above right]{a)};
\begin{scope}[yshift=-7.6cm]
\path (0,0) node[above right]{\includegraphics[scale=0.43]{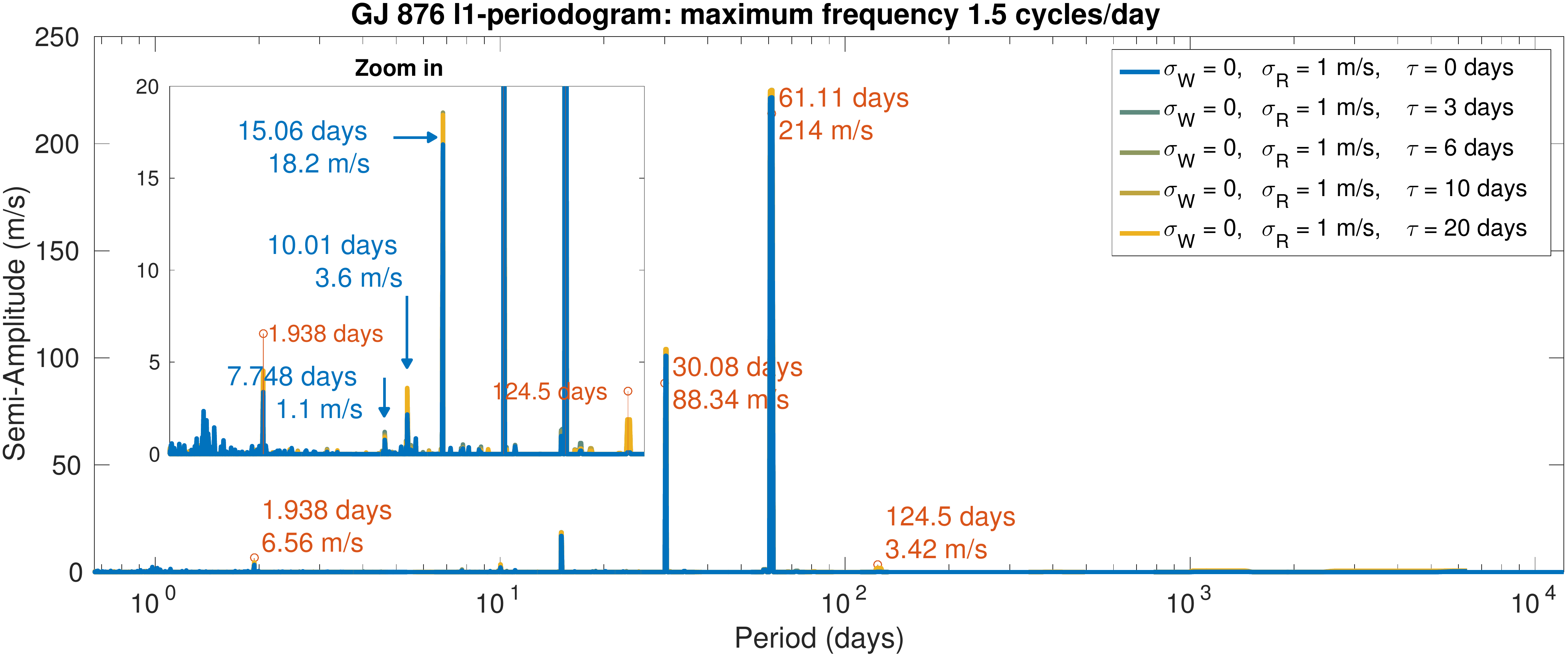}};
\path (1.2,7.2) node[above right]{b)};
\end{scope}
\begin{scope}[yshift=-15.05cm]
\path (0.2,0) node[above right]{\includegraphics[scale=0.43]{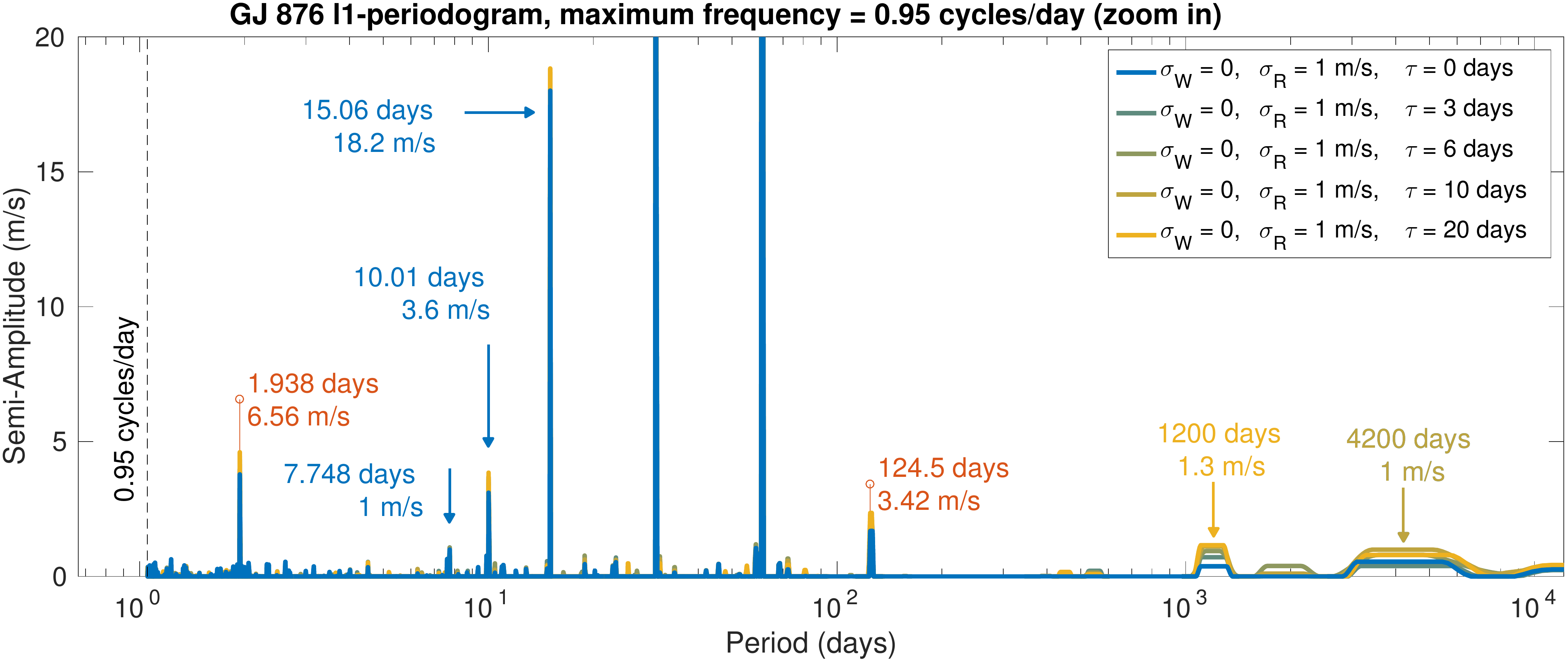}};
\path (1.2,7.2) node[above right]{c)};
\end{scope}
\end{tikzpicture}
\caption{GLS and $\ell_1$-periodograms of GJ 876 data set with means of KECK and HARPS measurement respectively subtracted. The red stems have the periods and amplitude of published planets. The other signals mentioned section~\ref{gj876} are indicated by the blue arrows.}
\label{rvsurvey_gj876}
\end{figure*}
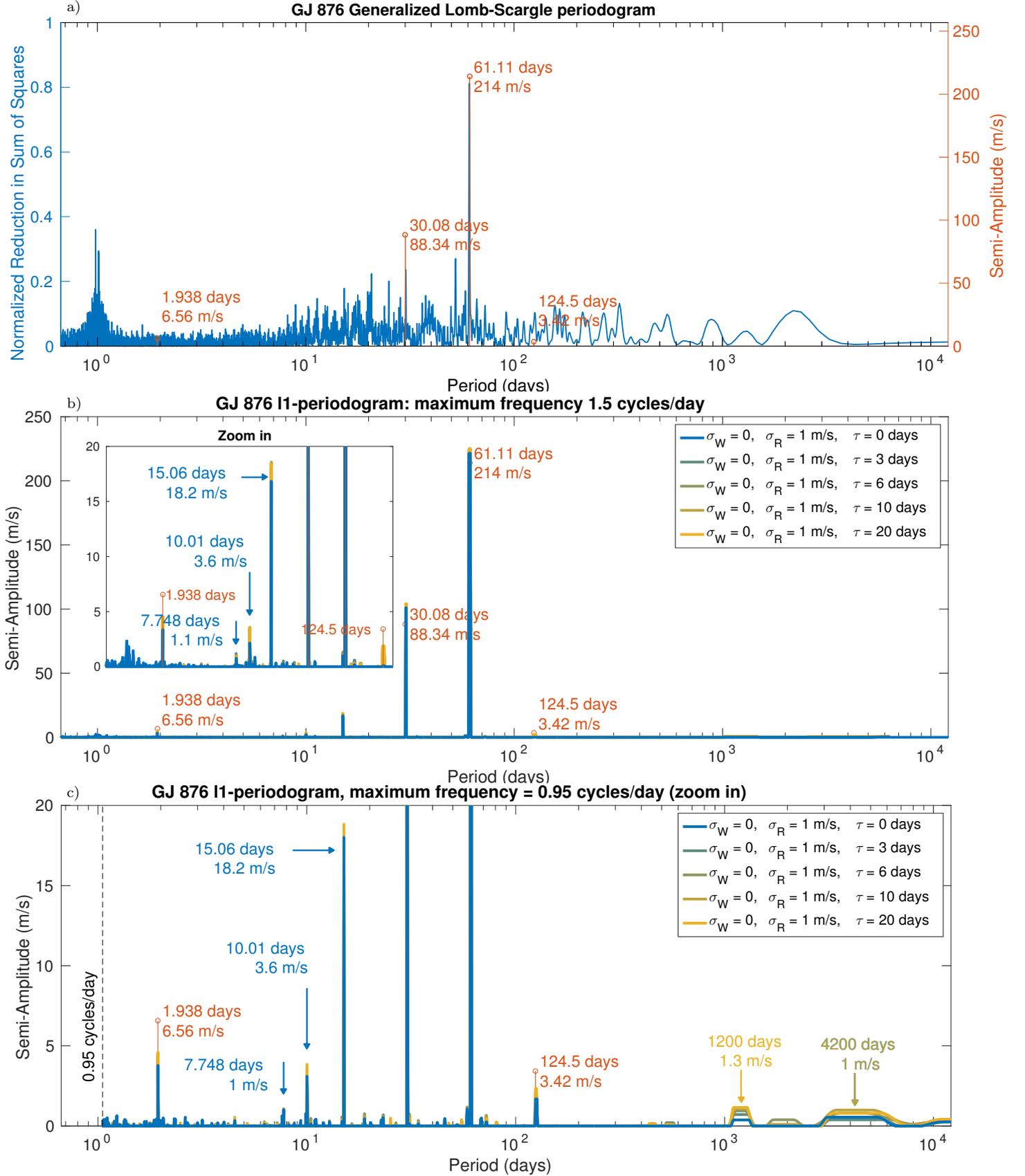

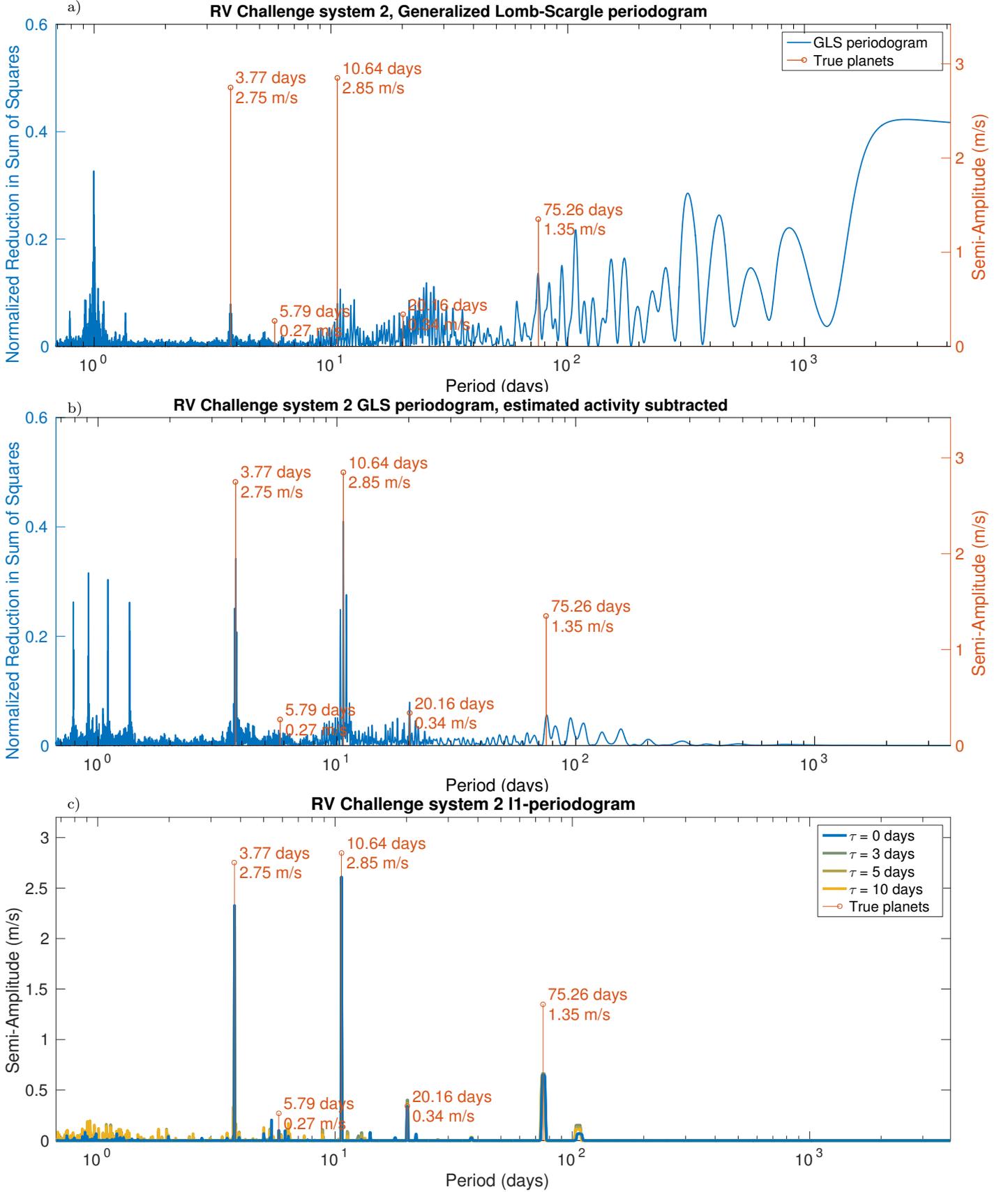
\begin{figure*}
\noindent
\centering
\hspace{-0.4cm}
\begin{tikzpicture}
\path (0.,0) node[above right]{\includegraphics[scale=0.43]{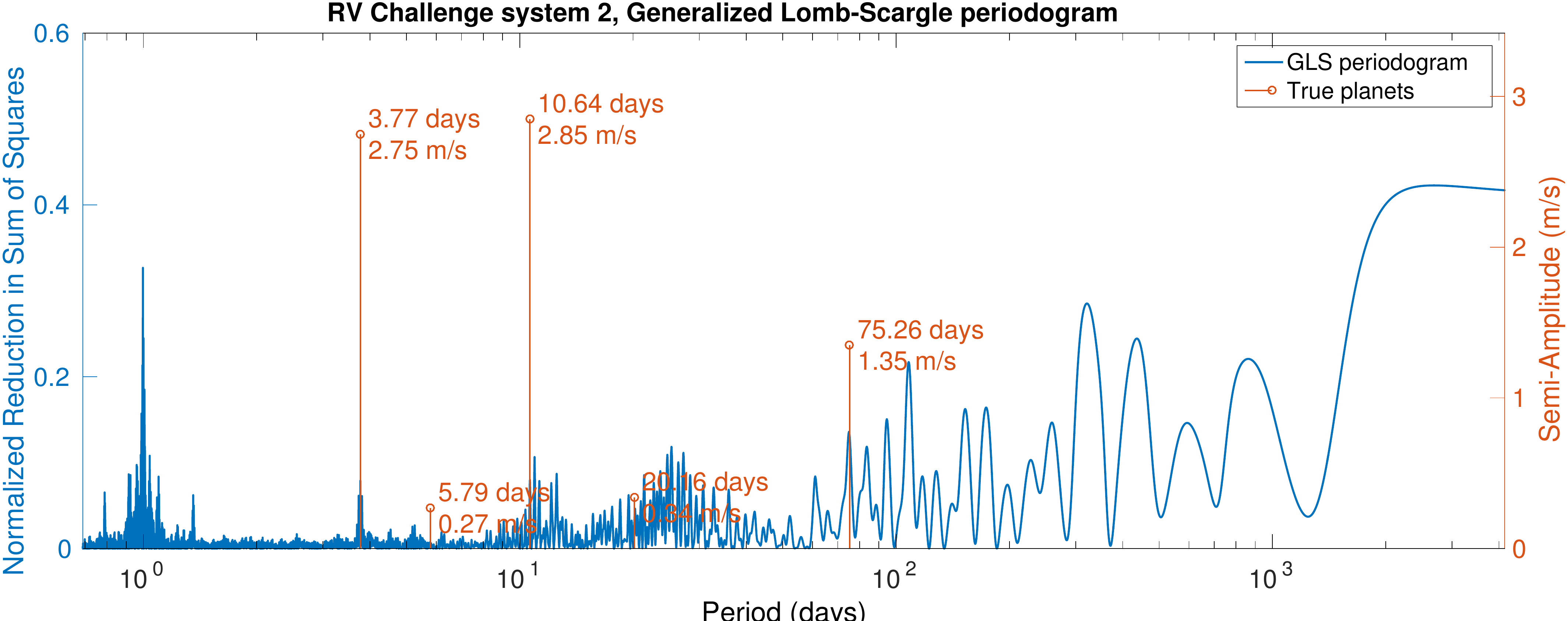}};
\path (1.2,7.15) node[above right]{a)};
\begin{scope}[yshift=-7.6cm]
\path (0,0) node[above right]{\includegraphics[scale=0.43]{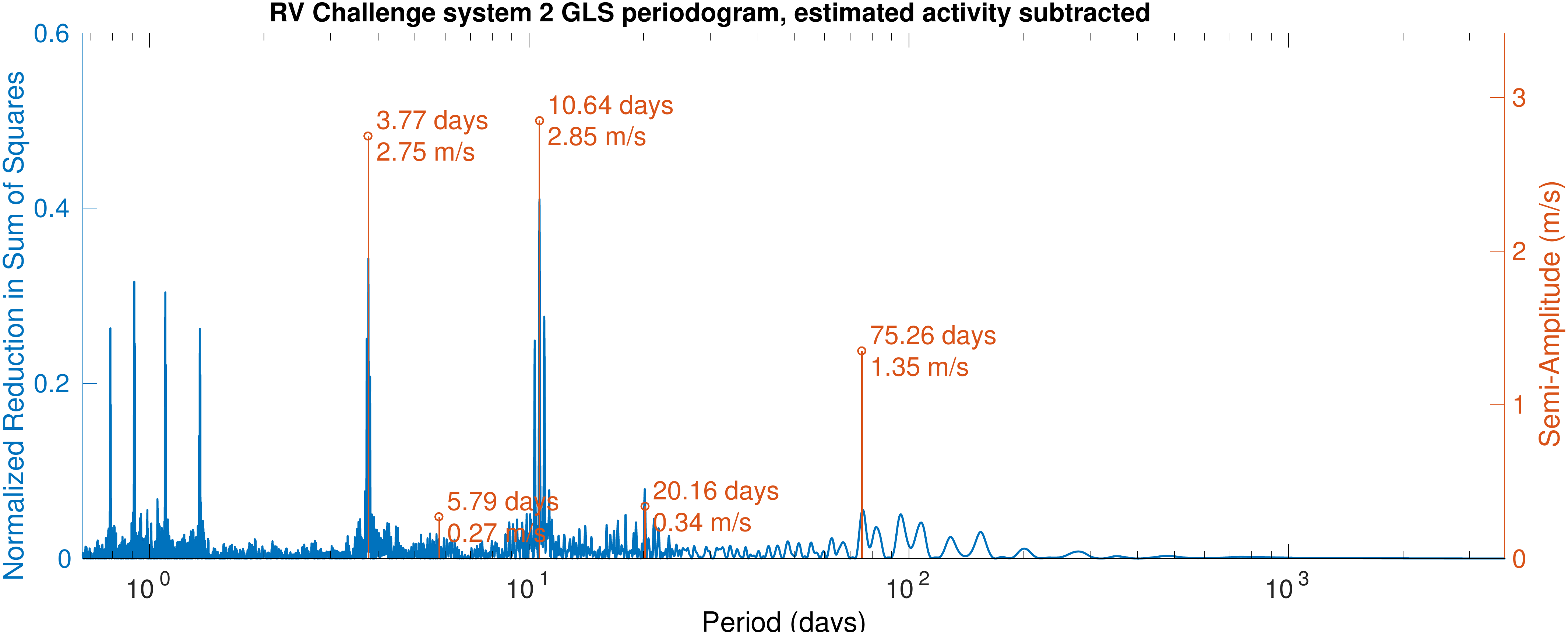}};
\path (1.2,7.12) node[above right]{b)};
\end{scope}
\begin{scope}[yshift=-15.2cm]
\path (0.,0) node[above right]{\includegraphics[scale=0.43]{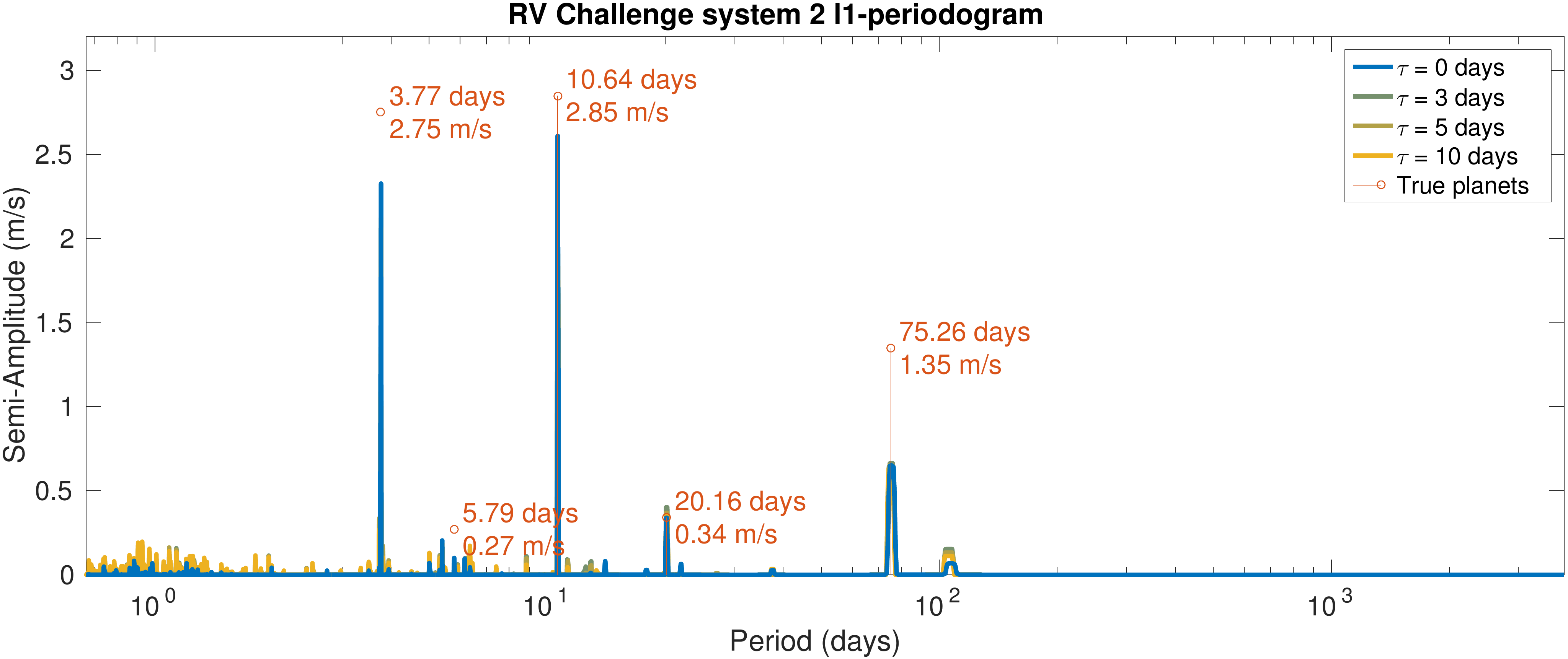}};
\path (1.2,7.2) node[above right]{c)};
\end{scope}
\end{tikzpicture}
\caption{Top: GLS of the RV Challenge system 1 (simulated signal). Top: GLS of raw data, middle: GLS after fitting ancillary measurements, bottom: $\ell_1$-periodogram after fitting ancillary measurements. True planets are represented by red lines.}
\label{rvsurvey_challenge}
\end{figure*}

\subsection{HD 69830}

In~\cite{lovis2006}, three Neptune-mass planets are reported around HD 69830 based on 74 measurements of HARPS spanning over 800 days. The precision of the measurements given in the raw data set (from now on called nominal precision)  is between 0.8 and 1.6 m.s\textsuperscript{-1}. The host star is a quiet K dwarf with a $\log R_{HK}' =-4.97$ and an estimated projected rotational velocity of $1.1^{+0.5}_{-1.1}$ m/s, therefore the star jitter should not amount to more than 1 m.s\textsuperscript{-1} ~\citep{lovis2006}. 

Our method consists in solving the minimization problem~\eqref{BPDNepsilonw} and average the solution as explained in~\ref{postprocessing}. The resulting array $x^\sharp(\omega)$ (see equation~\eqref{xsharp}) is plotted versus frequency, here giving figure~\ref{hd69830}.b and c. The tallest peaks are then fed to a Levenberg-Marquardt algorithm and the FAPs of models with an increasing number of planets are computed. We represent the FAPs of the signals when fitted from the tallest peaks to the lowest -- disregarding aliases -- figure~\ref{rvsurveyfap}.a. The FAP corresponding to a false alarm probability of $10^{-4}$ is represented by a dotted line.

 The values of most of the algorithm parameters defined section~\ref{tuning} are fixed in the previous section. Here we precise that the method is performed for two grid spans: 0 to 1.5 cycles per day and 0 to 0.95 cycles/day (figure~\ref{hd69830}.b resp. c). 

We first apply the method on a grid spanning between 0 and 1.5 cycles per day. The weight matrix is diagonal, $W_{kk} = 1/\sigma_k$ (not $1/\sigma_k^2$)  where $\sigma_k$ is the error on measurement $k$.
On figure~\ref{hd69830}.b, the peaks of published planets appear, as opposed to the generalized Lomb-Scargle periodogram~(\ref{hd69830}.a). However, there are still peaks around one day. The three main peaks in that region have periods of 0.9921, 0.8966 and 1.1267. The maximum of the spectral window occurs at $\omega_M=$ 6.30084 radian/day. Calculating $2\pi/(\omega - \omega_M)$ yields 194.06, 8.8877 and -8.6759 respectively for $\omega = 2\pi/0.9921, 2\pi/0.8966 $ and $2\pi/1.1267$, suggesting the short period peaks are aliases of the true periods. 

We now apply the method described in section~\ref{significance} to test the significance of the signal, obtaining figure~\ref{rvsurveyfap}.a. Taking 8.667, 31.56 and 197 days gives 
 a reduced $\chi^2$ of the Keplerian fit with three planets plus a constant (16 parameters) is 1.19, yet the stellar jitter is not included. As a consequence, finding other significant signals is unlikely.  
 
Looking only at figure~\ref{hd69830}.b, whether the signal at 197 days or its alias at 0.9921 days is in the signal is unsure. We perform two fits with the two first planets  plus one of the candidates. The reduced $\chi^2$ with 0.9921 days is 1.2548, suggesting the planet at 197 is in indeed the best candidate.

Now that there are arguments in favour of a white noise and three planets, let us examine what happens when using a red noise model. The frequency span is restricted to 0 - 0.95 cycles per day to avoid spurious peaks (figure~\ref{hd69830}.c). 
As said above, the star is expected to have a jitter in the meter per second range, so we take for the additional jitter $\sigma_W = 0$, $\sigma_R$ = 1 m/s and try several characteristic correlation time lengths $\tau =$ 0, 3, 6, 10 or 20 days with definitions of equation~\eqref{corrnoise}. In that case, as said section~\ref{complexnoisemodels}, the estimation of the power is expected to be biased. Figure~\ref{hd69830}.c shows that the peaks at high and low frequencies are respectively over-estimated and under-estimated.
We suggest the following explanation: the weighting matrix accounts for red noise that has more power at low frequencies. Therefore, the minimization of~\eqref{l1} has a tendency to ``explain'' the low frequencies by noise and put their corresponding energy in the residuals. 

When the signal is more complicated, there might be complex effects due to the sampling resulting in a less simple bias. This issue is not discussed in this work, but we stress that when using different matrices $W$, the tolerance $\epsilon$ must be tightened to avoid being too affected by the bias on the peak amplitudes. 

To illustrate the advantages of our method, in appendix~\ref{appendix_wrongpeak}, we generate signals with the same amplitude as the ones of the present example but with periods and phases randomly selected. We show that the maximum of the GLS periodogram does not correspond to a planet in $\approx$ 7\% of the cases, while the maximum peak of the $\ell_1$-periodogram is spurious in less than $0.5 $\% of the cases.

\subsection{HD 10180}

\cite{lovis2011} suggested that the system could contain up to seven planets based on 190 HARPS measurements, whose nominal error bars are between 0.4 and 1.3 m.s\textsuperscript{-1}. The star is such that $\log R_{HK}'=-5$
which lets suppose an inactive star with low jitter. In \cite{lovis2011}, the presence of the planets at 5.79, 16.35, 49.74, 122.7, 600  and 2222 days is firmly stated. Let us mention that there is a concern on whether a planet at 227 days could be in the 
signal instead of 600 days, as they both appear on the periodogram of the residuals and $1/227- 1/600 +1/365  \leqslant 1/ T_{\mathrm{obs}}$, where $T_{\mathrm{obs}}$ is the total observation time. The possibility of the presence of a seventh planet  planet is also 
discussed. After the six previous signals are removed with a Keplerian fit, the tallest peaks on the periodogram of the residuals are at 6.51 and 1.178 days~\citep{lovis2011}. They are such that $1/6.51 + 1/1.178 - 1\leqslant 1/T_{\mathrm{obs}}$, so one is probably the alias of the other. The dynamical stability of a planet at 1.17 days is discussed in~\cite{laskar2012}, and its ability to survive is shown. However in our analysis, the statistical significance is too low to claim the planet is actually in the system.

We compute the $\ell_1$-periodogram for a grid span of 0 to 1.5 cycle/day and 0 to 0.95 cycles per day, giving respectively figures~\ref{hd10180}.b and c (blue curve). In appendix~\ref{appendix_noise} we show that when $W$ correctly accounts for the red noise, signals might become apparent. Therefore, on the latter we also test different weight matrices. As explained appendix~\ref{appendix_noise} and previous section, in that case we have to decrease $\epsilon_{\mathrm{noise}}$ and here $ F_{\chi^2_m} (\epsilon_{\mathrm{noise}}^2) = 0.1$ was taken. Where $F_{\chi^2_m}$ is the cumulative distribution function of the $\chi^2$ distribution with $m$ degrees of freedoms, $m$ being the number of measurements, in accordance with the notations of section~\ref{tuning}. We note that there is a signal appearing at 15.2 days and that there is a small peak at 23 days, which is close to the stellar rotation period estimate of 24 days~\citep{lovis2011}. Whether this is due to random or not is not discussed here.

Alike the case of HD 69830, the aliases are over-estimated when the frequency span is 3 cycles per day. In that case the highest one at 0.9976 days corresponds to an alias of the 2222 days period. We will see that in the two next systems the aliases are not as disturbing, which is discussed section~\ref{limitations}.

We now need to evaluate the significance of the peaks. The FAP test is performed for the seven highest signals, that are the published planets plus 0.177 days or 15.2 days. 
The latter appears for a non-diagonal weight matrix $W$, therefore when performing a Keplerian fit the $\chi^2$ we take is $(y(t)-\hat{y}(t))^TW^2(y(t)-\hat{y}(t))$ with the same $W$, that is $\sigma_W=0$, $\sigma_R=1$ m.s\textsuperscript{-1} and $\tau=$25 days (with notations of equation~\eqref{corrnoise}). This analysis gives figure~\ref{rvsurveyfap}.c and d.  In both cases the signals are below the significance threshold. It is also not clear which seventh signal to choose (figure~\ref{hd10180}.c), but doing the analysis with other candidates as 6.51, 23 or 67.5  days does not spot significant signals either.  Let us note that when choosing a non diagonal $W$, the FAP of the 16.4 and 600 days planets respectively increase and decrease. We suggest the following explanation: the noise model is compatible with noises that have a greater amplitude at low frequencies. As a consequence, the minimization has a tendency to interpret low frequencies as noise and ``trust'' higher frequencies. Deciding if a signal is due to a low-frequency noise or a true planet could be done by fitting the noise and the signal at the same time.

\subsection{55 Cancri}
\label{55cnc}
\subsubsection{Data set analysis}

Also known as $\rho$ Cancri, Gl 324, BD +$28^\circ$1660 or HD 75732, 55 Cancri is a binary system. To date, five planets orbiting 55 Cancri A (or HR 552) have been discovered. The first one, a 0.8 $\rm{M_j}$ minimum mass planet at 14.7 days was reported by~\cite{butler1997}. Based on the Hamilton spectrograph measurements, \cite{marcy2002} found a planet with a period of approximately 5800 days and a possible Jupiter mass companion at 44.3 days. With the same obsevations and additional ones from the Hobby-Eberly Telescope (HET) and ELODIE, \cite{mcarthur2004} suggested a Neptune mass planet could be responsible for a 2.8 days period.  \cite{wisdom2005} re-analysed the same data set and found evidence for a Neptune-size planet at 261 days and suggested that the 2.8 period is spurious. This was confirmed by \cite{dawsonfabricky2010}, which showed that the 2.8 days periodicity is an alias and the signal indeed comes from a super-Earth orbiting at 0.7365 days. The transit of this planet was then observed by~\cite{winn2011} and~\cite{demory2011}, confirming the claim of~\cite{dawsonfabricky2010}. In the meantime, using previous measurements and 115 additional ones, \cite{fischer2008} confirmed the presence of a planet at 261 days of minimum mass  $M \sin i = 45.7$ $\rm{M}_{\oplus}$. They also point out that in 2004 they observed two weak signals at 260 and 470 days on the periodogram. The constraints on the orbital parameters were improved by \cite{endl2012} based on 663 measurements: 250 from the Hamilton spectrograph at Lick Observatory, 70 from Keck, 212  from HJST and 131 of the High-Resolution spectrograph (Eberly Telescope), giving planets at 0.736546 $\pm 3.10^{-6}$, 14.651$\pm 10^{-4}$, 44.38 $\pm 7.10^{-3}$, 261.2 $\pm 0.4$ and 4909 $\pm 30$ days.  This is the set of measurements we will work on in this section. Let us mention also that~\cite{baluev2015_55cnc} and~\cite{nelson2014} studied respectively 55 Cnc dynamics and noise correlations including additional measurements~\cite{fischer2008}.

\begin{figure*}
\centering
\noindent
\hspace{-0.35cm}
\begin{tikzpicture}
\path (0.15,0) node[above right]{\includegraphics[scale=0.43]{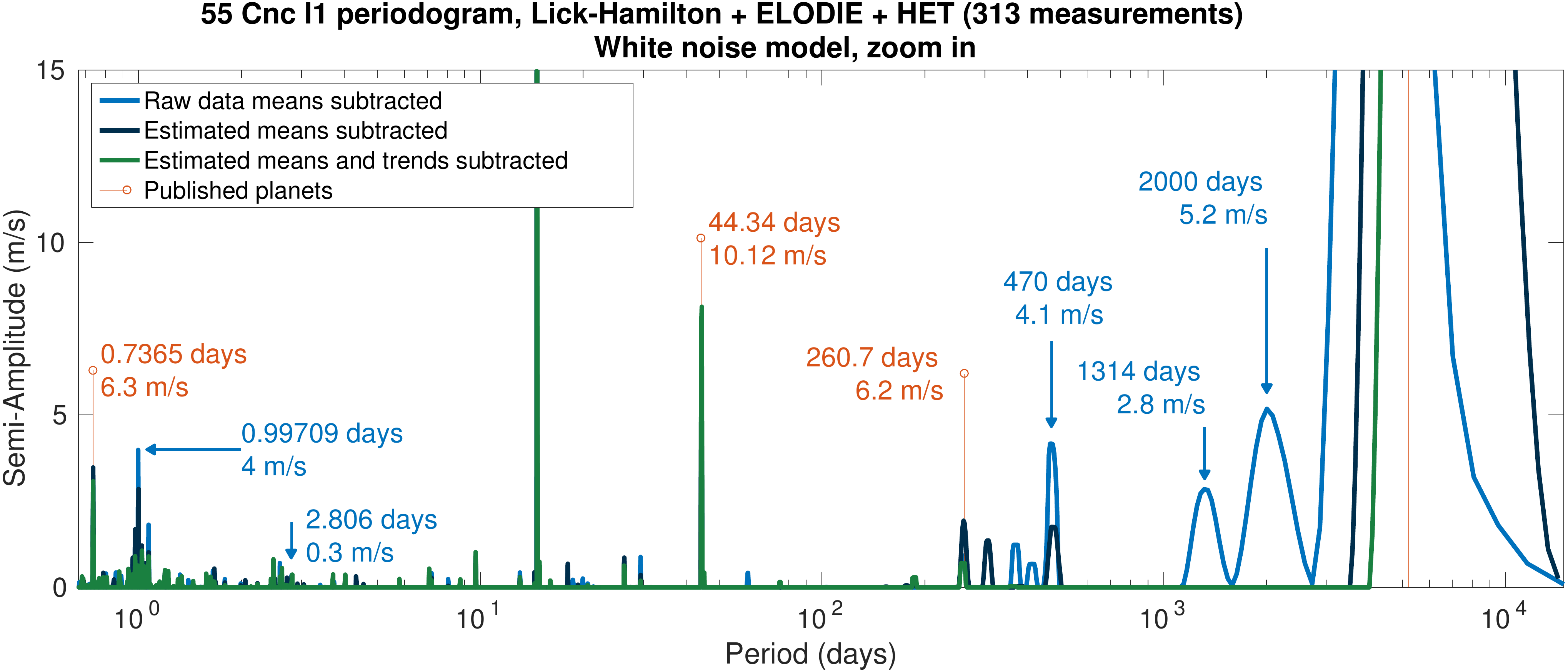}};%hd69830_fratio.png
\path (1.2,7.2) node[above right]{a)};
\begin{scope}[yshift=-7.6cm]
\path (0.15,0) node[above right]{\includegraphics[scale=0.43]{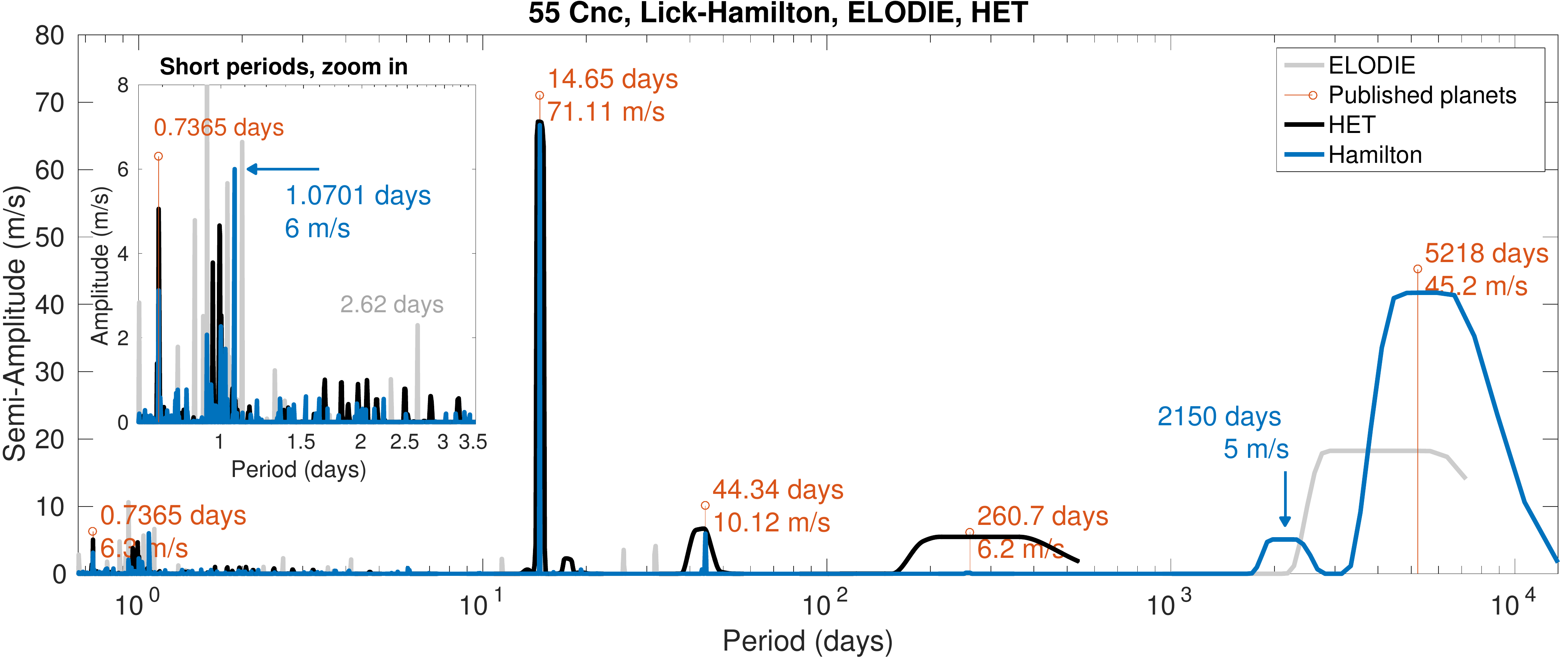}};
\path (1.2,7.2) node[above right]{b)};
\end{scope}
\begin{scope}[yshift=-15.1cm]
\path (0,0) node[above right]{\includegraphics[width=6.1cm]{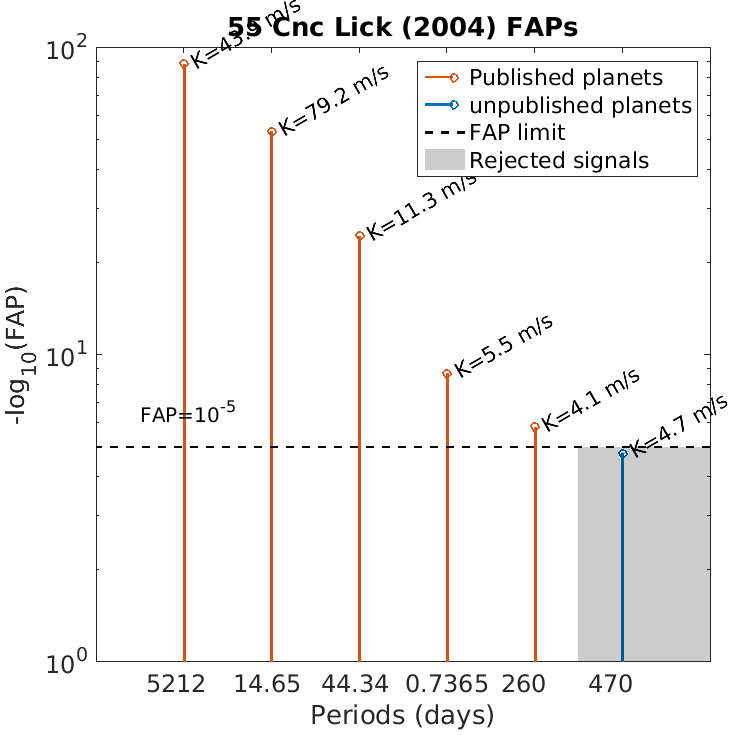}};
\path (1.2,5.7) node[above right]{c)};
\path (6,0) node[above right]{\includegraphics[width=5.9cm]{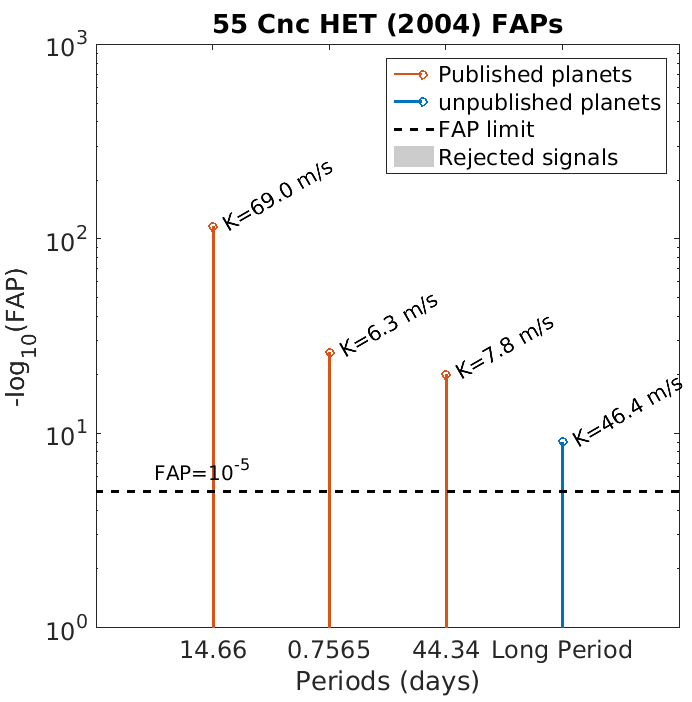}};
\path (7.2,5.7) node[above right]{d)};
\path (12.15,0) node[above right]{\includegraphics[width=6cm]{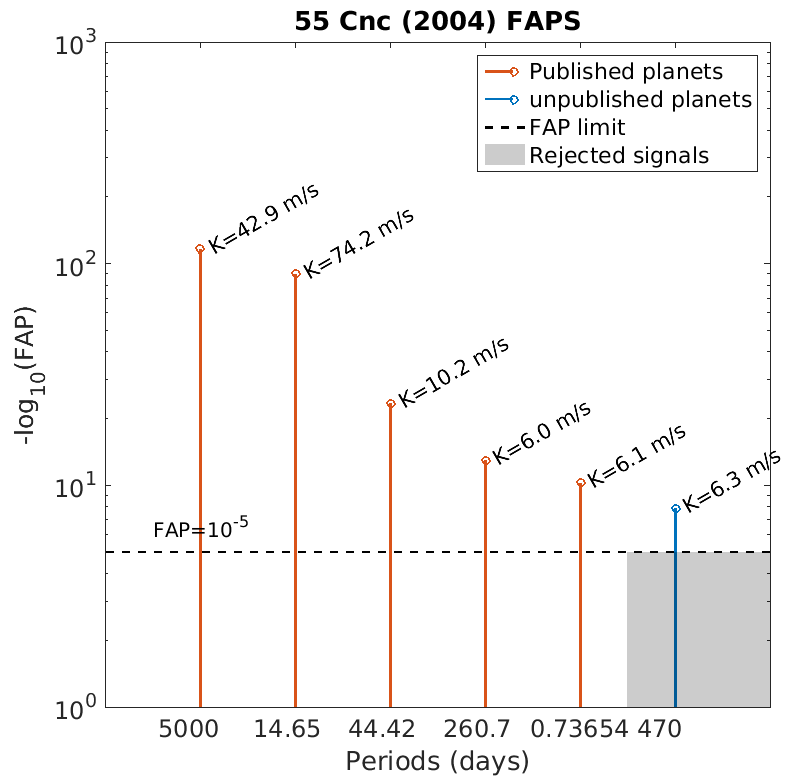}};
\path (13.2,5.6) node[above right]{e)};
\end{scope}
\end{tikzpicture}
\caption{$\ell_1$-periodogram of 55 Cnc, using measurements from the Lick-Hamilton, ELODIE spectrograph (Observatoire de Haute provence) and HET telescope.}
\label{55cnc313}
\end{figure*}
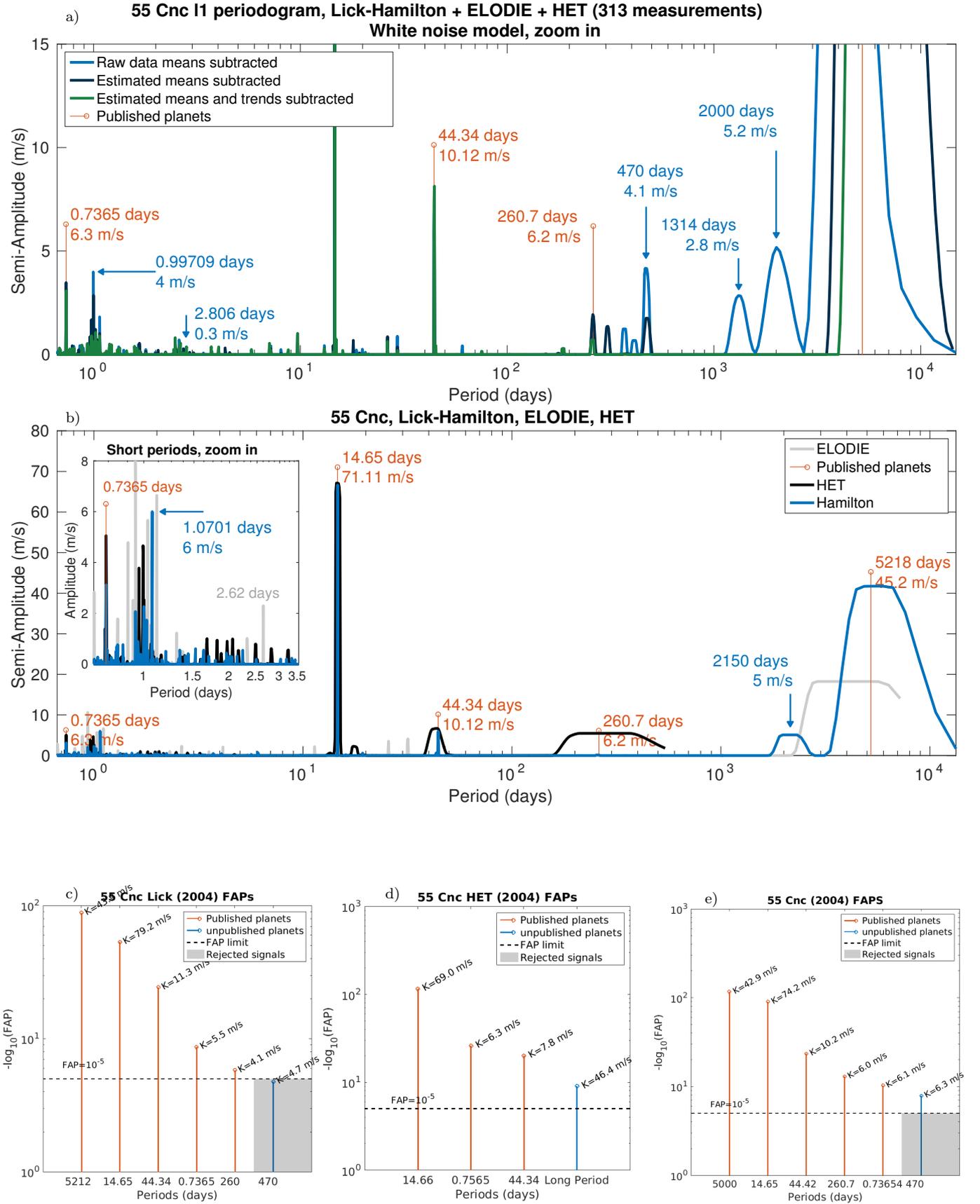

\begin{figure*}
\centering
\noindent
\hspace{-0.35cm}
\begin{tikzpicture}
\path (0.15,0) node[above right]{\includegraphics[scale=0.43]{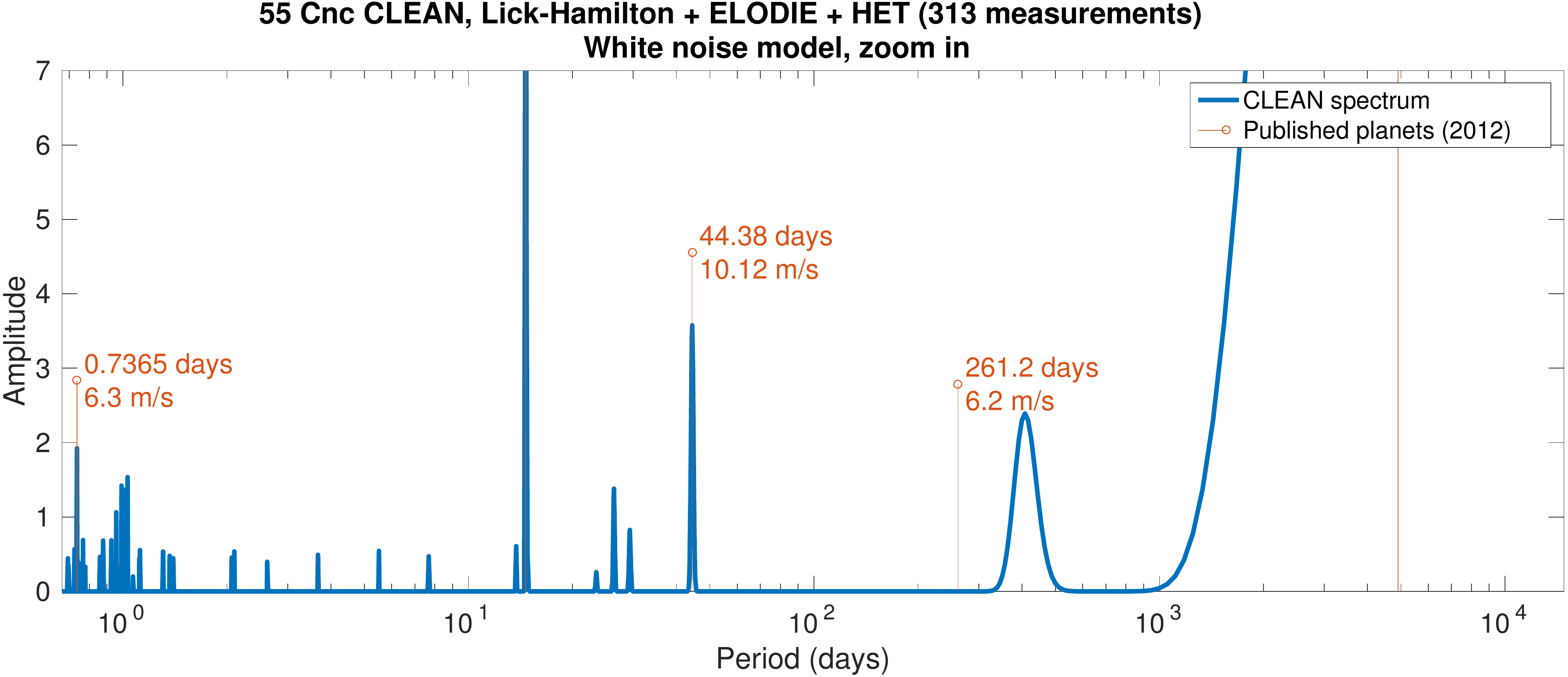}};%hd69830_fratio.png
\path (1.2,7.2) node[above right]{a)};
\begin{scope}[yshift=-7.6cm]
\path (0.15,0) node[above right]{\includegraphics[scale=0.43]{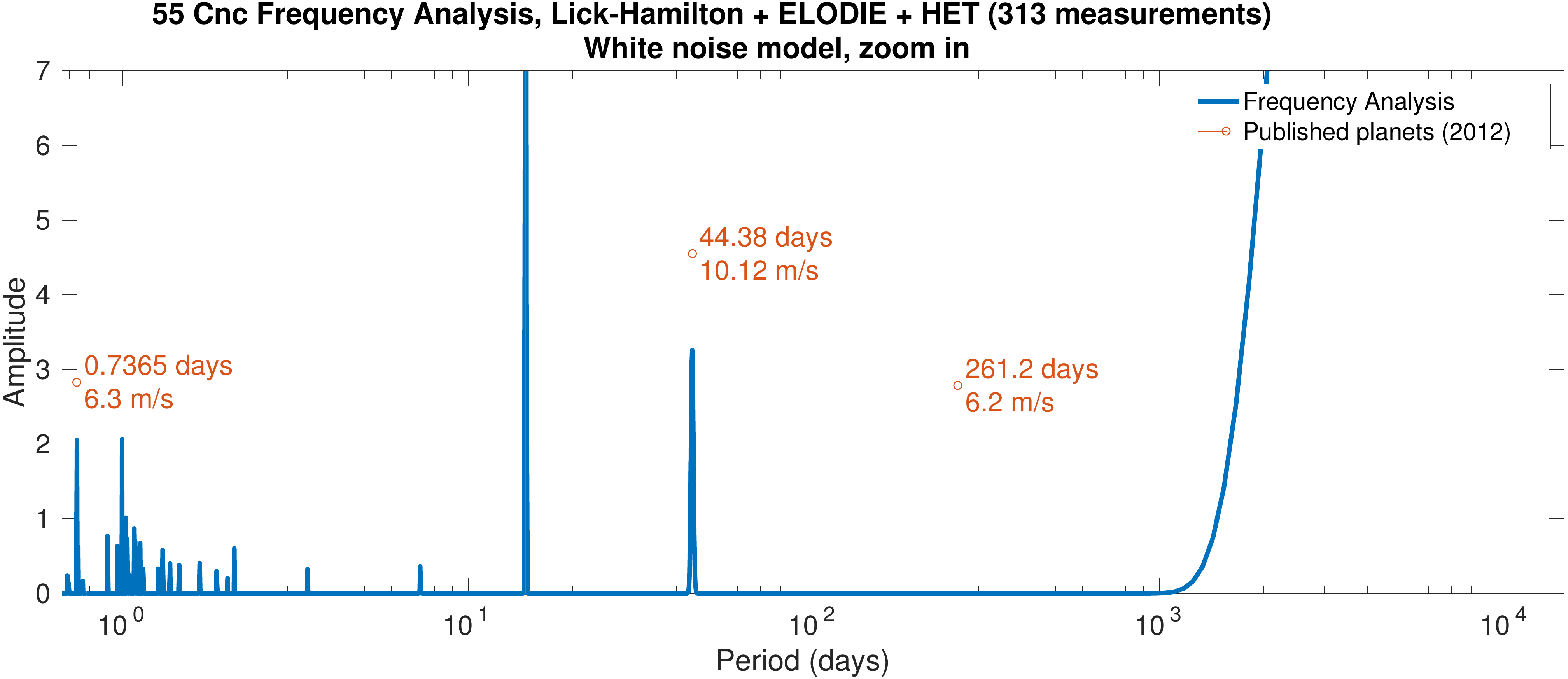}};
\path (1.2,7.2) node[above right]{b)};
\end{scope}
\end{tikzpicture}
\caption{a) CLEAN spectrum of 55 Cnc with the data available in 2004, b) Frequency Analysis of the same data}
\label{55cncclean}
\end{figure*}

Let us consider the set of 663 measurements from four instruments used in~\cite{endl2012}. The mean of each of the four data set is subtracted and the method described section~\ref{methods} is applied straightforwardly. 
 Here we only display the figure obtained for a white noise model as it is essentially unchanged when correlated noise is taken into account. Figure~\ref{rvsurvey55cnc}.b shows the $\ell_1$-periodogram and~\ref{rvsurvey55cnc}.c is the same figure with a smaller y axis range. The published signals appear without ambiguity. This is somewhat surprising, as the data comes from four different instruments and their respective mean was subtracted. Such a treatment is rather crude, so it shows that at least in that case the method is not too sensitive to the differences of instrumental offsets. When those are fitted with the planets found and corrected, a 365 days periodicity clearly appears on the $\ell_1$-periodogram.
 
 The FAPs computed following the method outlined section~\ref{significance} are significant (see figure~\ref{rvsurveyfap}.b). 
The sixth highest peak is at at 470 days, the FAP of which is too low to claim a detection. Interestingly enough, a signal at this period was mentioned by~\cite{fischer2008}. We will see next section that this one is already seen in 2004, and probably due to the different behaviour of the instruments at Lick and HET. The presence of a signal at 2.8 and 260 days in early measurements is also discussed.

\subsubsection{Measurements before 2004: no planet at 2.8 days nor 470 days but visible 55 Cnc e and f}
\label{55cnc28}
The 55 Cnc system has several features that are interesting to test our method. There has been some false detections at 2.8 days, and among candidate signals, one was confirmed (260 days) and one was not (470 days). We now have at least 663 reliable measurements that are very strongly in favour of five planets. As a consequence, the method can be applied on a shorten real data set with specific questions in mind, while being confident about what really is in the system. We will see that the use of the $\ell_1$-periodogram could have helped detecting the true planets based on the 313 measurements considered in~\cite{mcarthur2004}. These ones are from Hamilton spectrograph at the Lick Observatory, the Hobby-Eberly Telescope (HET) and ELODIE (Observatoire de Haute Provence). We also show that the signal at 0.7365 days (55 Cnc e) was detectable on the separate data sets from Lick or from HET available in 2004.

Our method is first applied to the three data sets at once, the means of which were subtracted, which gives the lighter blue curve on figure~\ref{55cnc313}.a. The true periods appear, although the 260 period is very small and there are peaks at 470, 1314 and 2000 days (the other features of the figure will be explained later). We then consider the three data sets separately, the figure obtained is displayed figure~\ref{55cnc313}.b.
 The fact that the $\ell_1$-periodograms of each three instruments span on different length is due to the fact that they don't have the same observational span. As the moving average on the result  of SPGL1 is $2\pi/3T_{\mathrm{obs}}$, it is wider when the total observation time $T_{\mathrm{obs}}$ is small. The 14.65 and long periods are seen for each data sets, but the 0.7365 and 44.34 days periodicities are not seen for the ELODIE data set. Interestingly, HET $\ell_1$-periodogram displays a periodicity close to 260 days. However, one cannot claim a detection at this period  in HET data, as those only span on 180 days, any period longer than the observation timespan is very poorly constrained. Furthermore, the period at 2.8 days is not seen in any data set. The closest one would be a peak at 2.62 days obtained with ELODIE data, which was checked not to be significant. The 470 days periodicity does not appear either. We show next paragraph that this is likely due to the velocity offset between Lick-Hamilton and HET data sets.
 Let us point that CLEAN~\cite{roberts1987} or Frequency analysis~\cite{laskar1988,laskar1992} (see figure~\ref{55cncclean}) also allow to retrieve the 0.7365 periodicity, which basically means that the strongest peak of the residual was already this one in 2004.

To compute the significance, the method of section~\ref{significance} is applied to the Lick and HET data separately. The FAPs are computed for circular models with an increasing number of planets whose periods correspond to the subsequent tallest peaks of the $\ell_1$-periodogram. Here, as the data comes from different instruments we add to the model three vectors $1_{\mathrm{Lick}}(t), 1_{\mathrm{Elodie}}(t)$ and $1_{\mathrm{HET}}(t)$ where $1_I(t)=1$ if the measurement at time $t$ is made by instrument $I$, $1_I(t)=0$ otherwise.
 In the case of Lick data, there is a peak of 6 m.s\textsuperscript{-1} at 1.0701 days, but this one can be discarded as it is an alias of the 14.65 days periodicity.  In both HET and Hamilton data, the 0.7365 periodicity is significant (figure~\ref{55cnc313}.c and d). Also, one sees a significant long period in both cases (respectively 8617 and 5212 days). The HET data set spans on 170 days, so in this case one can only guess that there is a long period signal. Finally, when combining the two data sets, the 470, 2150 and 1314 days periodicities become insignificant. 

The difference in zero points of the three instruments has a signature on the $\ell_1$-periodogram. Indeed, in problem~\eqref{BPDNepsilonw}, the signal is represented as a sum of sinusoids. The algorithm could then attempt to ``explain'' the bumps in velocity that occur when passing from one instrument to the other by sines. The previous analysis ensures the presence of four periodicities in the signal: at $\approx$ 14.65 day, 44.34, 5000 and 0.7365  days. The fit with these four periods plus the vectors $1_I(t)$ gives coefficients of the latter $\alpha_{\mathrm{Lick}}, \alpha_{\mathrm{Elodie}}$ and $\alpha_{\mathrm{HET}}$. 
The vector 
$\alpha_{\mathrm{Lick}} 1_{\mathrm{Lick}}(t)+ \alpha_{\mathrm{Elodie}} 1_{\mathrm{Elodie}}(t)+ \alpha_{\mathrm{HET}} 1_{\mathrm{HET}}(t)$ is subtracted from the raw data. The $\ell_1$-periodogram of the residuals is computed, which gives the dark blue curve figure~\ref{55cnc313}.a). The 2000 and 1314 
periods disappear and the 470 days peaks decreases. Interestingly enough the 5th tallest peak (except the 0.99709 days alias) becomes 260 days, which was suggested by~\cite{wisdom2005} and confirmed by~\cite{fischer2008} and~\cite{endl2012}, but it does not appear on the CLEAN spectrum nor the Frequency analysis (figure~\ref{55cncclean}.a and b).

We now fit the model with five planets along with the $1_I$ vectors and trends for 
each instrument, that are vectors $t_I$ such that $t_I(t) = t$ and 0 elsewhere if the measurement at time $t$ is done by the instrument $I$. The vector
 $\sum \alpha_k 1_{I_k} + \beta_k t_{I_k}$ is subtracted from the raw data, and we compute again the $\ell_1$-periodogram (figure~\ref{55cnc313}.a, green curve). This time, the 470 days periodicity disappears, suggesting -- though not proving -- it is due to a difference in behaviour between the instruments. The fact that the 470 days signal disappears just shows its presence depends on the models of the instruments.
 The same analysis on Lick and HET data altogether shows the same features at 470 days, therefore we exclude the possibility that it is due to the lesser precision of ELODIE.

The analysis by~\cite{wisdom2005} does not use $\ell_1$ minimization to unveil the 260 days periodicity (55 Cnc f). We tried to reproduce a similar analysis ``by hand''  on the same data set, namely the one of~\cite{mcarthur2004}. The rationale is to determine if it was easy to make 55 Cnc f appear with an analysis more conventional than the $\ell_1$-periodogram. Also, the short period planet can be injected at 0.7365 days, not $\approx$2.8 days as it was then. We found that the size of the peak in the residuals at 260 days depends on the initialization of the fits, both with classical and recursive periodograms. While in most cases the 260 periodicity does appear in the residuals, it sometimes coexists with peaks of similar amplitude.  Interestingly enough, an analysis of Lick-Hamilton and HET data sets by recursive periodograms suggests that the periods estimated by HET are shifted to longer ones with respect to Lick ones. We found that adding the periods 14.8, 15000 ($1/14.65 - 1/14.8 \approx 1/5000-1/15000$) to those of the four planets and a 2500 one (probably due to an harmonic of the 5000 days periodicity) makes the 400 (seen on the CLEAN spectrum figure~\ref{55cncclean}.a) and 470 periodicity disappear and the 260 days peak appears very clearly. As the data comes from an older generation of spectrographs one could expect complicated systematic errors. Again, this discussion focuses on the possibility of seeing the 55 Cnc f in 2004, we do not raise the question of its existence, well established by the subsequent measurements. 

% , with recursive periodograms on Lick and HET data sets separately and jointly. It seems like the periods estimated by HET are shifted to longer ones with respect to Lick ones. We found that adding the periods 14.8, 15000 ($1/14.65 - 1/14.8 \approx 1/5000-1/15000$) to those of the four planets and a 2500 one (probably due to an harmonic of the 5000 days periodicity) makes the 400 (seen on the CLEAN spectrum figure~\ref{55cncclean}.a) and 470 periodicity disappear and the 260 days peak appears very clearly. When including eccentricity in the fits, the results seem to be sensitive to the periods set in input of the Keplerian fits and the order in which they are performed. The 260 days periodicity appears in most cases but we obtain peaks with similar amplitudes elsewhere.  Again, this discussion focuses on the possibility of seeing the 55 Cnc f in 2004, we do not raise the question of its existence, well established by the subsequent measurements. 

Finally, we perform the FAP test on the data from the three instruments (see figure~\ref{55cnc313}.e). The model is made of Keplerians plus the $1_I$ vectors. The four significant signals in each data set are still significant. The 260 days periodicity is significant as well. This analysis shows that both the 0.7365 and 260 days periodicity were already present in the data. Long periods might be due to instrumental effects, therefore the planetary origin of the 260 period could have been subject to discussion. In contrary, it seems hard to explain a steady 0.7365 days periodicity with a non-planetary effect.

\subsection{GJ 876}
\label{gj876}
\subsubsection{Previous work}
 
The GJ 876 host star is one of the first discovered multiplanetary systems. First, two giant planets at 30 and 61 days were reported by~\citep{marcy1998,delfosse1998}. Subsequently,~\cite{rivera2005} finds a short period Neptune at 1.94 days and a Uranus-mass planet at 124 days~\citep{rivera2010}. 

The giant planets are close to each others and in 2:1 resonance, therefore we might expect visible dynamical effects. Indeed,~\cite{correia2010}, \cite{baluev2011}  and~\cite{nelson2016} perform 4-body Newtonian fits which give a $\chi^2$ of the residuals smaller than a Keplerian fit. The dynamical fits also allow to have constraints on the inclinations, therefore on the true masses of the planets. Furthermore,~\cite{baluev2011} shows that the maximum of a posterior likelihood including a noise model as the one used here (equation~\eqref{corrnoise}) occurs at $\sigma_W=1.31$ m.s\textsuperscript{-1}, $\sigma_R=1.8$ m.s\textsuperscript{-1} and $\tau$ = 3 days.

\cite{jenkins2014} takes a different approach and searches for sine functions in the signal. They claim six significant sinusoidal signals are in the data. The following discussion first confirms these results. Secondly, we investigate the origins of the additional two signals and find they are likely to be due to the interactions between the giant planets.

\subsubsection{Six significant sines}
\label{gj876_sixsines}

 \cite{jenkins2014} analyses the GJ 876 data by aiming at solving the problem~\eqref{p1}, which they call Minimum Mean Squared Error (MMSE). To do so, the phase space is explored with an iterative arborescent method. They find the following periods: 61.03$\pm$3.81, 30.23$\pm$0.19, 15.04$\pm$0.04, 1.94$\pm$0.001, 10.01$\pm$0.02 and 124.69$\pm$90.04 days. 
To compare our results with~\cite{jenkins2014}, the significance of the signals is tested with FAPs as previously. We use different weight matrix models according to equation~\eqref{corrnoise} and two grid spans: 1.5 cycles per day and 0.95 cycles/day (see figure~\ref{rvsurvey_gj876} b and c). On figure~\ref{rvsurvey_gj876}.c, we see that the six tallest signals correspond to the periods we expect. Depending on the noise model, the seventh tallest peak varies. We compute the FAP test for 7.748, 1200 or 4200 days as candidate 7th planets, respectively with the $W$ matrix yielding their greatest amplitude. On figure~\ref{rvsurveyfap}.e), we display the result for 7.748 days but in other cases the signals are not significant. Let us still point out that in the case of $\tau = $ 6 days, initializing a 4200 days periodicity, after the non-linear fit we obtain a 4862 days periodicity which has a FAP of $0.0007$. This one is close to the total observation timespan (4600 days). Therefore it is hard to determine what could be its cause.

Before discussing the origin of these signals, we wish to comment the behaviour of the $\ell_1$-periodogram towards the 124 days perodicity. Indeed, in the case of the 1.5 cycles per day, this one has the same order of magnitude as the tallest alias in the one day region (at 0.9812 days, alias of the 61 days periodicity). Furthermore, the peak becomes visible only for non diagonal weight matrix $W$, while a white noise model is sufficient to see it when using a shorter grid (figure~\ref{rvsurvey_gj876}.c). To understand this feature, we argue as follows. There are three effects against finding the correct planets: the red noise~\citep{baluev2011}, the uncertainties on the two instrumental means and the inner faults of our method. The persistence of aliases at one day indeed shows that the recovery of the true signals is more difficult when considering a grid $\Omega$ where some of the frequencies are very correlated. We also computed the $\ell_1$-periodogram when the mean of each instrument is corrected after the orbital parameters fit, as done section~\ref{55cnc28}. In that case the 124 days periodicity does appear and the aliases are reduced. We suggest the following explanation: when at least one of the three obstacle is correctly taken into account, the method is sufficient. When the three are ignored, their joint effect is deadly to our ability to recover the correct planets.

\subsubsection{Signals at 10 and 15 days}

 Now that the six sines are seen in the signal, we show that the peaks at 15.06 and 10.01 days are due do the dynamical interactions. 

We perform the same 4-body fit of GJ 876 with the same method as~\cite{correia2010}. This one includes 25 parameters: the mass of the star, a velocity offset, the mass of the planets, for the smallest planets: period, semi-amplitude, eccentricity, argument of periastron and initial mean anomaly. For the giant planets at 30 and 61 days the inclination is also a free parameter. 

A planetary system with the orbital elements found by the least square fit is simulated on 100 years for the two giant planets and the four planets at once. The frequency analysis~\citep{laskar1988,laskar1992,laskar1993} is then performed on the resulting time series of the star velocity along the x axis. We find that 15.06 and 10.01 periods appear and are a combination of the fundamental frequencies. Denoting by $\omega_{P}$ the frequency of a planet of period $P$, we have: 
$ \omega_{15} = 3 \omega_{30} - 2\omega_{60}$ and $ \omega_{10} = 5 \omega_{30} - 4 \omega_{60}$, both in the two planet and four planets cases. We also performed another test: if we adjust the two giant planets with a dynamical fit, then the peaks at 15.06 and 10.01 days are not seen on the residuals.
 This agrees with the analysis of~\cite{nelson2016}, where they discuss the possibility that the signals at 10.01 and 15.06 days could be due to additional planets, and find it unlikely. They compute the evidence ratio of Newtonian models with four and five planets, $\Pr\left\{y|5 \;\rm{planets} \right\}/ \Pr\left\{y| 4\; \rm{ planets} \right\} $,  and find it is not higher than the threshold we chose. 
  The difference between $ \omega_{15}$ and the first harmonic of the planet gives an estimate of the frequency of precession of the periastron of the inner orbit, we find 2$\pi$(1/$\omega_{15}$ - 2/$\omega_{30}$)$\approx$ 8.77 years, which is consistent with the estimate of~\cite{correia2010} ($g_2 = 8.73$ year, table 4). 

To obtain the expressions of $\omega_{15}$ and $\omega_{10}$, we used Frequency Analysis. This could be puzzling as the present work defines a method to retrieve the frequencies in the signal. The rationale is that we do the frequency analysis on a numerical integration, therefore we have tens of thousands of points available. Frequency analysis has been used in that situation for years and is known to be fast and robust. We double checked the results by computing the $\ell_1$-periodogram on a thousand points from the simulation (handling as many as the frequency map analysis is too long for now), the period at 15.06 and 10.01 do appear very clearly.

\subsection{Very active star (simulated signal)}
\label{activity}

The examples above concern rather quiet stars, where the noise can be modelled by Gaussian time series. However, in some cases the stellar activity has not a Gaussian signature. The method described here is not yet adapted to handle such situations. In this section we show that the problem can be circumvented, provided there are enough measurements.

We exploit the fact that stellar noise can be correlated with the bisector span~\citep{queloz2001}, the full width at half maximum (FWHM) and the $\log R_{HK}'$. This correlation has been used for example in~\cite{meunier2012}, which shows that the detection threshold limit improves by an order of magnitude by testing the correlation between the radial velocity and ancillary measurements. They compute the correlation of the periodograms of radial velocity measurements and bisector span, but a correlation in the frequency domain is also visible in the time domain, as the Fourier transform contains the same amount of information as the original time series. Here we take an approach similar to~\cite{melo2007},~\cite{boisse2009} and~\cite{gregory2016} insofar as we use the ancillary measurements as proxys for estimating the activity induced signal. Here, we simply fit and remove the three ancillary measurements from the data then use the method described above on the residuals. To compute the FAP we use a model of the form $A \rm{FWHM} + B\rm{bisector} + C \log R_{HK}' + \mathrm{Circ(k,h,P,D,E)}$, Circ denoting a circular model as defined section~\ref{modeldef}.  The validity of this approach is discussed in~\ref{appendix_stellarnoise}.

The data set used is taken from the RV Fitting Challenge~\citep{dumusque2016_1,dumusque2016}. In this challenge, fifteen systems were simulated with a red noise component taken from observations of real stars plus activity simulated via SOAP 2~\citep{dumusque2014}. Here we consider the system number two of the challenge. The data set is made of 492 measurements and the mean precision is 0.67 cm.s\textsuperscript{-1}. The first step of the processing is to fit a linear model made of the ancillary measurements, an offset, a linear and a quadratic trend (6 parameters).  Secondly, we compute the $\ell_1$-periodogram for different weight matrices, which gives figure~\ref{rvsurvey_challenge}.c. The Generalized Lomb-Scargle periodogram is also computed before and after the fit of the 6 parameters for comparison (figure~\ref{rvsurvey_challenge}.a and b).

We find without ambiguity the three planets whose semi-amplitude is above 1 m.s\textsuperscript{-1}, and also the 20.16 days  periodicity. The planet with the smallest amplitude does not appear clearly, but there is a peak at 5.4 days which seems to be significant. In fact, the spectral window is such that 5.4 days is an alias of 5.32 = 10.64/2 days, and corresponds to the first harmonic due to eccentricity. This feature seems to be due to an error in the noise model. When accounting for a red noise effect, the relative amplitude of 5.32 and 5.4 changes in favour of 5.32 days. This effect is also observed on the recursive periodograms which are not represented here for the sake of brevity. One can see a peak at 6.25 days which grows stronger as the characteristic correlation time of the noise model increases. This coincides with the fourth harmonic of the rotational period and is therefore not surprising.

%There is also a peak close to the 120 days signal but it is below the FAP limit of $10^{-4}$. In that case the peaks also appear clearly on the GLS periodogram computed after the fit of the ancillary measurements to the raw data (figure~\ref{rvsurvey_challenge}.b). This should not be too surprising, as there are 492 high precision measurements. The limiting factor is the stellar activity, which seems in that case to be reasonably well approximated by the ancillary measurements. 

\section{Discussion}
\label{discussion}
\subsection{Summary}
The present work was first devised to overcome the distortions in the residual that arise when fitting planets one by one. It is compatible with the assumption that the noise is Gaussian and correlated through the weighting matrix $W$. One of the main advantages of the method is that, as opposed to global $\chi^2$ minimization, the minimization problem~\eqref{BPDNepsilonw} is convex therefore quicker to solve. On our workstation (Intel Xeon CPU E5-2698 v3 at 2.30 GHz) it takes typically thirty seconds to ten minutes to obtain (resp. for HD 69830, 74 measurements and 55 Cnc, 663 measurements).  The speed here depends mainly on three parameters: the number of observations $m$, the number of columns of matrix $A$ (see section~\ref{dictionary}), $n$, and the precision wanted in output, tol (see section~\ref{optimizationroutine}). The SPGL1 algorithm used to solve~\eqref{BPDNepsilonw} relies on a Newton algorithm, therefore its complexity is $O(\log(p) F(p))$  where $p = 10^{-\mathrm{tol}}$ is the number of significant digits desired and $F(p)$ the cost of evaluating the objective function to $p$ digits accuracy. The most expensive steps of the evaluation are a matrix vector product and a projection onto a convex set~\citep[see][]{spgl1}, which have a respective complexity of $O(mn)$ and a worst case complexity of $O(n\log n)$. The post processing operation also is in $O(mn)$. This overall should amount asymptotically to complexity $O(mn)$, similarly to the Lomb-Scargle periodogram. Its complexity is in $O(mn)$ if there are $m$ measurements and $n$ frequency scanned. The constants are however different.

Furthermore, our method does not require the number of planets as input parameter and offers a graphic representation of the information content of the signal. However, the statistical properties of the solution are not as easy to interpret as in the case of a global least square minimisation. 
Considering that the method presented here is in its infancy, comparing its merits to other techniques is left for future work. Here, we will only stress that the $\ell_1$ and Generalized Lomb-Scargle periodogram are tools are of different levels, and we do not advocate to give up the latter.
 
We will confine ourselves to addressing some internal issues of our method.
Ultimately, we would like to know if is there a way to determine which peaks are to be associated to planets. As the present paper is concerned with unveiling the periodicities in the signal but not their origins, we will address a simpler question: assuming the signal is only made of sines plus a Gaussian noise, are there risks to see spurious peaks on the $\ell_1$-periodogram ?

Unfortunately the answer is yes, as we have seen in the previous examples. The method is in particular sensitive to the aliases due to the daily repetition of the measurements: spurious peaks are especially present around one day periods. To shed some light on this problem, the following questions will be briefly discussed in the two next sections:
\begin{enumerate}
\item Are spurious peaks to be expected from the theoretical properties of the method or from its implementation? 
\item If they are to appear anyway on the $\ell_1$-periodogram, is there a way to spot them ?
\end{enumerate}

\subsection{Mutual coherence}
\label{limitations}
To test if the algorithm behaves appropriately, we reason as follow. Considering a set of observational times $t=t_1...t_m$, a linear combination of $p$ pure sine signals $y(t_k) = a_1 \cos( \omega_1 t_k + \phi_1) + ... + a_p \cos( \omega_p t_k + \phi_p)$ is generated with uniformly distributed phases $\phi$ and various amplitudes. For any tolerance $\epsilon$, the SPGL1 algorithm must give a solution $x^\star$ (see equation~\eqref{BPDNepsilonw}) such that $\|x^\star \|_{\ell_1} \leqslant |a_1|+...+|a_p|$, as obviously $y(t)$ belongs to the set of signals $u$ verifying $\|u-y(t) \|_{\ell_2} \leqslant \epsilon$. To test if SPGL1 gives the best solution we take the measurement dates of HD 69830 and generate three pure cosine functions of amplitude one whose frequencies are in the grid. They are fed to the SPGL1 solver for $\epsilon=0.01$ and $W$ equal to the identity matrix. The solution $x^{\star}$ to~\eqref{BPDNepsilonw} must verify $\|x^\star\|_{\ell_1} < 3$ as the original signal is not noisy. The test is performed for a thousand set of three frequencies randomly selected on the grid. We find that the average $\ell_1$ norm of the solution is 3.26, suggesting the algorithm could be improved.

Secondly, in the discrete case (problem~\eqref{l1}) there are theoretical guarantees on the success of the recovery if the mutual coherence of the dictionary is sufficiently small~\citep{donoho2006_2}. This one is defined as the maximal correlation between two columns $a_j$ and  $a_k$ of the dictionary $A$.
\begin{equation}
\mu = \max\limits_{ \tiny \begin{array}{c}
k=1..n \\j=1..n \\ j\neq k
\end{array} } |\langle a_k,a_j \rangle |
\end{equation}
In the case of a dictionary such that $a_k=\e^{\ii \omega_k t}$, taking the convention $\langle a_k, a_j \rangle=a_k^{\ast} a_j$ where the superscript $\ast$ denotes the conjugate transpose,  
 \begin{align}
 |\langle a_k,a_j \rangle | & = \left| \sum\limits_{l=1}^m \e^{-\ii (\omega_k-\omega_j) t_l} \right| 
 \end{align}
 that is the spectral window in $\omega_k-\omega_j$. As a consequence, the method cannot resolve very close frequencies due to their high correlation. More importantly,  aliases are still a limitation -- though not as much as in iterative algorithms in general~\citep{donoho2006}, see also appendix~\ref{appendix_wrongpeak}. This feature is responsible for the aliases that still appear around one day, where there is generally a strong alias due to the sampling constraints. The problem tends to get worse as the maxima of the spectral window increase. Aliases are higher relative to the true peaks for HD 69830, HD 10180 and the separate sets of 55 Cnc than GJ 876 (see figures~\ref{hd69830},~\ref {hd10180},~\ref{rvsurvey55cnc},~\ref{rvsurvey_gj876},~\ref{55cnc313} and table~\ref{swindow_peaks}). 
 
 \begin{table}\centering
 \caption{Maximum amplitude of the spectral window in the 1 cycle/day and 1 cycle per year for the examples considered here }
 \begin{tabular}{l|c|c}
           & $\approx$ 1 cycle/day & $\approx$ 1 cycle/year  \\\hline
  HD 69830 & 0.926 & 0.600 \\ \hline
  HD 10180 & 0.949 & 0.703 \\ \hline
  55 Cnc   & 0.822 & 0.557 \\ \hline
  GJ 876   & 0.73246 & 0.501 \\ \hline
  RV Challenge 2 & 0.870 & 0.800 \\
   \end{tabular}
 \label{swindow_peaks}
 \end{table}
 
 \subsection{Spotting spurious peaks}
 
We know that the theoretical obstacle for a good recovery is correlation between the elements of the dictionary. If a frequency $\omega_0$ truly is in the signal, it is expected to cause significant amplitudes at $\omega_0 + \omega_k$ where the $\omega_k$ are maxima of the spectral window. So if two peaks at frequencies $\omega_1$ and $\omega_2$ are seen on the $\ell_1$-periodogram and the spectral window has a strong local maximum close to $\omega_1-\omega_2$, one can suspect that one of the two peaks is spurious.

\subsection{When to use the method ?}

We consider the general problem of finding the frequencies of a signal made of several harmonics (the multi-tone problem). It seems natural -- though not mandatory -- to try to find the global minimum for a given number of sinusoids, and possibly additional parameters such as the offset or a trend. 
 We do not know \textit{a priori} the number of sinusoids in the signal. Ideally, we would like to solve the global minimisation~\eqref{p1} for any number of sines inferior to the number of measurements and regarding their amplitudes, which ones seem to truly be in the signal. The approach consisting in using grids has a computational cost growing exponentially with the number of frequency. Therefore, strategies must be found to estimate a reliable solution to this problem.
The recursive periodogram~\citep{angladaescude2012}, the treillis approach \citep{jenkins2014} or the super-resolution methods~\citep{candesfernandez2012, tangbhaskar2013} can be viewed as a way to approximate~\eqref{p1} and selecting the relevant number of frequencies at the same time. These ones have the advantage of not being bothered by the $\ell_1$ norm minimization, which biases downwards the amplitude of the signal. Even more, the bias becomes more complicated when using a correlated noise model.

The most interesting use of the $\ell_1$-periodogram seems to be as a complement to the classical periodogram: it gives a much clearer idea of the number of spikes and their significance. If the peaks spotted by the $\ell_1$-periodogram yield a $\chi^2$ of the residuals consistent with the noise assumptions as in HD 69830, then it is likely that there is not many more signals. To check that there are not very high correlations between signals one can use the spectral window.
Furthermore, we have exhibited appendix~\ref{wrongpeak} examples where the main peak of the classical periodogram is spurious while $\ell_1$ minimization~\eqref{l1} avoids selecting the first spurious peak. Such an example was also presented in~\cite{bourguignon2007}. Those findings are consistent with the claims of~\cite{donoho2006}: the $\ell_1$ method are more reliable in general than orthogonal matching pursuit. 
 A failure of the $\ell_1$-periodogram is also informative, as shown figure~\ref{l1failure} appendix~\ref{wrongpeak}. If there still is a forest of peaks below a certain amplitude it might indicate that the signal is noisy, possibly that noise is higher than expected or non Gaussian. This means that the set of observations requires a more careful analysis. To sum up, the $\ell_1$-periodogram can yield an estimation of the difficulty of the system, in some cases it is a short-cut to random searches and its use decreases the chance of being mislead by a spurious tallest peak. 
\section{Conclusion}
\label{conclusion}
The aim of the present paper was to produce a tool for analysing radial velocity that can be used as the periodogram but without having to estimate the frequencies iteratively. To do so, we used the theory of Compressed Sensing, adapted for handling correlated noise, and went through the following steps:

\begin{enumerate}
\item Selecting a family of normalized vectors where the signal is represented by a small number of coefficients.
\item Approximating a solution to~\eqref{ANDNwlambda}; for example by discretizing the dictionary, and ensuring the grid spacing is consistent with the noise power (see eq~\eqref{omegabound}) then solving~\eqref{BPDNepsilonw} with SPGL1 and take the average power. The introduction of the weight matrix $W$ accounts for correlated Gaussian noises. 
\item Estimating the detection significance, which we do by computing subsequent FAPs of the models with an increasing number of planets.
\end{enumerate}

We showed that the published planets for each systems could be seen directly on the same graph, and that taking into account the possible correlations in the noise could make a signal appear. This was established in the case of radial velocity data but the method could be adapted to other types of measurements, such as astrometric observations.

The use of the Basis Pursuit/$\ell_1$-periodogram we suggest is as follows. This method can be used as a first guess to see if the signal is sparse or not, in that extent it constitutes an evaluation of the difficulty of the system and possibly a short-cut to the solution. It can bring attention to signal features that are hidden in the classical periodogram, which can still be used for an analysis ``by hand''. Secondly, for confirming the planetary nature of a system we advocate to use in a second time statistical hypothesis testing.

The perspective for future work are two-fold. First, we saw that the algorithm itself could be improved. Also, there might be significance tests more robust than the FAP and the effect of introducing a weight matrix $W$ must be studied into more depth. 
Secondly,  let us recall that our method uses an \textit{a priori} information, that is the sparsity of the signal, but still does not handle all the information we have. To improve the technique we wish to broaden its field of application by:
\begin{itemize}
\item Adapting the method for very eccentric orbits, through the addition of Keplerian vectors to the dictionary for example.
\item Using precise models of the noise, especially magnetic activity, granulation, p-modes. Possibly include an adaptive estimation of the noise, especially one could extend the dictionary to wavelets.
\item Handling several types of measurements at once (e.g. radial velocity, astrometry and photometry).
\end{itemize}

\section{Acknowledgements}

The authors wish to thank the anonymous referee for his insightful suggestions. N. Hara thanks Evgeni Grishin for pointing out the Matched Filter technique to him. A. Correia acknowledges support from CIDMA strategic project UID/MAT/04106/2013.

\bibliography{biblio.bib} 
\bibliographystyle{mnras}

\appendix

\section{Minimum grid spacing}
\label{mindeltaomega}
Let us consider a signal made of $p$ pure harmonics sampled at times $t = (t_k)_{k=1..m}$, $y = \sum\limits_{j=1}^p c_j \e^{\ii \omega_j t} $. We denote by $\omega_j'$ and $\Delta \omega$ two a real numbers such that for each $j$ 
\begin{align}
\label{deltaomega1}
\Delta \omega & < \frac{4}{T} \\
|\omega_j - \omega_j'|& < \Delta \omega \; ,
\label{deltaomega2}
\end{align}
where $T=t_m-t_1$.
For each $t_k$ and each $j$,
\begin{align*}
|c_j||e^{\ii \omega_j t_k} - e^{\ii \omega_j' t_k} | & = |c_j| \left| \e^{\ii\frac{\omega_j +\omega_j'}{2} t_k} \left(\e^{\ii \frac{\omega_j -\omega_j'}{2}t_k} - \e^{-\ii \frac{\omega_j -\omega_j'}{2} t_k} \right)\right| \\
                  & = 2|c_j| \left|\sin  \left( \frac{\omega_j-\omega_j'}{2} t_k\right) \right|
\end{align*}

So denoting $y'= \sum\limits_{j=1}^p c_j \e^{\ii \omega_j' t}  $,
\begin{align*}
|y_k-y_k'| & = \left| \sum\limits_{j=1}^p c_j \left( \e^{\ii \omega_j t_k} - \e^{\ii \omega_j' t_k} \right) \right| 
\\
&\leqslant 2 \sum\limits_{j=1}^p \left|c_j \sin  \left( \frac{\omega_j-\omega_j'}{2}t_k \right) \right|
\end{align*}
Without loss of generality the origin of time is shifted to $-T/2$, therefore
\begin{align}
\label{deltaomega3}
2 \sum\limits_{j=1}^p \left|c_j \sin  \left( \frac{\omega_j-\omega_j'}{2}t_k \right) \right| \leqslant \sin\frac{\Delta \omega T}{4} \sqrt{\sum\limits_{j=1}^p |c_j|^2}
\end{align}

Finally, a condition for $y'$ to be an acceptable solution is

\begin{align*}
\|W(y-y')\|^2_{\ell_2} &  \leqslant \|W \|^2 \|y-y' \|_{\ell_2}^2 \\
                       & \leqslant  \|W \|^2 \sum\limits_{k=1}^m |y_k-y_k'|^2 \\
                       & \leqslant 4\|W \|^2  \sum\limits_{k=1}^m  \left( \sum\limits_{j=1}^p \left| c_j\sin \left( \frac{\omega_j-\omega_j'}{2} t_k\right) \right| \right)^2 \\ \text{given (\eqref{deltaomega3}), }\\                       
                       & \leqslant 4 m \| W\|^2 \sin^2 \frac{\Delta \omega T}{4} \sum\limits_{j=1}^p |c_j|^2                        
\end{align*}
where $\|W\| = \sup \limits_{x \in \mathbb{C}^m} \frac{\|Wx\|_{\ell_2}}{\|x\|_{\ell_2}}$.
When the matrix $W$ is diagonal, the formula can be improved:
\begin{align*}
\|W(y-y')\|^2_{\ell_2} & = \sum\limits_{k=1}^m \frac{|y_k-y_k'|^2}{\sigma_k^2} \\
                       & \leqslant 4 \sum\limits_{k=1}^m \frac{1}{\sigma_k^2} \left( \sum\limits_{j=1}^p \left| c_j\sin \left( \frac{\omega_j-\omega_j'}{2} t_k\right) \right| \right)^2  \\ \text{ given (\eqref{deltaomega3})}, \\
                       & \leqslant 4 \sin^2 \frac{\Delta \omega T}{4} \sum\limits_{j=1}^p |c_j|^2 \sum\limits_{k=1}^m   \frac{1}{\sigma_k^2}
\end{align*}
So $\epsilon_{\mathrm{grid}}$ can be chosen as:
\begin{equation}
\epsilon_{\mathrm{grid}} = 2 \sqrt{\sum\limits_{j=1}^p |c_j|^2} \sqrt{\sum\limits_{k=1}^m   \frac{1}{\sigma_k^2}}  \sin \frac{\Delta \omega T}{4}
\end{equation}
And conversely given an $\epsilon$, the grid spacing that ensures that there exists a vector that has the correct $\ell_0$ norm is:
\begin{equation}
\Delta \omega = \frac{4}{T} \arcsin \frac{\epsilon}{ 2\sqrt{\sum\limits_{j=1}^p |c_j|^2} \sqrt{\sum\limits_{k=1}^m   \frac{1}{\sigma_k^2}}  }
\end{equation}

\section{Digging in red noise with non-diagonal $W$}
\label{appendix_noise}
\subsection{Short period buried in the noise}

Our method uses the tools of compressed sensing, especially the algorithms to minimize $\ell_1$ norms with the constraint that the reconstructed signal is not too far from the observations (see equation~\eqref{l1}). To the best of our knowledge, the case where the noise is correlated has been considered only in~\cite{arildsen2014}, and is not specialized for Gaussian processes. Here, we introduce a weight matrix and obtain problem~\eqref{BPDNepsilonw}, reproduced here:
\begin{equation*}
  \tag{$\text{BP}_{\epsilon,W}$}
  x^\star =\underset{x \in \mathbb{C}^n}{ \arg \min} \quad \|x\|_{\ell_1} \quad \text{s. t.} \quad \|W(Ax-y)\|_{\ell_2}\leqslant \epsilon
\end{equation*}

To illustrate the interest of choosing an appropriate weight matrix, we will show an example where acknowledging the red noise makes a planet visible.
Let us first consider a data set constructed as follow:
\begin{itemize}
\item The measurement times are those of HD 69830 (74 measurements);
\item The true signal is $y(t)= 1\cos(\frac{2\pi}{7.5} t) + 2\cos(\frac{2\pi}{40}t+2) + 2\cos(\frac{2\pi}{120}t+1)$ m.s\textsuperscript{-1}.
\item The noise is red, with parameters $\sigma_W = 0$, $\sigma_R = 2$ m.s\textsuperscript{-1} and $\tau = 12$ days, where $\sigma_W, \sigma_R$ and $\tau$ are the parameters of the autocorrelation function $R$ defined equation~(\eqref{corrnoise}) reproduced here:
\begin{align*}
\begin{split}
R(\Delta t) &= \sigma_R^2 \e^{-\frac{|\Delta t|}{\tau}}, \quad \Delta t \neq 0 \\
R(0) &= \sigma_W^2 +\sigma_R^2 
\end{split}
\end{align*}
\end{itemize}
The noise defined above is such that its correlation with low frequencies is higher than with high frequencies.

\begin{figure*}
\includegraphics[scale=0.43]{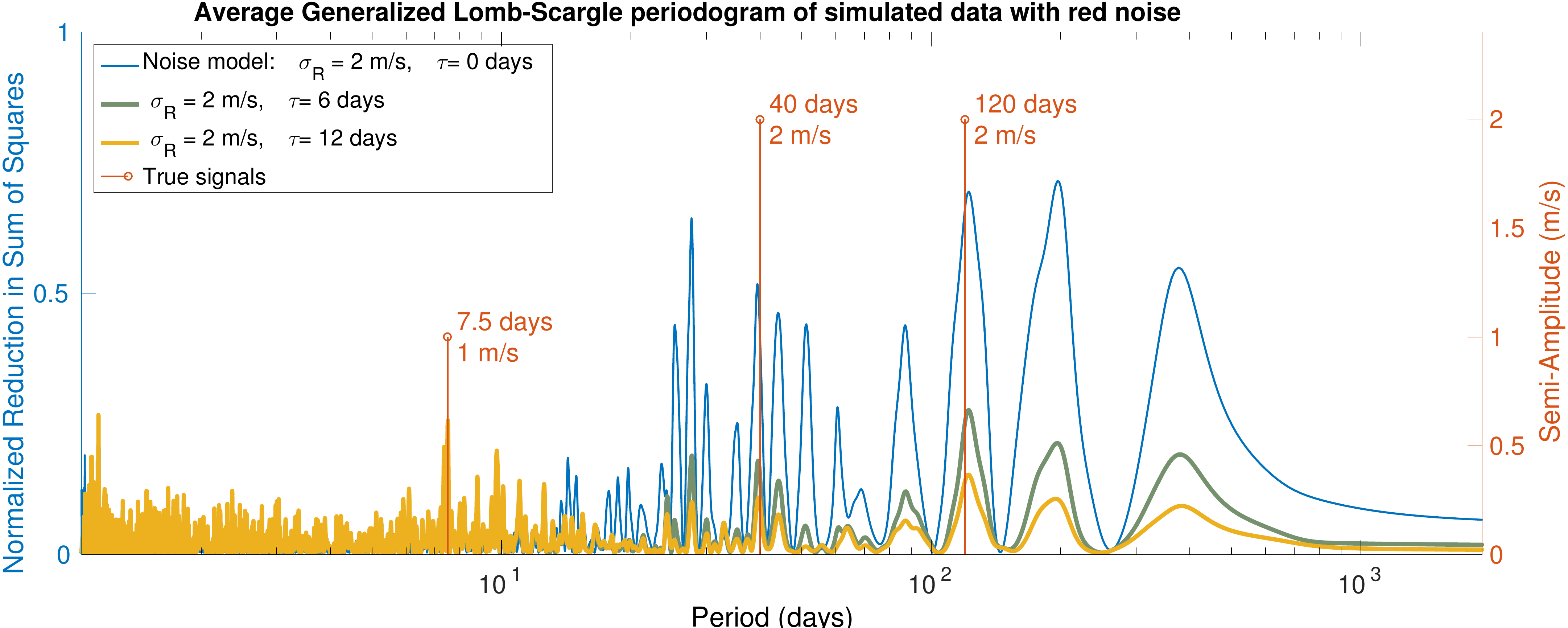}
\noindent
\includegraphics[scale=0.43]{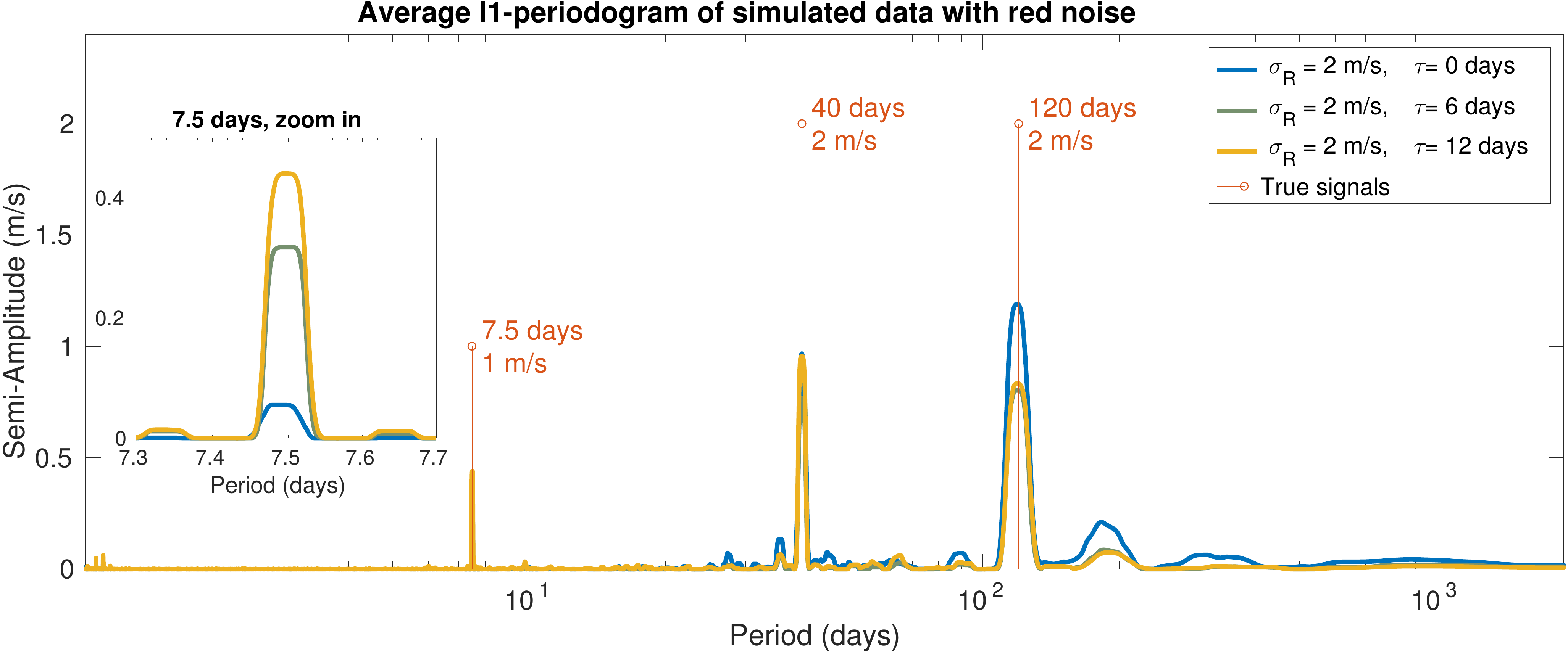}
\caption{Average $\ell_1$-periodogram for 50 data sets generated with red noise of characteristics $\sigma_W = 0, \sigma_ R = 2 $  m.s\textsuperscript{-1} and $\tau = 12$ days according to model~\eqref{corrnoise}. The curves correspond to the solutions of~\eqref{BPDNepsilonw} with different weight matrices $W$ whose parameters are $\sigma_W =0, \sigma_R = 2m/s$ and $\tau = 0$ , 6 or 12 days ( respectively the blue, green and yellow curves).}
\label{bruitcorr_example}
\end{figure*}

We test if changing the weight matrix could allow us to find signals that would not be seen otherwise. To do so, fifty noise time series $(n_k(t))_{k=1..50}$ are generated and the method is applied to each $y_k(t) = y(t) + n_k(t)$ for three different weight matrices, all other parameters being fixed. In each case they are defined according to model~\eqref{corrnoise} with $\sigma_W = 0$, $\sigma_R = 2$ m.s\textsuperscript{-1}  and $\tau$ = 0, 6 or  12 days.
 The grid goes between 0 and 0.95 cycles/day and $\epsilon$ verifies $F_{\chi^2_m} (\epsilon_{\mathrm{noise}}^2) = 0.1$.
The resulting $\ell_1$-periodograms are averaged (see figure~\ref{bruitcorr_example}.b).

 To compare with a classical approach, we also compute classical periodograms for the same signals $y_k(t)$ and average them. For the comparison to be fair, we fit the model parameters $A$, $B$, $C$  in $A \cos \omega t + B \sin \omega t +  C$ to $y(t)$ with the same weight matrices as the ones used above. This gives figure~\ref{bruitcorr_example}.a. If the weight matrix is left diagonal, then the low frequency terms dominate. Using the appropriate noise model gradually reduces the spurious low frequencies.

We stress two features: as the noise model becomes accurate, the short period becomes apparent, which justifies the try of different noise matrices on real radial velocity data sets to see if a peak appears. Secondly, when $W$ is defined with an exponential autocorrelation function, the estimation of the peaks becomes biased: some frequencies will have a tendency to be interpreted by the algorithm as noise. The amplitude of the 120 days periodicity is then under-estimated. This bias could prevent from finding small amplitudes when using non diagonal weight matrices. When the number of frequency in the signal increases, the bias becomes more complicated. In order to mitigate this effect, we suggest to decrease the value of $\epsilon$ when testing different noise model. Thus the model ``sticks'' to the observations and if a periodicity truly is in the data the chance of it being too under-estimated decreases. This is why we took $\epsilon_{\mathrm{noise}}$ such that  $F_{\chi^2_m} (\epsilon_{\mathrm{noise}}^2) = 0.1$ and not $F_{\chi^2_m} (\epsilon_{\mathrm{noise}}^2) = 0.999$, which would reject more signals in the residual.

\subsection{No automatic procedure so far}

Here the improvement due to an appropriate handle of the noise is seen by eye. One could wonder if a simple criterion could allow to chose an appropriate weight matrix automatically. In all cases when the algorithm has converged we have $\|W(Ax-y) \|_{\ell_2} = \epsilon$ to a certain tolerance, or $x=0$. Looking at the $\chi^2$ of the residuals as usual is then not appropriate. 

As in all cases the columns of matrix $WA$ and the weighted observations $Wy$ are normalized. Therefore the problem always comes down to minimizing 
\begin{equation}
x^\star =  \arg \min\limits_{x \in \mathbb{R}^n} \|x \|_{\ell_1} \quad s.t. \quad \|A'x-y'\|_{\ell_2} \leqslant \epsilon
\end{equation}
 where $A'$ has normed columns and $y'$ is a unitary vector. It is then tempting to see if there is a correlation between the $\ell_0$ or $\ell_1$ norm of $x^\star$ and the success of the method. Unfortunately, this is not the case. Whether there is an automatic way to select the appropriate weight matrix remains an open question.  

\section{Spurious tallest peak of the GLS periodogram}
\label{appendix_wrongpeak}

In this section we show examples where the initial highest peak of the periodogram is spurious due to aliasing. We take the 74 measurement dates of HD 69830 and generate 500 systems with three circular orbits with the following properties:
\begin{itemize}
\item The amplitudes are those of the three Neptunes of HD 69830 (2.2, 2.66 and 3.51 m.s\textsuperscript{-1}).
\item The periods $P_1$, $P_2$, $P_3$, are selected uniformly in $\log P$ in the range 1.2 to 2000 days
\item The phases are uniformly distributed on $[0,2\pi]$.
\item The noise standard deviation is 0.6 m.s\textsuperscript{-1}
\end{itemize}
We compute the number of times the maximum peak of the GLS and $\ell_1$-periodogram are spurious. The criterion we take for failure is when the frequency of the highest peak and any of the three true frequencies is greater than the inverse of the total observation time, that is $\left| 1/P_{1,2,3}-1/P_{\mathrm{max}}\right| > 1/T_{\mathrm{obs}}$.

\begin{figure*}
\noindent
\centering
\begin{tikzpicture}
\path (0,2) node[above right]{\includegraphics[width=8.5cm]{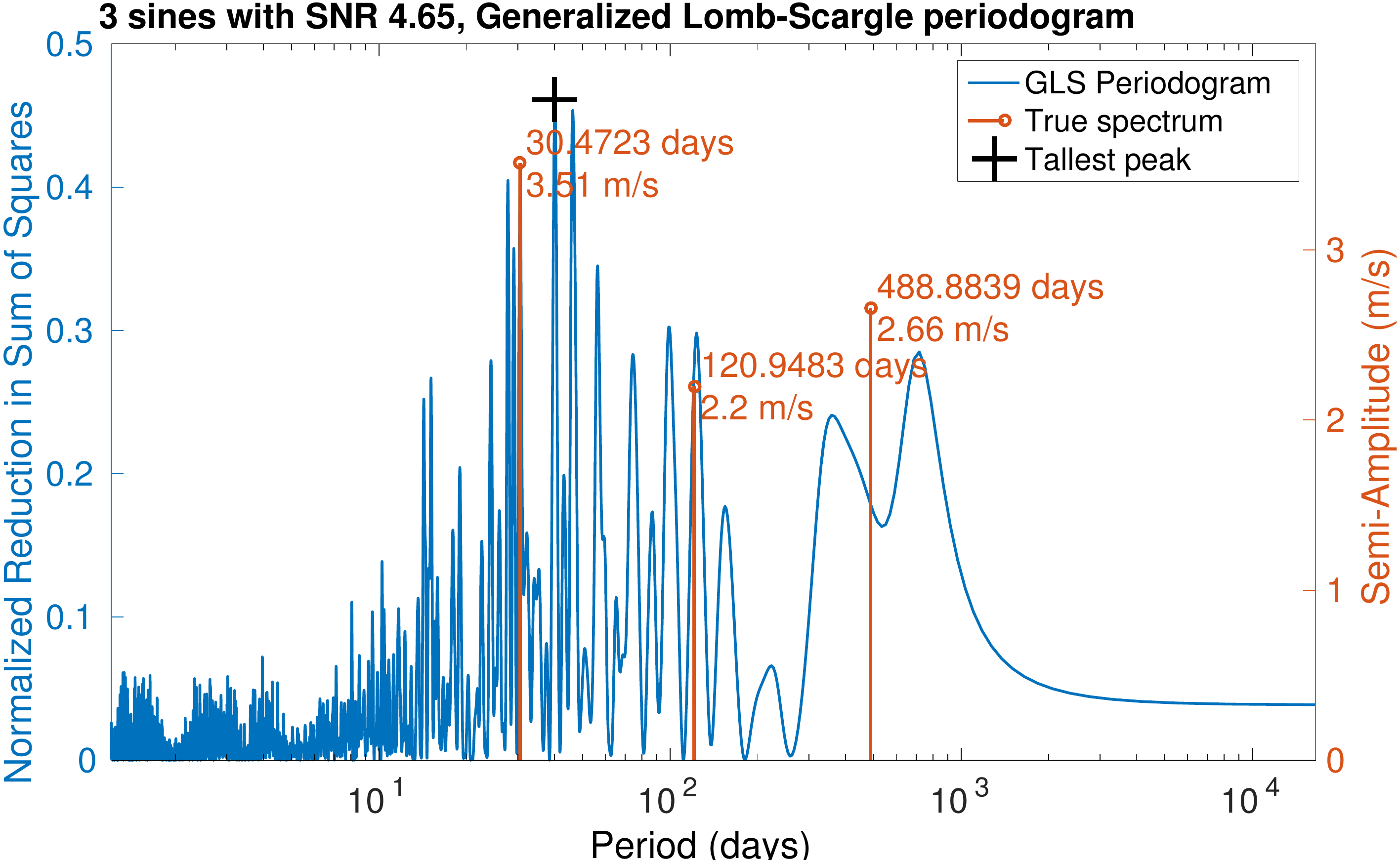}};%hd69830_fratio.png
\path (1.2,6.5) node[above right]{a)};
\path (9,2) node[above right]{\includegraphics[width=8.5cm]{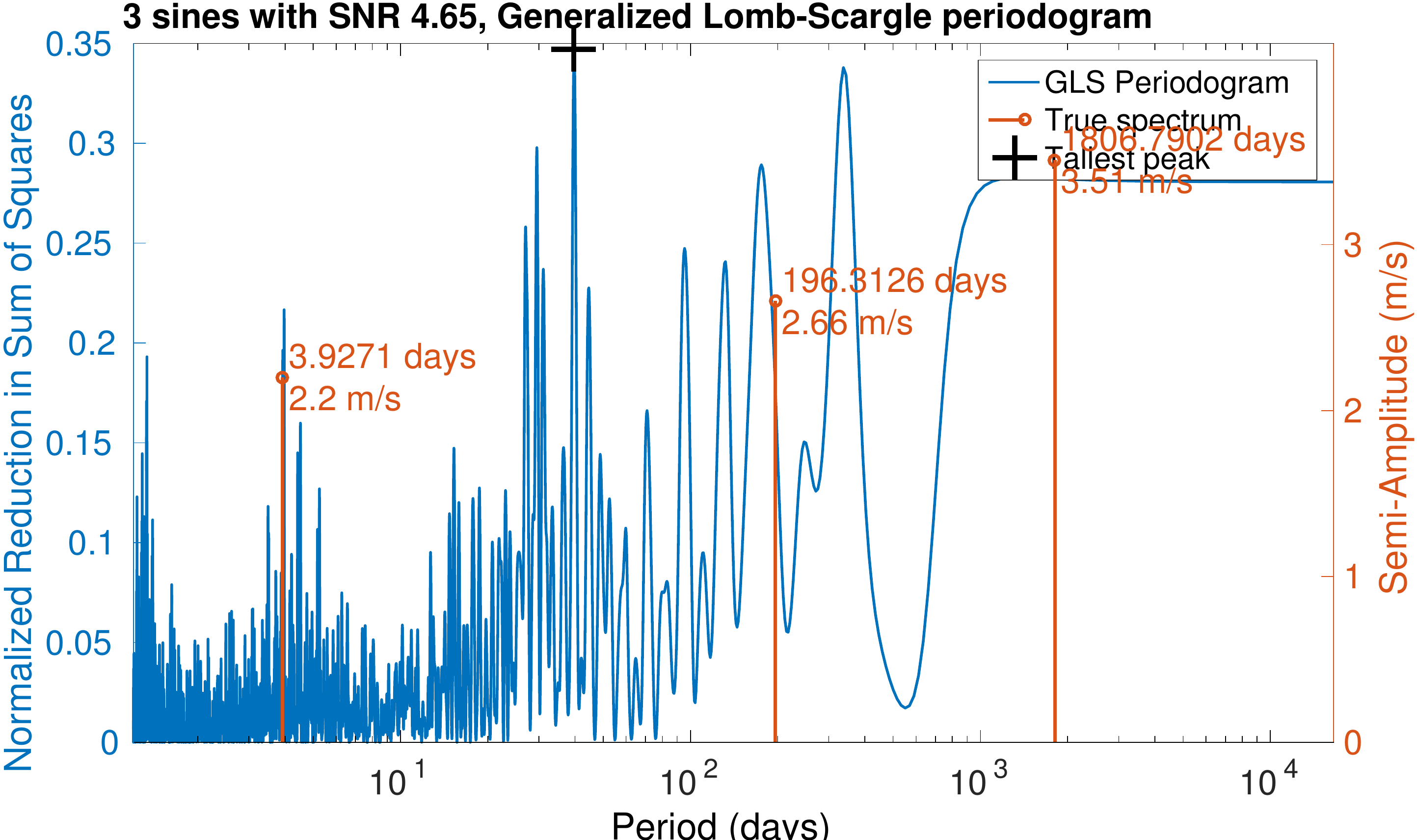}};%55cnc_fratio.png
\path (10.2,6.4) node[above right]{b)};

\begin{scope}[yshift=-5cm]
\path (0,2) node[above right]{\includegraphics[width=8.1cm]{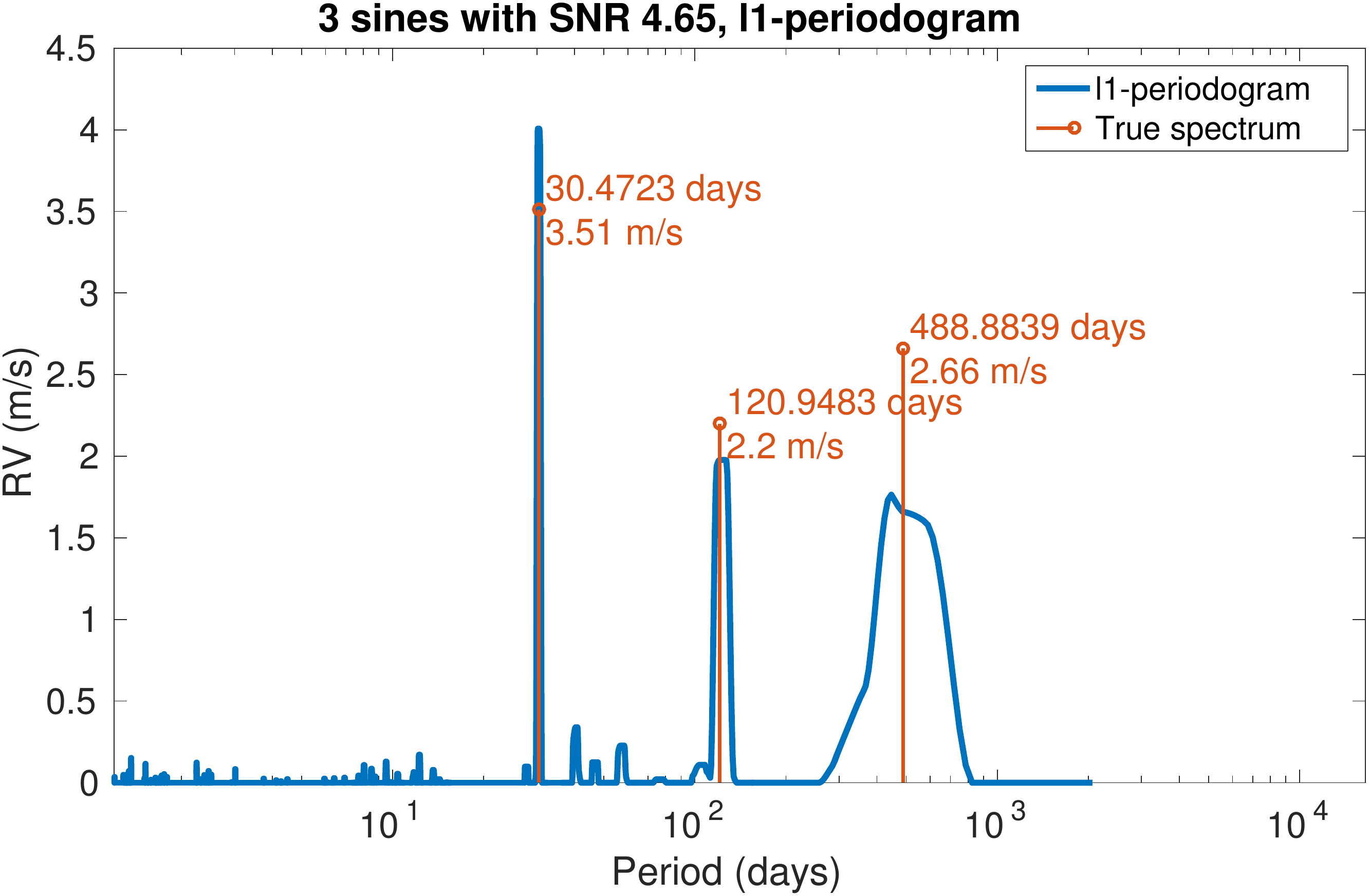}};%hd69830_fratio.png
\path (9,2) node[above right]{\includegraphics[width=8.1cm]{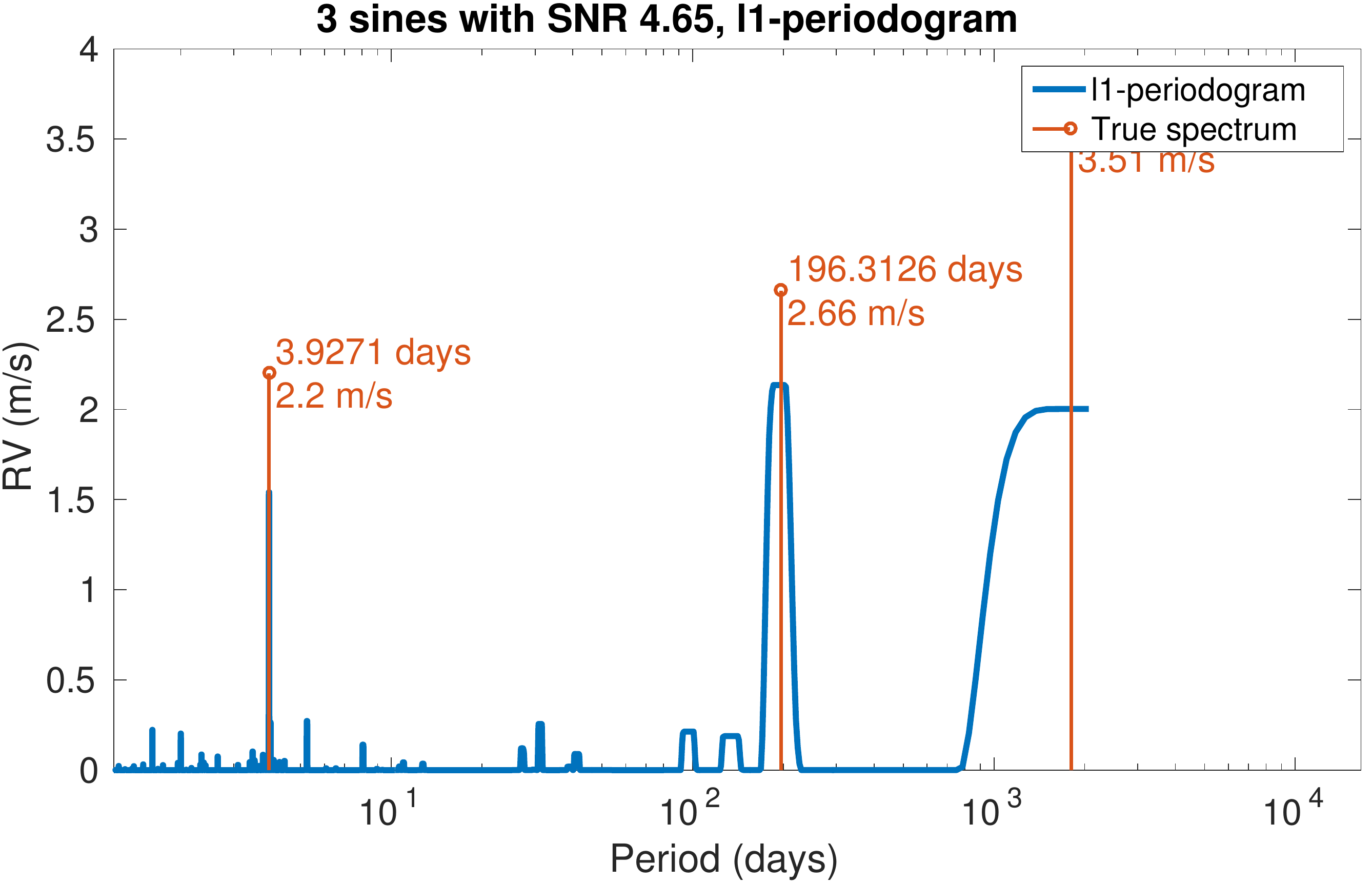}};%55cnc_fratio.png
\end{scope}
\begin{scope}[yshift=-12cm]
\path (0,2) node[above right]{\includegraphics[width=8.5cm]{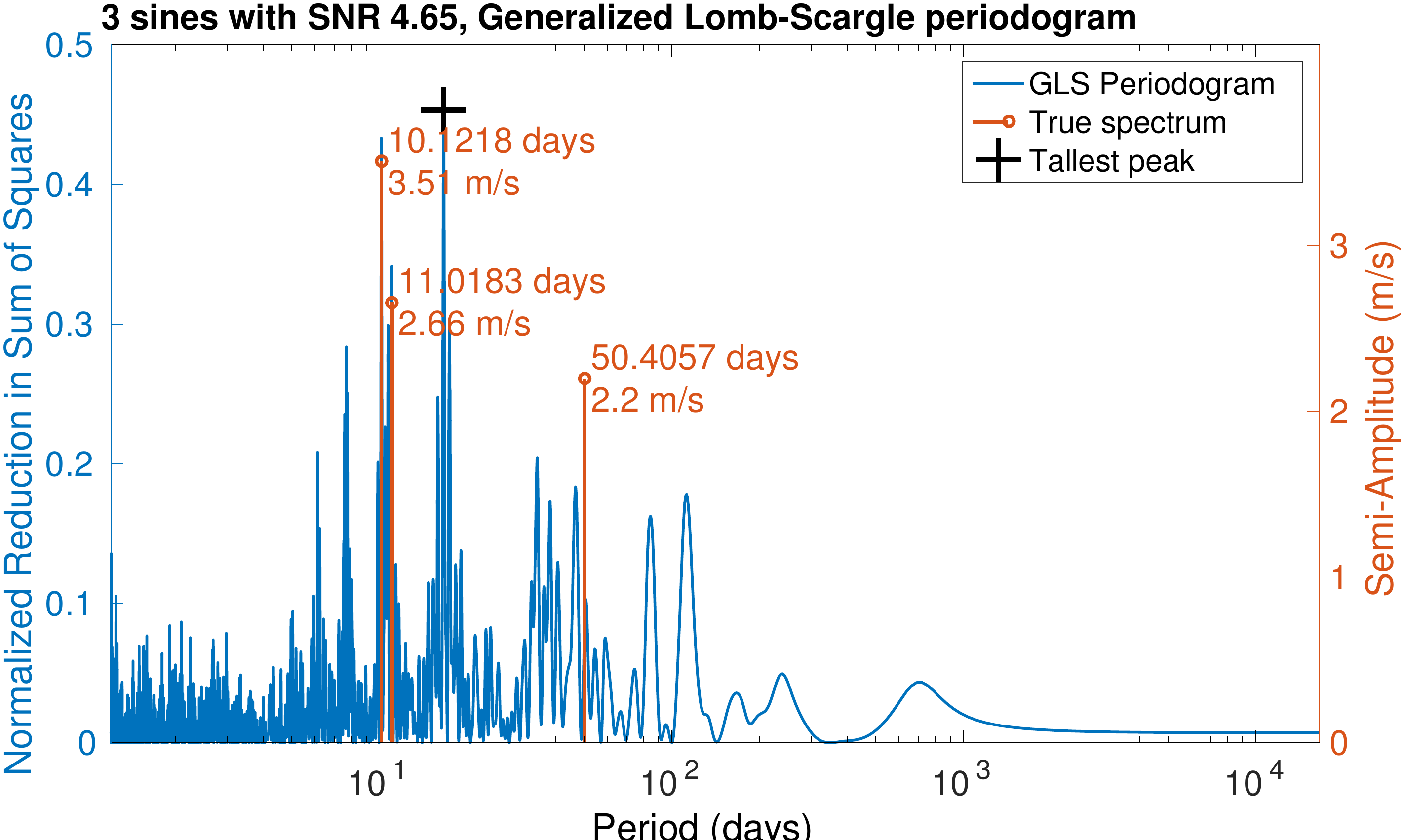}};%hd69830_fratio.png
\path (1.2,6.4) node[above right]{c)};
\path (9,2) node[above right]{\includegraphics[width=8.5cm]{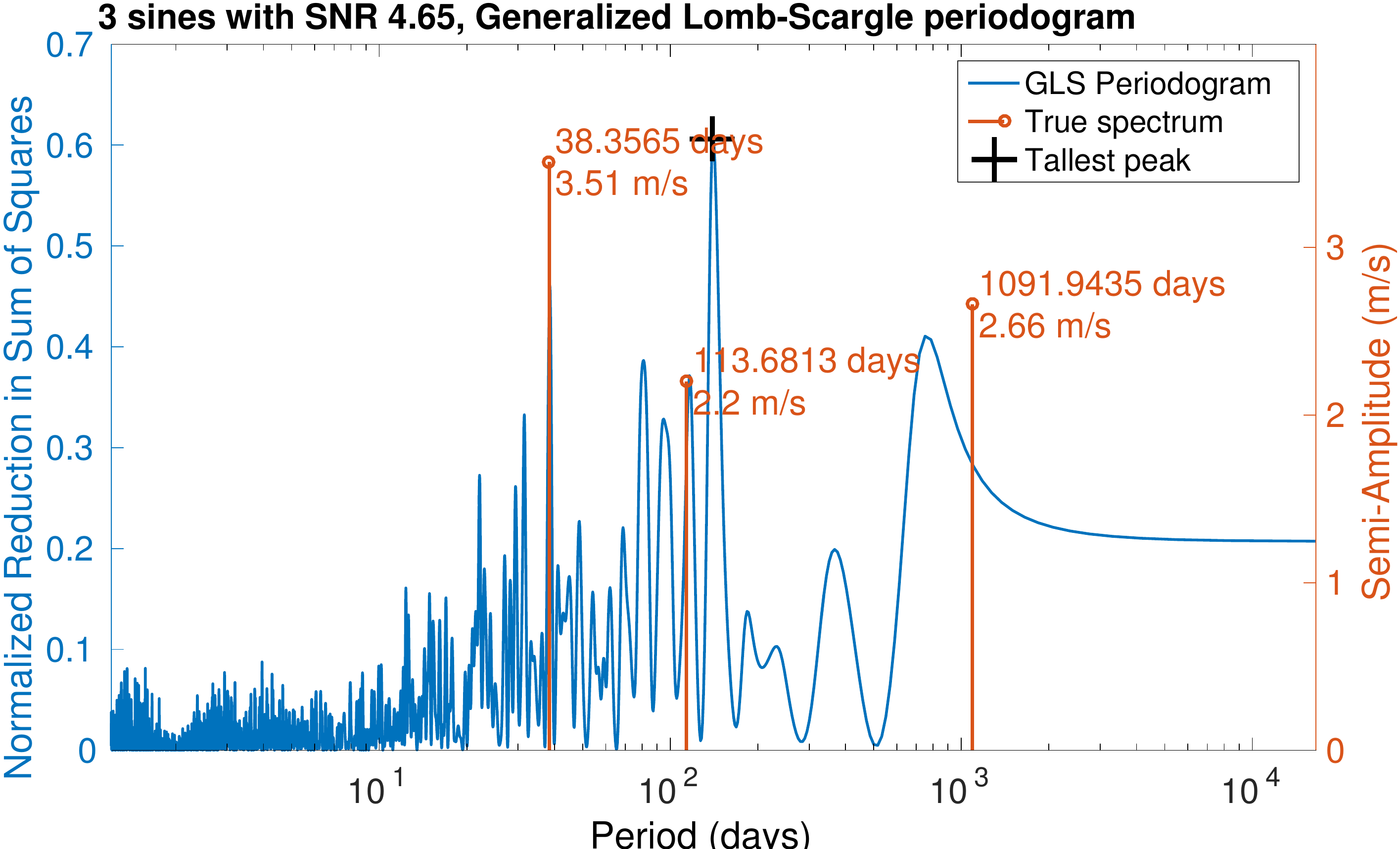}};%55cnc_fratio.png
\path (10.2,6.4) node[above right]{d)};
\end{scope}
\begin{scope}[yshift=-17cm]
\path (0,2) node[above right]{\includegraphics[width=8.1cm]{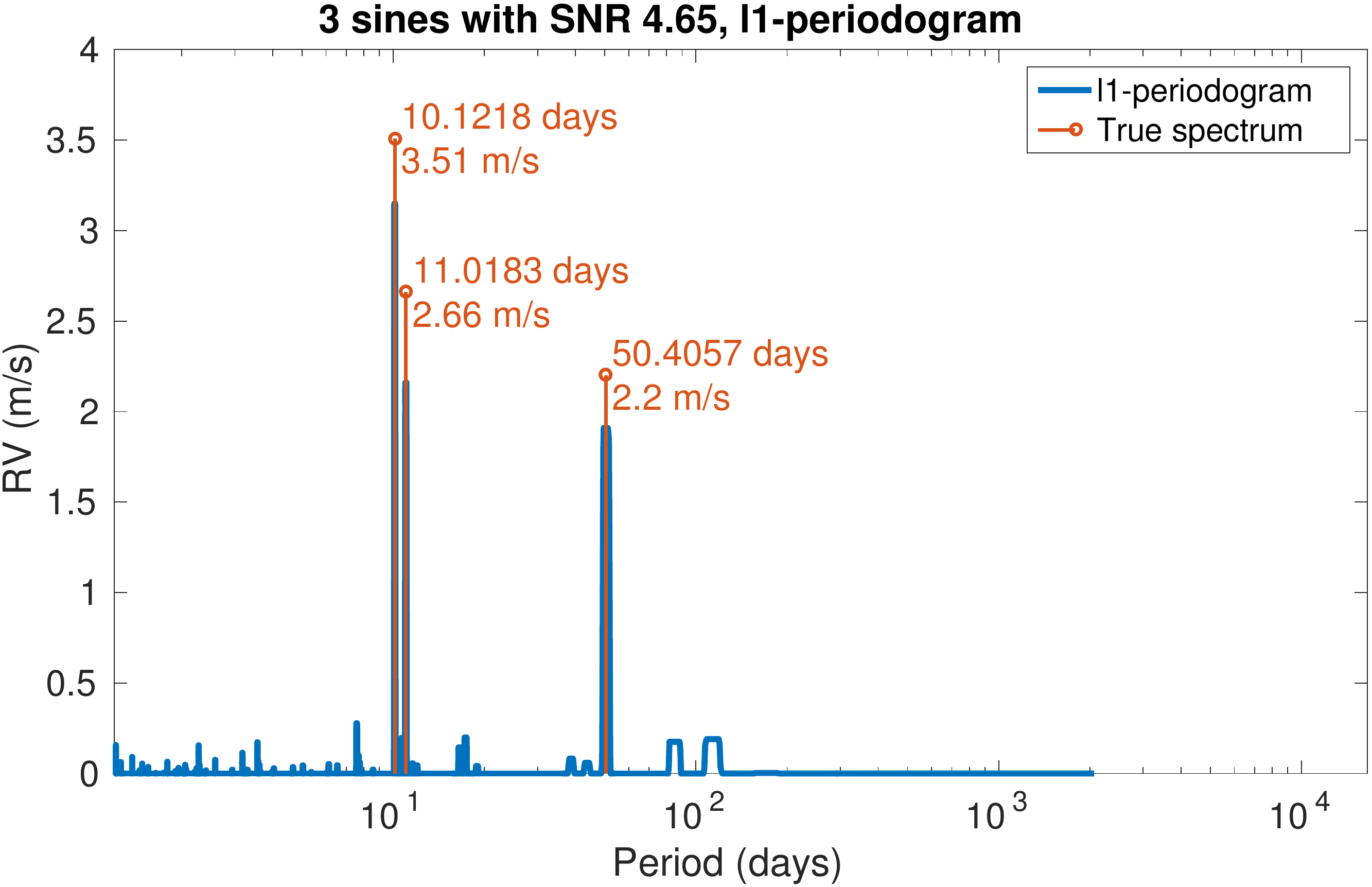}};%hd69830_fratio.png
\path (9,2) node[above right]{\includegraphics[width=8.1cm]{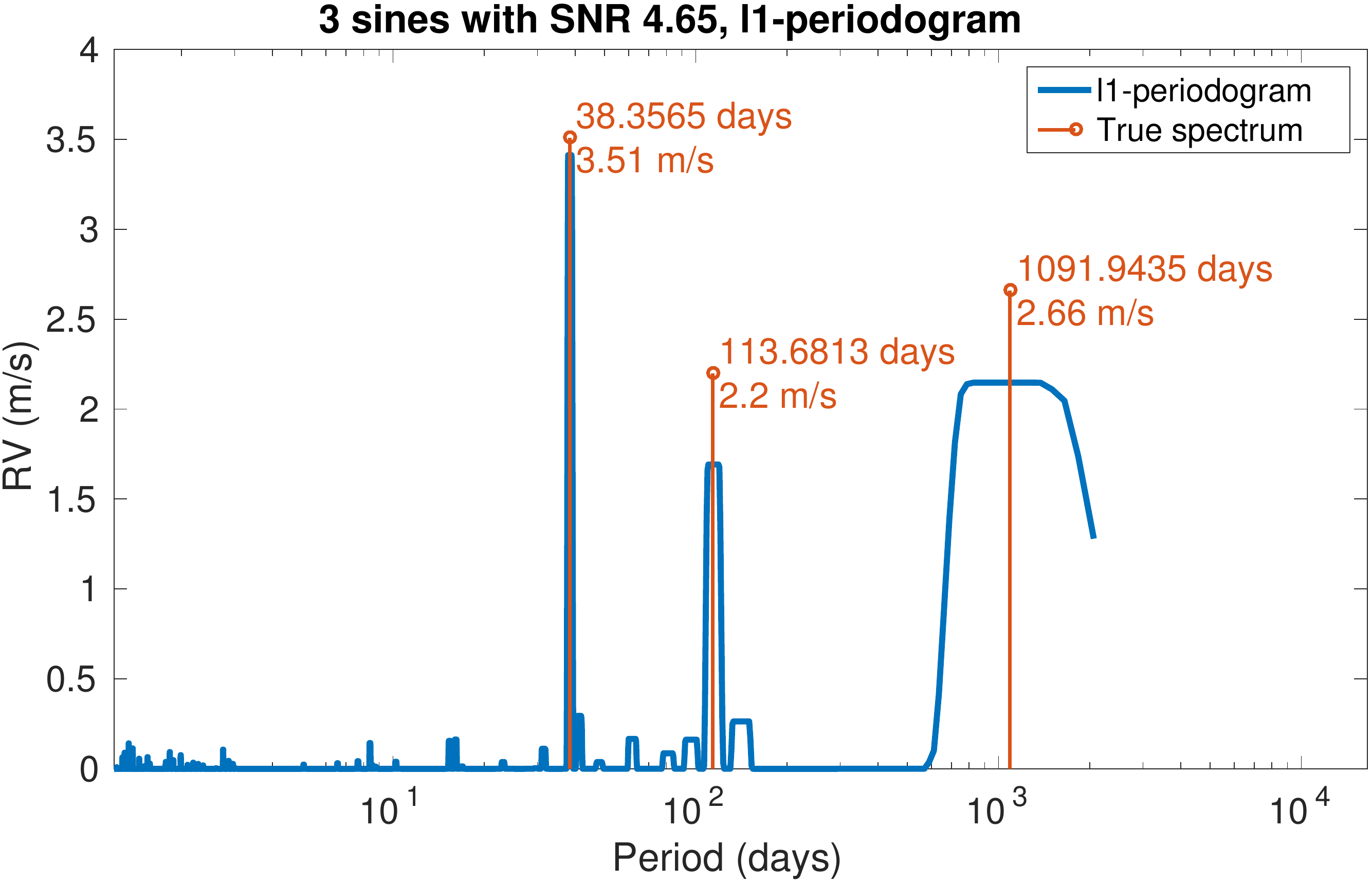}};%55cnc_fratio.png
\end{scope}
\end{tikzpicture}
\caption{Peak amplitudes and associated FAPs for the four systems analysed}
\label{wrongpeak}
\end{figure*}

Figure~\ref{wrongpeak} shows the GLS periodogram and $\ell_1$-periodogram of representative cases where the highest peak of the GLS periodogram is spurious. In these conditions, when searching for periods in the 1.2-2000 days with the periodogram, we find that the strongest peak is spurious in 33  cases out of five hundred simulations, while the tallest peak of the $\ell_1$-periodogram only was incorrect in two cases. In those, the GLS periodogram was also failing. 

\begin{figure*}
\centering
\noindent
\hspace{-0.35cm}
\begin{tikzpicture}
\path (0.15,0) node[above right]{\includegraphics[scale=0.43]{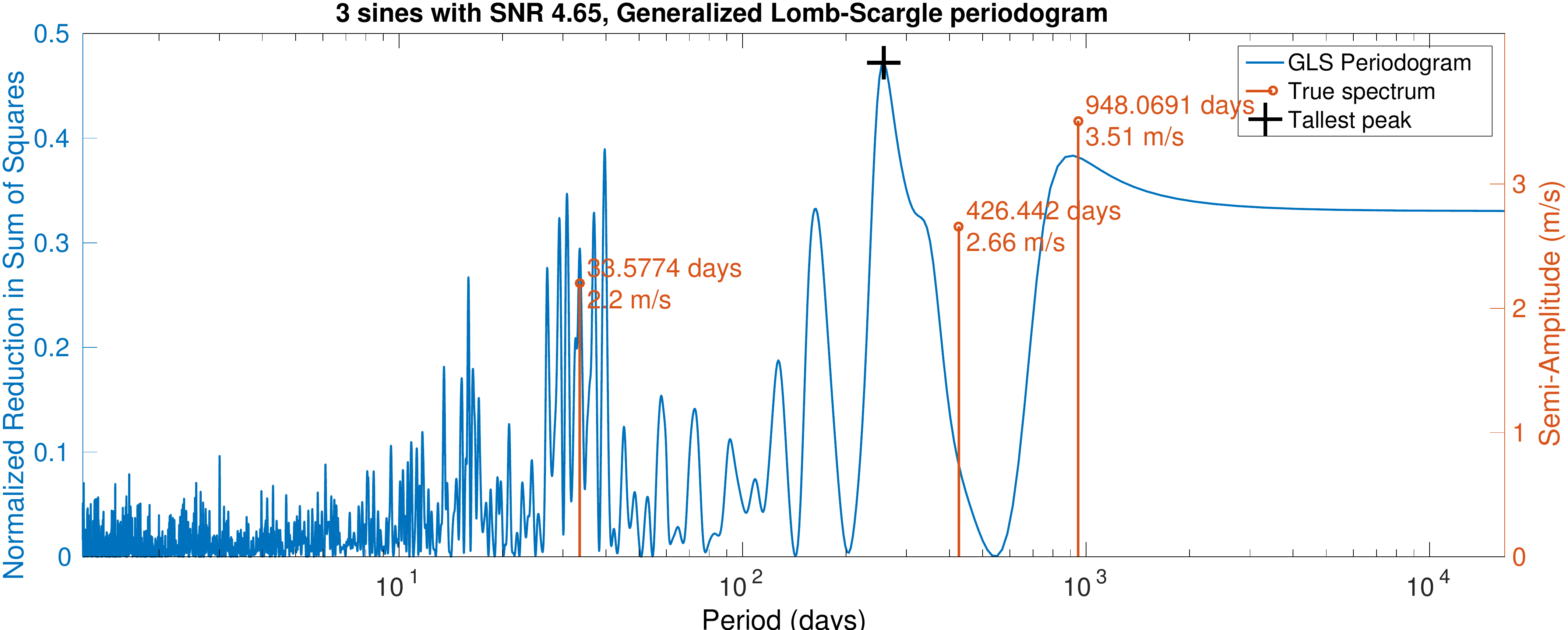}};%hd69830_fratio.png
\path (1.2,7.2) node[above right]{a)};
\begin{scope}[yshift=-7.6cm]
\path (0.15,0) node[above right]{\includegraphics[scale=0.43]{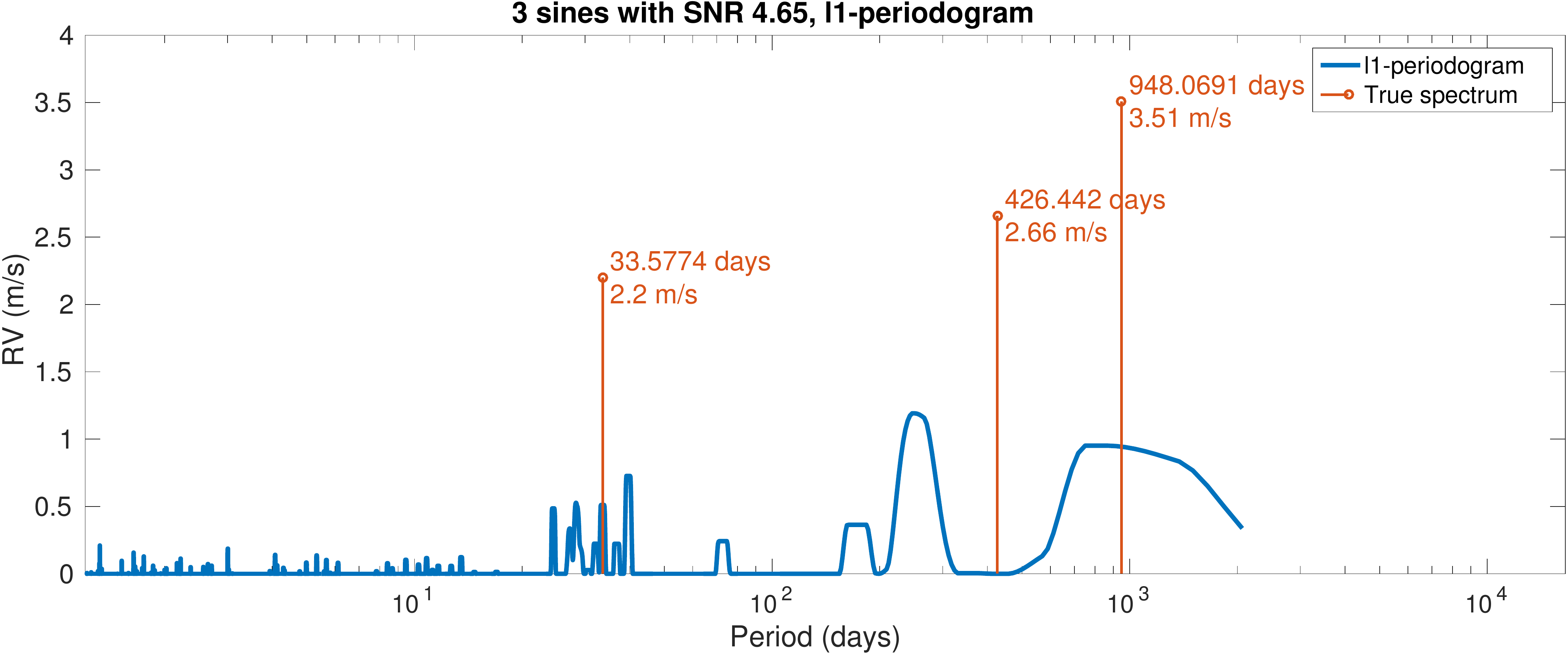}};
\path (1.2,7.2) node[above right]{b)};
\end{scope}
\end{tikzpicture}
\caption{Failure of the GLS (a) and $\ell_1$ (b) periodograms.}
\label{l1failure}
\end{figure*}

An interesting feature of the cases where the $\ell_1$-periodogram fails is that one can see that the solution is not sparse. This is a very useful property we observed empirically: we haven't found any occurrence of $\ell_1$-periodogram that looks clean, with well separated clear peaks, where one of the peaks was completely spurious. We display one of the two failures of the $\ell_1$-periodogram on figure~\ref{l1failure}. First of all in neither the GLS nor the $\ell_1$-periodogram leads the observer completely astray. Secondly, we see that as opposed to the $\ell_1$-periodogram of the systems studied here, the figure is not clean, which should invite the analyst to a certain suspicion.


\section{Fitting the ancillary measurements}
\label{appendix_stellarnoise}
\begin{figure}
\flushleft
\includegraphics[scale=0.30]{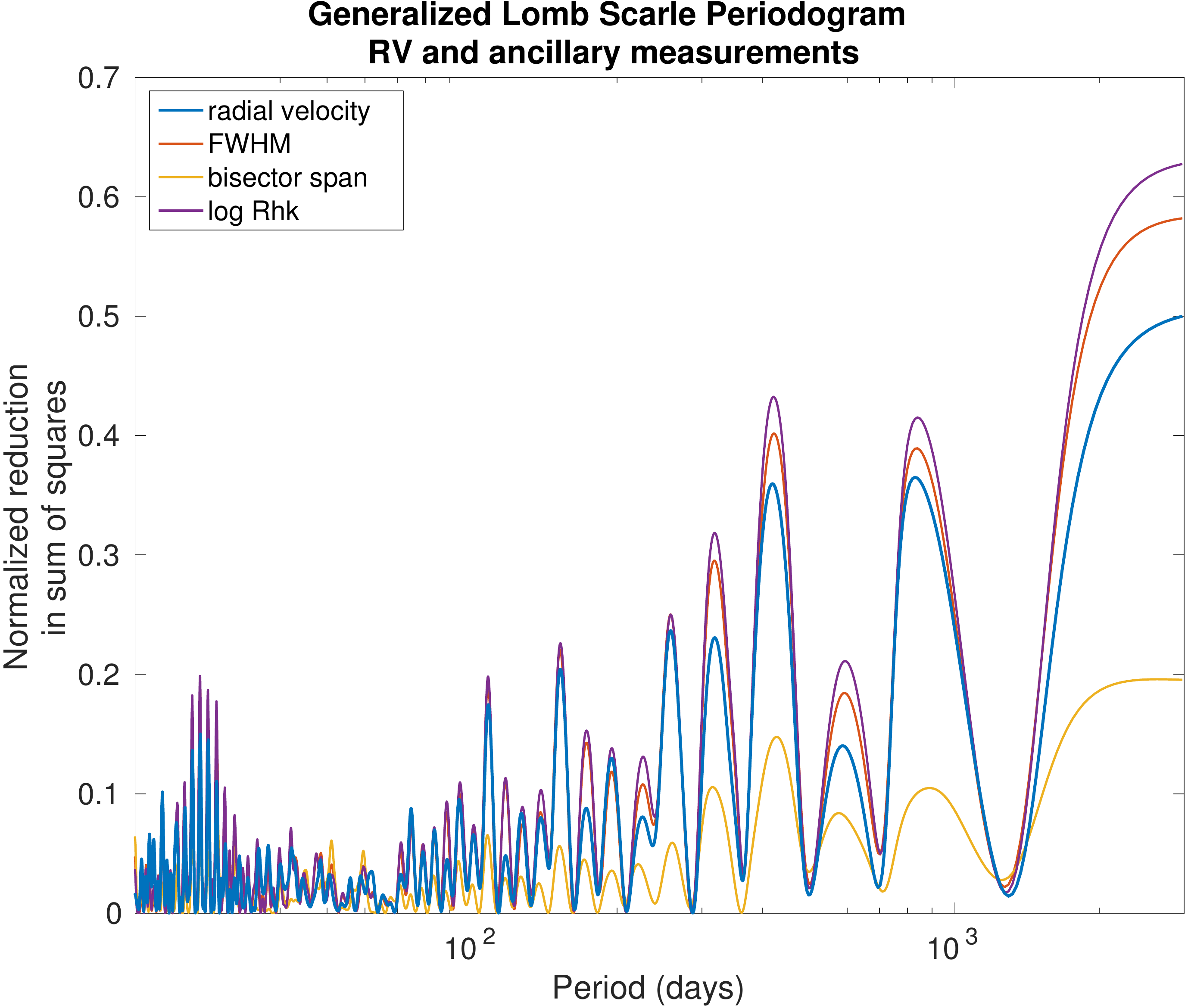}
\caption{Generalized Lomb-Scargle periodogram of radial velocity and ancillary measurements at low frequencies }
\label{stellar}
\end{figure}

In section~\ref{activity} we suggest to fit the activity indicators to the radial velocity time series. The present discussion wishes to give a justification to this approach. The idea is to exploit the possible correlations between radial velocity and ancillary measurements when the star is active. For instance, on the first system of the RV Fitting Challenge~\citep{dumusque2016} where activity dominates the signal, the radial velocity, FWHM, bisector span and $\log R_{HK}'$ exhibit very similar features at low frequency (see figure~\ref{stellar}).

Let us approximate the error made when fitting an ancillary indicator. We consider the radial velocity signal $y(t)=P(t)+a(t)+\epsilon(t)$ where $P(t)$ is due to a planetary companion,  $a(t)$  is a deterministic signal due to activity and  $\epsilon$ is a Gaussian noise of covariance matrix $V$. We also consider an ancillary measurement $z(t) = a(t) + \epsilon'$ where $\epsilon'(t)$ is another Gaussian  noise of covariance matrix $V$. If we fit $z(t)$ to $y(t)$, we obtain (dropping the $t$ notation):
\begin{align}
y_{\mathrm{detrend}}& = y - y_{\mathrm{fit}}  = y - \frac{z^T V^{-1} y}{ z^T V^{-1} z} z \\
y_{\mathrm{detrend}}& = y - \frac{(a+\epsilon')^T V^{-1} (P+a+\epsilon)^T}{(a+\epsilon')^T V^{-1}(a+\epsilon) } (a+\epsilon').
\end{align}
 We assume that the noise is small compared to $a$, which allows to develop the denominator at first order in $\epsilon$ and $\epsilon'$
%\begin{align}
%\mathbb{E}\{y_{\mathrm{fit}}\} = & \mathbb{E}\{ \frac{a+\epsilon}{a^TWa}   a^T W^{-1}a + a^T W P + a^T W \epsilon' \\ & \epsilon^T W a + \epsilon^T W P   \epsilon^T  W^{-1} \epsilon'\}
%\end{align}
\scriptsize
\begin{align*}
y_{\mathrm{fit}} \approx \frac{(a+\epsilon')^T V^{-1} (P+a+\epsilon)}{a^T V^{-1} a} \left( 1- \frac{\epsilon'^TV^{-1}a}{a^TV^{-1}a } -\frac{\epsilon ^TV^{-1}a}{a^TV^{-1}a } \right)(a+\epsilon')
\end{align*}
\normalsize
%As $\epsilon$ is Gaussian of covariance matrix $V=W^{-1}$,
%After simplifications, 
%\begin{align}
%\mathbb{E}\{y_{\mathrm{fit}}\} \approx \frac{a^T W a + a^T W P }{a^TWa }a
%\end{align}
%\begin{align}
%µ\mathbb{E}\{y_{\mathrm{fit}}\} \approx \left( \frac{a^T W a + a^T W P }{a^TWa } - \frac{2}{a^TWa} \right)a
%\end{align}
After developing that  expression at first order in $\epsilon$ and $\epsilon'$, we compute its mathematical expectancy taking into account only the zero order, $\epsilon^2$ and $\epsilon'^2$ coefficients.  In the simple case where the noise is i.i.d of variance $\sigma^2$ we obtain:
%\begin{align}
%\mathbb{E}\{y_{\mathrm{fit}}\} \approx \frac{a^T a/ \sigma^2 + a^T P/ \sigma^2}{a^Ta/\sigma^2 %+1}a =  \frac{\| a\|_{\ell_2} + a^T P}{\| a\|_{\ell_2} + \sigma^2} a
%\label{expectancy_fit}
%\end{align}
\begin{align}
\mathbb{E}\{y_{\mathrm{fit}}\} & \approx \frac{\sigma^2}{\|a \|^2} P + \\ & \left(1+ \frac{ a^T P}{\| a\|_{\ell_2}^2 } -\frac{2\sigma^2}{\| a\|_{\ell_2}^2} -\frac{\|P \|_{\ell_2} \sigma^2}{\|a\|_{\ell_2}^3} -\frac{a^T P \sigma^2}{\|a\|_{\ell_2}^4} \right) a 
\label{expectancy_fit}
\end{align}
%\left(\frac{a^T a/ \sigma^2 + a^T P/ \sigma^2}{a^Ta/\sigma^2 } - 2\frac{1}{a^Ta/\sigma^2}\right) a \\ 
We would like $y_{\mathrm{fit}}$ to be as  close to $a$ as possible. This will be better satisfied as the correlation $a^TP$ and as the signal to noise $\sigma^2/a$ decrease. The fact that a term $a^T P$ appears in the equation above should not be surprising. The mutual coherence defined section~\ref{limitations} grasps that the correlation between the parts of the model is an obstacle to recovery of the true signals. 

For the RV Fitting Challenge, not only have we fitted one activity indicator but several. We point out that this approach is consistent with~\cite{rajpaul2015}. Indeed, they consider that the activity-induced variations of the measurements depend linearly on an underlying zero-mean Gaussian process $G(t)=F^2(t)$ and its derivative  $\dot{G}(t)$, where $F(t)$ is the fraction of the sphere covered with spots. The evolution of the indicators is modelled by formulae (14-16), reproduced below.
\begin{align}
&\Delta RV = V_c G(t) + V_r \dot{G}(t);  \\
&\log R_{HK}' = L_c G(t) \\
&\mathrm{BIS} = B_c G(t) + B_r \dot{G}(t)
\end{align}
for some constants $V_c,V_r,L_c,B_c,B_r$.
This means that for a given realization $(g,g')$ of $(G(t),\dot{G}(t))$, the subspace generated by the $\log R_{HK}'$ and the bisector span BIS is the same as the space generated by $g,g'$. So according to that model, projecting the radial velocity onto $(\log R_{HK}'$,BIS) is equivalent to projecting onto  $(g,g')$.

\begin{figure}
\flushleft
\includegraphics[scale=0.33]{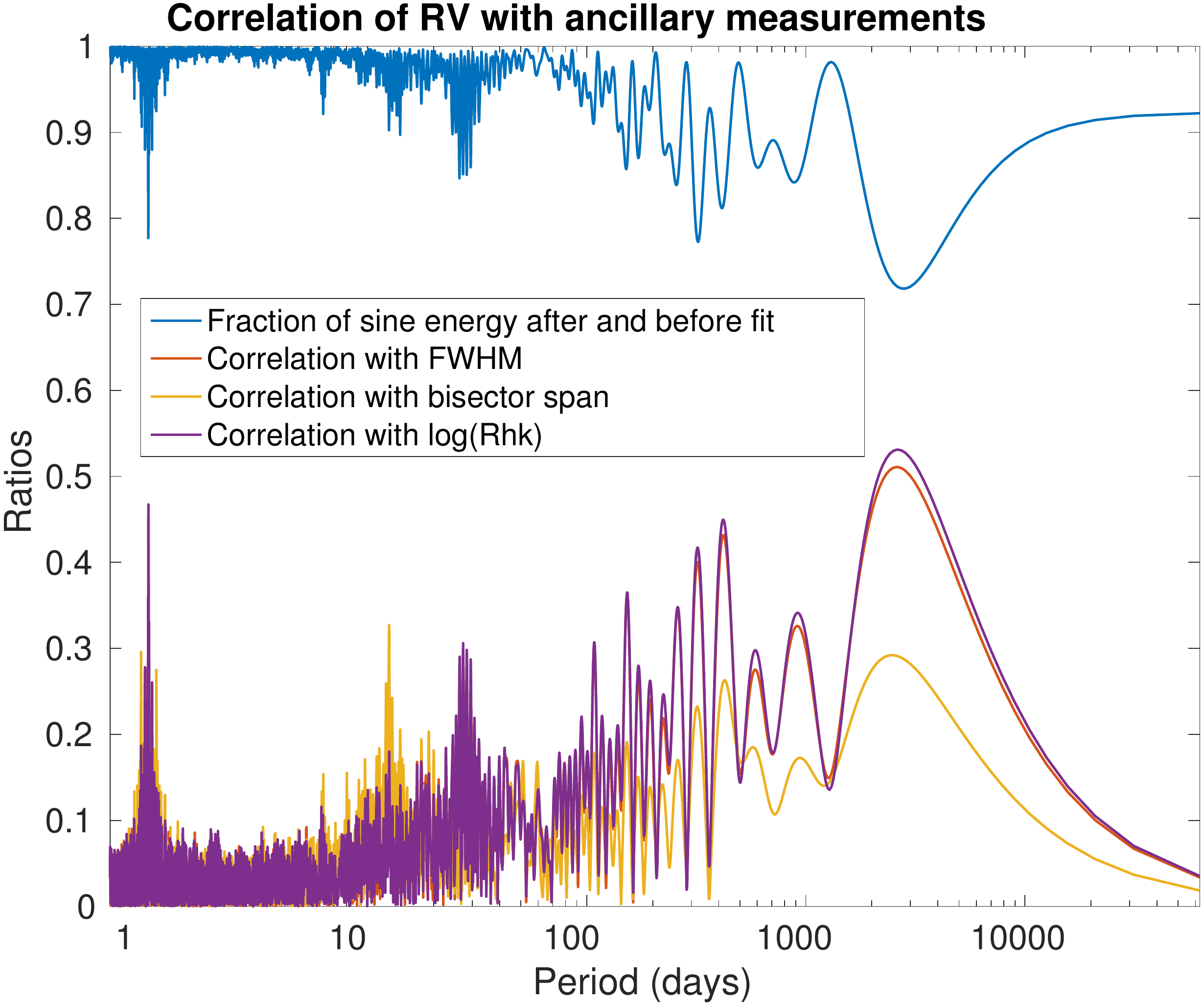}
\caption{Energy of a cosine function after the fit of the FWHM, bisector span, $\log R_{hk}'$ and a constant}
\label{fraction}
\end{figure}

However, there is an uncertainty on the behaviour of the ancillary measurements and additional noise. We have to decide if fitting an uncertain model is better than working with the raw data. One thing that could happen is that fitting the combination of the three ancillary measurements would greatly change the spectral content of the radial velocity time series by absorbing some frequencies, potentially due to planets.
To estimate this risk, we first compute the term $a^T P / \| a\|^2_{\ell_2}$ in equation~\eqref{expectancy_fit}, assuming the signal $y=P=\e^{\ii \omega t}$ is a pure harmonic of amplitude 1 m/s. Here $a$ designates the FWHM, Bisector span or $\log R_{HK}'$ respectively the red, yellow and purple curves figure~\ref{fraction}. We also compute the fraction of the energy of the signal  before and after the fit of the three ancillary measurements simultaneously, that is:
\begin{align}
\rm{Fraction}(\omega) = \frac{ (y_{\omega}-y_{\mathrm{fit}})^T V^{-1} (y_{\omega}-y_{\mathrm{fit}})}{y_{\omega}^T V^{-1} y_{\omega}}
\end{align}
this one is represented by the blue curve figure~\ref{fraction}.
for the system analysed section~\ref{activity}. Only 15\% of the energy is absorbed in general, with a maximum of 27\% at a period of 2000 days. The peaks at 25 and 12.5 days correspond to the rotation period of the star and its first harmonic, which are expected to be correlated with the radial velocity and ancillary measurements.

This discussion does not intend to provide strong statistical arguments, but rather to show that the spectral content should not be too affected by fitting the FWHM, bisector span and $\log R_{hk}'$.

% Don't change these lines
\bsp	% typesetting comment
\label{lastpage}
\end{document}